\documentclass[lettersize,journal]{IEEEtran}
\usepackage{amsmath,amsfonts}
\usepackage{algorithm}
\usepackage{array}
\usepackage[caption=false,font=normalsize,labelfont=sf,textfont=sf]{subfig}
\usepackage{textcomp}
\usepackage{stfloats}
\usepackage{url}
\usepackage{verbatim}
\usepackage{graphicx}
\usepackage{cite}
\usepackage[table]{xcolor}
\usepackage{tabularx}
\usepackage{multirow}
\usepackage{makecell}
\usepackage{bigstrut}
\usepackage{booktabs}
\usepackage{colortbl}
\usepackage{pifont}
\usepackage{amssymb}
\usepackage{algpseudocode}
\usepackage{threeparttable}

\newtheorem{definition}{Definition}
\floatname{algorithm}{Algorithm}

\definecolor{GoogleRed}{RGB}{244, 66, 60}
\definecolor{GoogleGreen}{RGB}{10, 168, 88}
\definecolor{GoogleBlue}{RGB}{45, 133, 240}
\definecolor{GoogleYellow}{RGB}{255, 187, 50}
\definecolor{Gray}{RGB}{128,128,128}

\usepackage{comment}
\usepackage{pifont}
\usepackage{newfloat}
\usepackage{listings}

\begin{document}

\title{MEASER: Malware embedding attacks on open-source LLMs}

\author{Ming~Tan*, Wei~Li*, Hailong~Ma, Tao~Hu, Aodi~Liu, Qian~Chen, and Zilong~Wang,~\IEEEmembership{Member,~IEEE}
\thanks{*These authors contributed equally to this work.}
\thanks{Manuscript received xxx, xxxx; revised xxx, xxxx. \textit{(Corresponding author: Hailong~Ma and Tao~Hu )}}
\thanks{Ming~Tan, Hailong~Ma and Tao~Hu are with Information Engineering University, Zhengzhou 450002, China, also with Laboratory of Cyberspace Endogenous Security, Henan Province, Zhengzhou 450000, China, also with National Key Laboratory of advanced communication networks, Shijiazhuang 050000, China, and also with Key Laboratory of Cyberspace Security Ministry of Education of China, Zhengzhou 450000, China. (e-mail: t4nming@163.com; longmanclear@163.com; hutaondsc@163.com.)}
\thanks{Aodi~Liu are with Henan Province Key Laboratory of Information Security, Information Engineering University, Zhengzhou 450004, China (e-mail ladyexue@163.com).}
\thanks{Wei~Li, Qian~Chen and Zilong~Wang are with the State Key Laboratory of Integrated Service Networks, Institute of Mathematics and Interdisciplinary Sciences, Xidian University, Xi'an, 710126, China (e-mail: wei.li@stu.xidian.edu.cn; chenqian02@xidian.edu.cn; zlwang@xidian.edu.cn).}
}
\markboth{IEEE Transactions on Dependable and Secure Computing,~Vol.~xx, No.~x, xxx~20xx}%
{Tan \MakeLowercase{\textit{et al.}}: MEASER: MEAs on open-source LLMs}

%


\maketitle

\begin{abstract}
Open-source large language models (LLMs) have demonstrated considerable dominance over proprietary LLMs in resolving neural processing tasks, thanks to the collaborative and sharing nature. Although full access to source codes, model parameters, and training data lays the groundwork for transparency, we argue that such a full-access manner is vulnerable to MEAs, and their ill-effects are not fully understood. In this paper, we conduct a systematic formalization for MEAs on open-source LLMs by enumerating all possible threat models associated with adversary objectives, knowledge, and capabilities. Therein, the threat posed by adversaries with internal knowledge, who inject payloads and triggers during the model sharing phase, is of practical interest. We go even further and propose the first MEA against open-source LLMs, dubbed MEASER, which wields impacts through identifying targeted parameters, embedding payloads, injecting triggers, and executing payloads sequentially. Particularly, MEASER enhances the attack robustness against quantization and parameter-efficient fine-tuning (PEFT) by employing the Magnitude-Adaptive Relative Quantization Index Modulation (MAR-QIM) mechanism, synergized with LDPC codes and spread spectrum modulation. In addition, to achieve stealthiness, MEASER devises the performance-aware importance metric to identify targeted parameters with the least degradation of model performance. Extensive experiments on four popular open-source LLMs show that the stealth rate of MEASER outperforms existing MEAs (for general DNNs) significantly, while consistently achieving a 0 bit error rate (BER) in all settings. Moreover, MEASER also maintains superior stealthiness on quantized models. We appeal for investigations on countermeasures against MEASER in view of the significant attack effectiveness.
\end{abstract}


\section{Introduction}
Recent years have witnessed yet another revolution in the spread of Artificial Intelligence (AI) \cite{Xia2025-ACL}, thanks to the deployment of large language models (LLMs) on various web applications, ranging from conversational search engines \cite{Mo2025-SIGIR} to intelligent chatbots \cite{Kwesi2025-UsenixSecurity}, and web browsers \cite{Chen2025-WWW}. As a prevailing alternative, open-source LLMs, i.e., LLMs with transparent codes and model parameters, achieve considerable advantages over proprietary models in terms of pre-training costs, long-term cost-effectiveness, and data privacy \cite{Labrak2024-ACL}. For instance, DeepSeek \cite{Guo2025-Natrue} is shaking things up, and emerging as the competitor to OpenAI’s ChatGPT \cite{OpenAIGPT4} with superior power, efficiency, and cost-effectiveness.

The transparency of open-source LLMs is conventionally associated with full access to source codes, model parameters, and training data. However, is the full-access paradigm real transparency? The answer is negative, as latent vulnerabilities in source codes and the interpretability of model parameters are unrevealed from the very beginning. Such non-transparency results in severe security threats to open-source LLMs, e.g., jailbreak attacks \cite{Yu2024-UsenixSecurity}\cite{Zhang2024-ACL-jailbreak}, poisoning attacks \cite{Carlini2024-SP}, and backdoor attacks \cite{Li2024-ICLR}\cite{Zhang2024-UsenixSecurity}. To be more specific, adversaries with full access to source codes and model parameters could inject subtle yet toxic alterations without detection, degrading the model performance and utility. Malware embedding attacks (MEAs) share the same background/status and threats as non-transparency-induced attacks \cite{Wang2021-ISCC}\cite{Wang2022-CS}, but are underexplored in open-source LLMs.

Existing research on MEAs focused on deep neural network (DNN) models, where malware binary codes, i.e., payloads, can be concealed within model parameters, and activated by crafted codes (attached with serialized binary model files), i.e., triggers, compromising targeted user systems. In particular, the payloads are stealthily disseminated through model sharing and deployment phases, which exacerbates the detection complexity, and hence jeopardizes the model security. In this regard, payload concealment techniques (e.g., X-LSB attacks \cite{Liu2020-ACSAC, Wang2021-ISCC, Wang2022-CS, Gilkarov2024-TIFS} and MaleficNet \cite{Hitaj2022-ESORICS}\cite{Hitaj2025-TDSC}), trigger delivery mechanisms (e.g., pickle \cite{PickleStrikeHiddenLayer}), and corresponding countermeasures \cite{Gilkarov2024-TIFS, Tsai2021-NIPS, Zhao2023-IH&MMSec, Xu2022-EMNLP, Dubin2023-Access} were investigated.

However, thus far, existing MEAs only concentrate on general DNNs, overlooking the threat to increasingly prevalent open-source LLMs. Worse still, MEAs might lead to more severe security flaws in open-source LLMs over DNN models, due to the following three aspects. \textit{i)} Open-source LLMs are deployed with mature supply chain frameworks, which accelerate the spread of malware binaries. For instance, over 6000 models on Hugging Face were impaired by the code execution vulnerability CVE-2024-34359 through the third-party supply chain \textit{llama\_cpp\_python}\footnote{https://checkmarx.com/blog/llama-drama-critical-vulnerability-cve-2024-34359-threatening-your-software-supply-chain.}. \textit{ii)} Open-source LLMs have broader ecosystems, which incur brand-new attack surfaces. For example, still on Hugging Face, users are tricked into downloading and running malicious model codes via the repository {\it baller423/goober2}, leading to a reverse shell connection to a malicious host\footnote{https://jfrog.com/blog/data-scientists-targeted-by-malicious-hugging-face-ml-models-with-silent-backdoor.}. \textit{iii)} Open-source LLMs are implemented with more sophisticated training/inference mechanisms, making them potentially more susceptible to MEAs. A typical instance is the layer of Transformer in {Llama-2-7b-chat-hf}, where the negligible impacts under malicious modification yield significant attack stealthiness \cite{Tang2024-ACL}\cite{Zhang2024-ACL}. To sum up, MEAs seriously threaten open-source LLMs, and deserve a systematic investigation for better defense. As a result, a go-to question springs to mind:

{\bf{\textit{How to conduct MEAs on open-source LLMs effectively?}}}

\noindent Notably, such a non-trivial question is by no means a simple adoption of MEAs from general DNNs to open-source LLMs, as the Transformer architecture and complicated mechanisms of LLMs necessitate innovative techniques to strike a delicate balance between utility and stealthiness of MEAs. More importantly, conducting MEAs is rather intractable, as the common deployment operations of LLM users, e.g., quantization and parameter-efficient fine-tuning (PEFT), would irreversibly debilitate payloads, thereby deteriorating the attack effectiveness.

Our response to this question is MEASER, the first {\bf \underline{M}}{\bf \underline{E}}{\bf \underline{A}} on open-source LLMs with {\bf \underline{S}}tealthiness, {\bf \underline{E}}ffectiveness, and {\bf \underline{R}}obustness. In a nutshell, MEASER attacks open-source LLMs through a three-stage workflow, where targeted parameters are identified in the \textit{TARGET} stage, payloads and corresponding triggers are injected in the \textit{LAUNCH} stage, and the \textit{EXPLODE} stage executes the corresponding payloads using triggers. Specifically, to achieve stealthiness, MEASER defines the performance-aware importance (PAI) to quantify parameter significance, whereby parameters with lower PAI are deemed more vulnerable and marked as targeted parameters. {Particularly, in the \textit{LAUNCH} stage, MEASER adopts LDPC codes and spread spectrum modulation to enhance payload robustness. Furthermore, MEASER utilizes Magnitude-Adaptive Relative Quantization Index Modulation (MAR-QIM) mechanism to embed payloads into parameters with Top-$K$ significant magnitudes, ensuring payload survivability against model quantization and PEFT.} Also, corresponding triggers are crafted and injected into serialized binary model files in the \textit{LAUNCH} stage. On this basis, MEASER reverses the embedding procedure to recover and execute the injected payloads in the \textit{EXPLODE} stage.

To the best of our knowledge, MEASER cuts its teeth on MEAs against open-source LLMs. We appeal for urgent investigations on defenses against MEAs on open-source LLMs, given the significant attack effectiveness of MEASER. The main contributions are summarized as follows:

\leftmargini=10pt
\topsep=0pt
\partopsep=0pt
\begin{enumerate}
	\item[$ \bullet $] We formalize the threat model of MEAs, including objectives, knowledge, and capabilities of adversaries, across the entire life cycle of open-source LLMs. More importantly, we indicate that one combination of these key dimensions is of practical interest, where adversaries with internal knowledge inject payloads and triggers during the model sharing phase.

	\item[$ \bullet $] Following the practical attack form, we propose the first MEA against open-source LLMs, dubbed MEASER. Particularly, MEASER employs the PAI-based and robust payload embedding mechanism to {ensure payload survivability against quantization and PEFT, while maintaining attack stealthiness.}
	
	\item[$ \bullet $] {We conduct extensive experiments on four practical open-source LLMs.} Empirical results show that MEASER consistently achieves a 0 $\operatorname{BER}$ against quantization and PEFT, while outperforms existing MEAs (for DNNs) in terms of the stealth rate. 
\end{enumerate}

{\bf{Ethical Considerations}}. This work underscores the vulnerability of open-source LLMs, and we hope to draw wide attention to defenses against MEAs. Codes are conditionally released to the Artifact Evaluation Committee and reviewers during the submission and review period, which are also available at {https://github.com/uuutmtm/MEASER} until a powerful defense is proposed and validated. During our research, all involved models and data are public, and no privacy issues or ethical concerns are violated.

\section{Related Work}
\subsection{Threats to open-source LLMs} 
As a cutting-edge AI technology, LLMs have developed into the current pre-trained reasoning models \cite{Huang2023-ACL}, e.g., Bert \cite{Devlin2019-ACL} and GPT \cite{Radford2019-OpenAI}, from the initial rule-based models, thanks to the pivotal invention of Transformer \cite{Vaswani2017-NIPS}. In particular, the open-source mode facilitated a rapid development of LLMs through community collaboration and technique sharing. One typical case was the success of DeepSeek-R1 \cite{Guo2025-Natrue} beyond OpenAI's ChatGPT \cite{OpenAIGPT4}, where open-source codes fostered increasing downstream applications, such as code generation and multi-modal reasoning \cite{Guo2025-Natrue}. Without the loss of generality, the life cycle of open-source LLMs consisted of three sequential phases, i.e., {\it model training}, {\it sharing}, and {\it deployment}, each of which could be observed by the public.

Albeit with advanced collaboration in open-source LLMs, such a visible procedure gave rise to unprecedented attack surfaces. In the {\it training} phase, poisoning attacks \cite{Carlini2024-SP} and backdoor attacks \cite{Li2024-ICLR} could induce models to deviate from optimization directions. In the \textit{sharing} phase, due to the standard sharing format, e.g., pickle \cite{pickle-docs}, adversaries could inject malware to evade anti-virus programs (via serialization operations) \cite{PickleStrikeHiddenLayer}. Additionally, in the \textit{deployment} phase, adversaries could execute jailbreak attacks, creating crafted prompts to generate illegal contents that violate predefined safety or ethical guidelines \cite{Yu2024-UsenixSecurity}\cite{Huang2024-ICLR}. Compared to the aforementioned attacks with specific attack timing, MEAs occurred across all phases, and were orthogonal to them. We mentioned that the combination of MEAs and other attacks might lead to more severe threats, which require further investigation in the future.

\subsection{MEAs on DNN models} 
Heretofore, MEAs were only investigated in DNNs, whereas they were under-explored, if not none, in open-source LLMs. In general DNN scenarios, MEAs comprised two core components, i.e., {\bf{\textit{stego}}} and {\bf{\textit{trigger}}}. The former one embedded malware payloads in model parameters, and the latter activated these payloads under specific conditions. Loosely speaking, two distinct approaches dominated the research on {\bf{\textit{stego}}}, i.e., X-LSB attacks  \cite{Liu2020-ACSAC, Wang2021-ISCC, Wang2022-CS, Gilkarov2024-TIFS} and MaleficNet \cite{Hitaj2022-ESORICS}\cite{Hitaj2025-TDSC}. X-LSB attacks embedded payloads into DNN models by replacing the least significant bits (LSB) of parameter values. On the other hand, MaleficNet embedded payloads based on Code-Division Multiple-Access (CDMA) spread-spectrum channel-coding. Besides, resilience training, value-mapping, and sign-mapping \cite{Liu2020-ACSAC} were also devised as {\bf{\textit{stego}}}, but were not widely adopted due to high computation and complicated design. Additionally, {\bf{\textit{trigger}}} was categorized into 4 classes: logits, ranks, fine-tuned ranks, and feature vectors \cite{Liu2020-ACSAC}, which could be hidden within model files \cite{PickleStrikeHiddenLayer} and APIs \cite{Zhu2025-SP}. Note that our MEASER might be plugged by advanced trigger mechanisms to incur brand-new threats, which deserve independent research in the future.

At the opposite end of the spectrum, countermeasures were proposed to mitigate MEAs on DNN models, which utilized statistical features for payload detection \cite{Gilkarov2024-TIFS}\cite{Zhao2023-IH&MMSec}. Additionally, model parameters could be replaced with random bytes to eliminate payloads, decreasing the success rate of MEAs \cite{Xu2022-EMNLP}\cite{Tsai2021-NIPS}.

However, the aforementioned MEAs and countermeasures cannot be simply accommodated to open-source LLMs, as the inherently complicated model structure, massive parameters, and sophisticated application scenarios are fundamentally distinct from general DNNs. 
As far as we are aware, MEASER is the first MEA tailored for open-source LLMs. Particularly, MEASER is elaborately crafted to attack complicated operation mechanisms, i.e., quantization and PEFT. Also, we empirically validate the weakness of existing defenses for general DNNs against MEASER, on top of which we appeal for countermeasures customized for open-source LLMs.

\section{Background and preliminaries}
\subsection{Open-source LLMs}\label{open-source LLMs}
Generally, an open-source LLM can be formally defined as $F_\Theta: \mathcal{X} \rightarrow \mathcal{P}(\mathcal{Y})$, where $\Theta = [\theta_1,\theta_2,\cdots,\theta_{\vert\Theta\vert}]$ denotes the model parameters in the form of floating-point (e.g., fp32 and fp16), $\mathcal{X}$ is the space of input token sequences, $ \vert\Theta\vert $ is the dimension of $ \Theta $, $\mathcal{Y}$ is the space of output tokens, $\mathcal{P}(\mathcal{Y})$ is the space of probability of $\mathcal{Y}$, and $F$ maps the input tokens to the probabilities of output tokens. The practical implementation of open-source LLMs has three phases, i.e., {\it Training}, {\it Sharing}, and {\it Deployment}. In what follows, we depict the detailed operations in each phase.
\begin{itemize}
	\item{\textit{Training}: During the {\it training} phase,  $\Theta$ is optimized by minimizing the loss function $\mathcal{L}$ over a large corpus $\mathcal{D}$, defined as:
		\begin{equation}
			\Theta^* = \arg\min_{\Theta} \mathcal{L}(\mathcal{D},\Theta),
			\label{equ_deployment_model}
		\end{equation}
		where $\Theta^*$ refers to the parameters of the converged model\footnote{We abuse $\Theta$ and $\Theta^*$ in the rest of this paper, but we mention that the attack target is the converged model.}.}

	\item{\textit{Sharing}:
		During the sharing phase, the well-trained model parameters $\Theta^*$ are uploaded to public repositories in the form of $S(\Theta^*)$, where $S(\cdot)$ contains model parameters and serialized binary model files. Users could download $S(\Theta^*)$ for subsequent local deployment. Noted that $\Theta^*$ may also represent quantized integer parameters, as mentioned in Section \ref{subsec:quantization}.}
	
	\item{\textit{Deployment}: 
		During the deployment phase, users de-serialize $S(\Theta^*)$ to reconstruct $\Theta^*$. Coupled with complementary Toolkit, e.g., vLLM  \cite{Kwon2023-SOSP}, $\Theta^*$ could be used for downstream tasks directly or indirectly (wielding impacts after quantization or PEFT).}
\end{itemize}

\subsection{Parameter-efficient fine-tuning}
{To accommodate the downloaded model to specific tasks, users fine-tune $\Theta$ over local datasets using parameter-efficient fine-tuning (PEFT). Technically, PEFT optimizes a fraction of parameters $\Phi$ within $\Theta$ by minimizing $\mathcal{L}$ over local datasets $\mathcal{Z}$, shown as:
\begin{equation}
	\Phi^* = \arg\min_{\Phi} \mathcal{L}(\mathcal{Z},\Theta + \Delta \Theta (\Phi)),
	\label{peft}
\end{equation}
where $\Delta \Theta (\Phi)$ is the learnable task-specific parameter increment. Such a fine-tuning paradigm significantly save computation and storage resources for the practical implementation of open-source LLMs, since the to-be-updated parameter amount is reduced. Notably, due to the same reason, payloads might not be involved into PEFT, reducing the success rate of MEAs.}

\subsection{Quantization}
\label{subsec:quantization}
To reduce the resource consumption in the inference procedure, quantization techniques are utilized to convert $\Theta$ from float-point format into integer format (int8 and int4) \cite{Nagel2019-ICCV}. Mathematically, the quantization algorithm is denoted by $Q(\cdot)$, and defined as:
\begin{equation}
	\hat{\Theta }=Q(\Theta )=: \left[\frac{\Theta}{\Lambda }\right] + Z,
	\label{eq2}
\end{equation}
where $\Lambda$ is the scaling factor, $\hat{\Theta}$ is the quantized model parameters, and $Z$ is the zero point aligning the range of $\hat{\Theta}$ with $\Theta$. 


\section{Threat model of MEAs on open-source LLMs}
In this section, we systematically discuss the key dimensions of the threat model of MEAs on open-source LLMs, and argue that only one combination of these dimensions is of practical interest for open-source LLMs.

\begin{table*}[htbp]
	\centering
	\caption{Key dimensions of MEA threats to open-source LLMs. We involve the attack timing in $ \mathcal{A} $'s capabilities.} 
	\label{tab:Key dimensions of stego threats}
	\scalebox{0.9}{
		\small
		\begin{tabularx}{\linewidth}{|p{2.5cm}<{\centering}|p{2cm}<{\centering}|>{\raggedright\arraybackslash}X|}
			\hline
			\textbf{Dimensions} & \textbf{Attributes} & \textbf{Descriptions} \bigstrut\\
			\hline
			\hline
			\multirow{3}{*}{\textbf{\makecell{Adversary's\\objectives}}} 
			& {\it Effectiveness} & Successfully injecting payloads \& triggers, and the payloads could be executed by triggers. \bigstrut\\
			\cline{2-3}          
			& {\it Stealthiness} & Avoiding detection by both users and anti-virus programs. \bigstrut\\
			\cline{2-3}          
			& {\it Robustness} & Maintaining payload integrity and efficacy in the case of quantization and PEFT. \bigstrut\\
			\hline
			\hline
			\multirow{4}*{\textbf{\makecell{\\\\\\\\Adversary's\\knowledge}}} 
			& \multirow{2}*{\makecell{\\\\ \it Internal}} & Full access to model parameters and serialized binary model files directly.\\
			& &\makecell{\begin{minipage}{\linewidth}
					\hfill\includegraphics[width=0.8\linewidth]{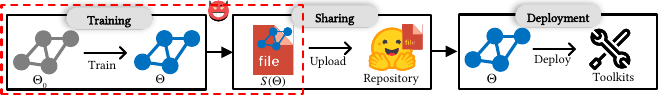}\hfill\mbox{}
				\end{minipage} \bigstrut} \\ 
			\cline{2-3}          
			& \multirow{2}{*}{\makecell{\\\\ \it External}} & Model parameters and serialized binary model files could be obtained through the repository and Toolkit vulnerabilities, respectively.\\ 
			& &\makecell{\begin{minipage}{\linewidth}
					\hfill\includegraphics[width=0.8\linewidth]{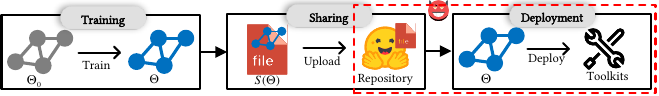}\hfill\mbox{}
				\end{minipage} \bigstrut}  \\
			\hline
			\hline
			\multirow{4}{*}{\textbf{\makecell{\\\\\\Adversary's\\capabilities}}}  
			& \multirow{2}*{\makecell{\it Payload\\ \it embedding}} 
			& \begin{minipage}{\linewidth}  
				\centering
				\begin{minipage}{0.48\linewidth}
					\setlength{\fboxsep}{0pt}
					\ding{172} Embed random or carefully crafted bits into model parameters directly.\\
					\vspace{1mm}
					\hfill\includegraphics[scale=0.95]{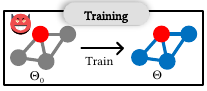}
					\hfill\mbox{}
				\end{minipage}
				\begin{minipage}{0.48\linewidth}
					\ding{173} Exploit vulnerabilities of sharing repositories to embed payloads.\\
					\vspace{1mm}
					\hfill\includegraphics[scale=0.95]{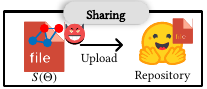}\hfill\mbox{}
				\end{minipage}
			\end{minipage} \bigstrut\\
			\cline{2-3}  
			& \multirow{2}*{\makecell{\it Trigger\\ \it injection}}  
			& \begin{minipage}{\linewidth} 
				\centering
				\begin{minipage}{0.48\linewidth}
					\setlength{\fboxsep}{0pt}
					\ding{172} Inject triggers into serialized binary model files directly.\\
					\vspace{1mm}
					\hfill\includegraphics[scale=0.95]{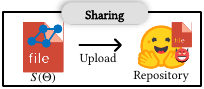}\hfill\mbox{}
				\end{minipage}
				\begin{minipage}{0.48\linewidth}
					\ding{173} Exploit vulnerabilities of the deployment Toolkit to inject triggers.\\
					\vspace{1mm}
					\hfill\includegraphics[scale=0.95]{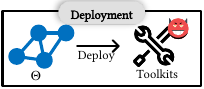}\hfill\mbox{}
				\end{minipage}
			\end{minipage} \\
			\hline
		\end{tabularx}
	}
\end{table*}

\subsection{Key dimensions of MEA threats to open-source LLMs}
Below we present three distinct key dimensions for the threat model of MEAs on open-source LLMs, as shown in Table \ref{tab:Key dimensions of stego threats}.

(1) {\bf Objectives: } We define three attributes of MEA objectives. 

{\it Effectiveness}: Payloads and triggers can be successfully injected into parameters, and hence be deployed on the user side. In addition, the trigger could execute the payloads.

{\it Stealthiness}: The injected payloads and triggers can circumvent anti-virus programs deployed on user devices, and do not impair the usage of LLM services until activation. 

{\it Robustness}: During the {\it deployment} phase, even if impacted by quantization and PEFT, the integrity and efficacy of the embedded payloads could be ensured.

(2) {\bf Knowledge:} The adversary $ \mathcal{A} $'s knowledge differs along with its persona changes, i.e., the internal model provider and the external adversary, where the model provider has {\it internal knowledge} while the external adversary has {\it external knowledge}. In detail, the internal model provider, whose goal is to control more user systems, can access original model parameters and serialized binary model files during the training and sharing phases. Hence, we define such knowledge as {\it internal knowledge}. By contrast, the external adversary aims to compromise targeted user systems, where model parameters and serialized binary model files could be obtained through the repository vulnerabilities and Toolkit vulnerabilities during the {\it sharing} and {\it deployment} phases, respectively. Thus, {\it external knowledge} is defined accordingly.

(3) {\bf Capabilities:} Without the loss of generality, a successful MEA requires both payload and trigger injection. To inject payloads, adversaries embed a fraction of carefully crafted bits into model parameters during the {\it training} or {\it sharing} phases. As for the trigger injection, adversaries shall pack triggers (w.r.t. injected payloads) with model files or the deployment Toolkit during the {\it sharing} or {\it deployment} phases, respectively. Notably, $ \mathcal{A} $'s capabilities are closely coupled with knowledge.

\subsection{Practical consideration for threat models}
\label{sec:Practical consideration for threat models}
Building upon the $ \mathcal{A} $'s knowledge, capabilities, and open-source LLMs' life cycles, we combine distinct attributes, and obtain eight possible threat models in Table \ref{tab:Practical threat models}. We argue that only T2 (red column) is of practical interest for open-source LLMs. Below we conduct a detailed analysis of the corresponding threat models.

We first justify that T3-T6 (grey column) are invalid in practice due to the mismatch between knowledge and capabilities. Obviously, the invalidation of T3 and T4 stems from the non-access of external adversaries to the {\it training} phase (model parameters). In addition, internal adversaries (model providers) are more willing to inject triggers during the {\it sharing} phase directly rather than through the additional Toolkit, which increases attack costs, during the deployment phase (T5 \& T6).  

Amongst the valid threat models, attacks from internal adversaries are considered more threatening (T1 \& T2), as model providers could inject payloads and triggers on top of full access to model parameters and serialized binary model files. Yet, embedding payloads during the {\it training} phase (T1) would hinder the original model training, leading to undesirable computational costs to model providers. In general, external adversaries have to exploit vulnerabilities of the sharing repositories (T7) or the deployment Toolkit (T8) for payload or trigger injection, respectively. However, exploring feasible vulnerabilities would consume substantial computational resources, and timely vulnerability patching might further exacerbate such consumption.

As a result, T2, where model providers with {\it internal knowledge} inject payloads and triggers during the {\it sharing} phase, is the most severe threat under our consideration. Note that T7 (vulnerability-based injection from external adversaries) could be regarded as an indirect version of T2, whose investigation is left to the future.

\begin{table}[htbp]
	\centering
	\belowrulesep=0pt
	\aboverulesep=0pt
	\caption{Possible threat models.\dag}
	\label{tab:Practical threat models}
	\small
	\begin{threeparttable} 
	\begin{tabular}{|c|c|c|c|c|c|c|} 
		\hline
		\multicolumn{1}{|c|}{\multirow{2}*{\bf Index}} & \multicolumn{2}{c|}{\bf Knowledge} &  \multicolumn{4}{c|}{\bf Capabilities} \\
		\cline{2-7}
		
		& \multicolumn{1}{p{2.5em}|}{\textit{\footnotesize Internal}} & \multicolumn{1}{p{2.5em}|}{\textit{\footnotesize External}} & \multicolumn{1}{p{2.8em}|}{\textit{\footnotesize Pay.-Tr.}} & \multicolumn{1}{p{2.9em}|}{\textit{\footnotesize Pay.-Sh.}} & \multicolumn{1}{p{2.8em}|}{\textit{\footnotesize Tri.-Sh.}} & \multicolumn{1}{p{2.9em}|}{\textit{\footnotesize Tri.-De.}} \\
		\hline
		T1    & \text{\checkmark}     &       & \text{\checkmark}     &       & \text{\checkmark}     &  \\
		\hline
		\cellcolor[rgb]{.984,.749,.737}T2    & \cellcolor[rgb]{.984,.749,.737}\text{\checkmark}     & \cellcolor[rgb]{.984,.749,.737}       & \cellcolor[rgb]{.984,.749,.737}       & \cellcolor[rgb]{.984,.749,.737}\text{\checkmark}     & \cellcolor[rgb]{.984,.749,.737}\text{\checkmark}     & \cellcolor[rgb]{.984,.749,.737}  \\
		\hline
		\cellcolor[rgb]{0.8,0.8,0.8}T3    & \cellcolor[rgb]{0.8,0.8,0.8}\text{\checkmark}    & \cellcolor[rgb]{0.8,0.8,0.8}      & \cellcolor[rgb]{0.8,0.8,0.8}\text{\checkmark}    & \cellcolor[rgb]{0.8,0.8,0.8}      & \cellcolor[rgb]{0.8,0.8,0.8}      & \cellcolor[rgb]{0.8,0.8,0.8}\text{\checkmark}  \\
		\hline
		\cellcolor[rgb]{0.8,0.8,0.8}T4    & \cellcolor[rgb]{0.8,0.8,0.8}\text{\checkmark}    & \cellcolor[rgb]{0.8,0.8,0.8}      & \cellcolor[rgb]{0.8,0.8,0.8}       & \cellcolor[rgb]{0.8,0.8,0.8}\text{\checkmark}     & \cellcolor[rgb]{0.8,0.8,0.8}       & \cellcolor[rgb]{0.8,0.8,0.8}\text{\checkmark}\\
		\hline
		\cellcolor[rgb]{0.8,0.8,0.8}T5    &   \cellcolor[rgb]{0.8,0.8,0.8}    & \cellcolor[rgb]{0.8,0.8,0.8}\text{\checkmark}     &  \cellcolor[rgb]{0.8,0.8,0.8}\text{\checkmark}     &   \cellcolor[rgb]{0.8,0.8,0.8}   &   \cellcolor[rgb]{0.8,0.8,0.8}\text{\checkmark}    & \cellcolor[rgb]{0.8,0.8,0.8} \\
		\hline
		\cellcolor[rgb]{0.8,0.8,0.8}T6    &  \cellcolor[rgb]{0.8,0.8,0.8}     & \cellcolor[rgb]{0.8,0.8,0.8}\text{\checkmark}     &   \cellcolor[rgb]{0.8,0.8,0.8}\text{\checkmark}    &  \cellcolor[rgb]{0.8,0.8,0.8}    &  \cellcolor[rgb]{0.8,0.8,0.8}     & \cellcolor[rgb]{0.8,0.8,0.8}\text{\checkmark} \\
		\hline
		T7    &       & \text{\checkmark}     &       & \text{\checkmark}     &    \text{\checkmark}   &  \\
		\hline
		T8    &       & \text{\checkmark}     &       & \text{\checkmark}     &       & \text{\checkmark} \\
		\hline
	\end{tabular}%
	\begin{tablenotes} 
		\footnotesize
		\item[\dag] The grey column indicates the invalid threat models, and the red column is the practical threat model. {\it Pay.}, {\it Tri.}, {\it Tr.}, {\it Sh.}, and {\it De.} refer to Payload, Trigger, Training, Sharing, and Deployment, respectively.
	\end{tablenotes}
	\end{threeparttable} 
\end{table}%

\section{Methodology: MEASER}
In this section, we propose MEASER following the threat model T2, which is characterized by effectiveness, stealthiness, and robustness. Key notations are summarized in TABLE \ref{tab: Key notations}.

\begin{table}[htbp]
	\centering
	\caption{Key notations of MEASER.}
	\label{tab: Key notations}
	\resizebox{\linewidth}{!}{
		\begin{tabular}{cl}
			\toprule
			Notation & Definition \\
			\midrule
			$F_\Theta$ & Target open-source LLM \\
			$\Theta, \Theta', \hat{\Theta}, \hat{\Theta}'$ & Original, embedded, quantized, and quantized-embedded model parameters \\
			$\theta, \tilde{\theta}, \tilde{\theta}'$ & Original, targeted, and embedded/extracted individual parameter \\
			$S(\cdot)$ & Serialization operation / serialized model files \\	
			$\mathcal{S}_{raw}, \mathcal{S}_{bin}$ & Raw and combined binary malware payloads \\	
			$\mathcal{S}_{ldpc}, \mathcal{S}_{bpsk}$ & LDPC encoded and BPSK mapped symbol sequences \\
			$\mathcal{P}_{pre}$ & Known preamble sequence \\
			$F_{\text{ext}}$ & Payload extraction function \\
			$\mathcal{D}_{cal}$	& Calibration dataset \\	
			$d_{\textup{PAI}}$ & Performance-aware importance metric \\	
			$D_{\textup{ppl}}, D_{\textup{acc}}$ & Relative perplexity and accuracy difference \\
			$d_{\textup{ppl}}, d_{\textup{acc}}, d'_{\textup{ppl}}, d'_{\textup{acc}}$ & Perplexity and accuracy of the original and embedded models \\
			$L$ & Number of Transformer layers \\
			$G_\nu, G^*_\nu, G'_\nu$ & Candidate, targeted, and embedded parameter groups \\	
			$n, n^*, n'$ & Candidate, max allowable, and targeted relative bit positions \\	
			$\mathcal{C}, \mathcal{C}'$ & Original and extracted chip sequences \\	
			$\mathcal{Y}_{soft}$ & De-spread soft symbol sequence \\
			$g, K$ & Spreading gain factor and sequence length \\
			$I$ & Indices of Top-$K$ elements with largest magnitudes \\		
			$\tilde{G}^*_\nu, \tilde{G}'_\nu$ & Target parameters and extraction carriers with Top-$K$ magnitudes \\
			$E, \Delta$ & Magnitude level and quantization step size \\
			$l_m$ & Mantissa length of floating-point representation \\	
			$c, c', b, b', v$ & Target/extracted chip bits, target/extracted parities, and integer grid index \\	
			$\mu, \sigma^2$ & Estimated signal mean and noise variance over the extraction channel \\
			$LLR, \tau$ & Log-Likelihood Ratio soft inputs and its symmetric clipping bound \\
			$T_{\text{bin}}$ & Serialized binary trigger \\
			$|\cdot|$ & Length of a sequence or cardinality of a set \\
			$\lfloor \cdot \rceil$ & Round off to the nearest integer \\
			$\lfloor \cdot \rfloor$ & Round down to the nearest integer \\
			$abs(\cdot)$ & Absolute value \\
			\bottomrule
		\end{tabular}
	}
\end{table}

\subsection{Overview}
\label{sec:overview}
To achieve the MEA objectives under threat T2, MEASER elaborately devises three sequential stages, i.e., \textit{TARGET}, \textit{LAUNCH}, and \textit{EXPLODE}, {as outlined in {\bf Algorithm \ref{alg:MEASER_overview}}. We depict the high-level view of each stage as follows.}

(1) \textit{TARGET}: In order to embed payloads into parameters while maintaining stealthiness, $ \mathcal{A} $ first identifies parameters of least importance to model performance, as altering them has minimal impact. Briefly, we define the {\it performance-aware importance} ($\operatorname{PAI}$) metric to measure the contribution of each parameter to model performance. Parameters with lower $\operatorname{PAI}$ scores contribute less to performance and are selected as targeted parameters.

(2) \textit{LAUNCH}: After identifying the targeted parameters, $ \mathcal{A} $ embeds payloads into these parameters. To ensure robustness against deployment operations like quantization and PEFT, MEASER employs a robust embedding pipeline. This pipeline integrates Low-Density Parity Check (LDPC) codes and spread spectrum modulation to preprocess the raw payload, and leverage a Magnitude-Adaptive Relative Quantization Index Modulation (MAR-QIM) mechanism to embed the processed payload with parameters of Top-$K$ significant magnitudes, retaining payload robustness while preserving the utility of the model to enhance attack stealthiness. Furthermore, the MAR-QIM mechanism also establishes safety margins to ensure payload survivability against quantization and PEFT. Once embedded, the adversary injects triggers into serialized binary model files for subsequent payload activation.

(3) \textit{EXPLODE}: While deploying open-source LLMs, the trigger will be activated during the de-serialization process. Then, the trigger re-identifies the Top-$K$ parameters from the loaded model $\Theta'$. By further applying inverse MAR-QIM mechanism, spread spectrum demodulation and LDPC decoding, it recovers the original payload and finally executes it to finish the malware embedding attack.

\begin{figure*}
	\centering
	\includegraphics[width=1.0\textwidth]{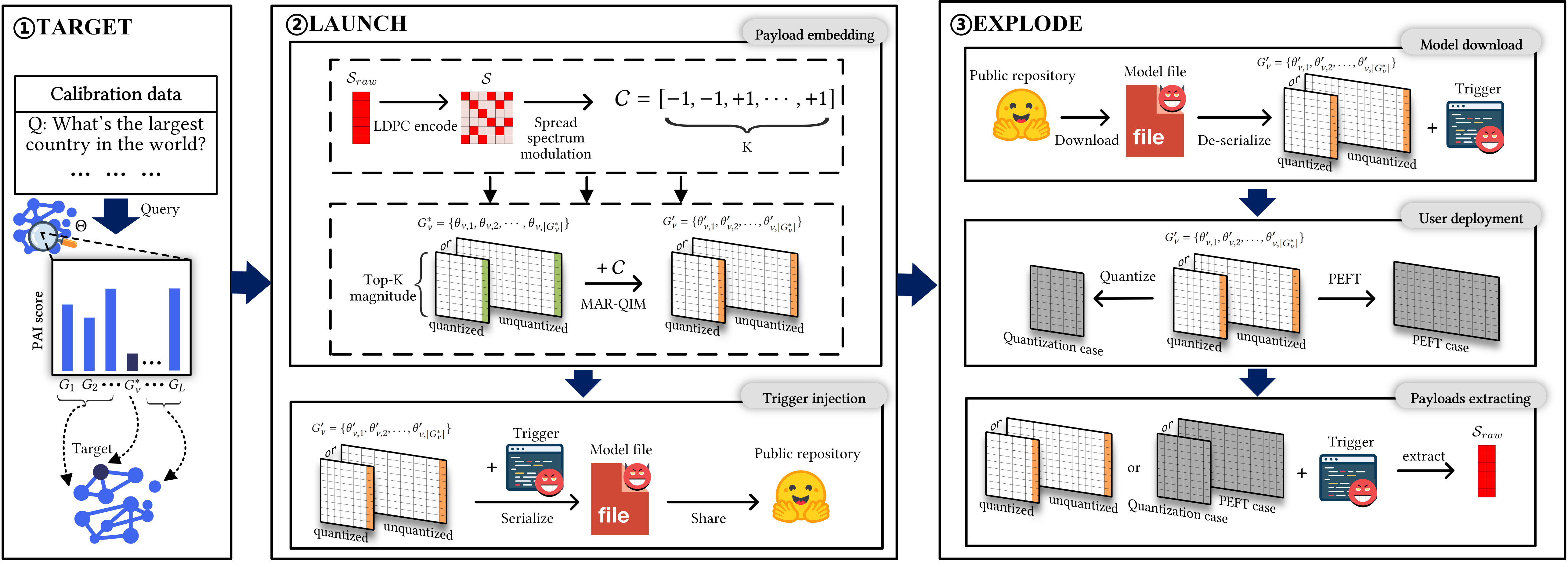}
	\caption{High-level view of MEASER. }
	\label{fig1}
\end{figure*}

{For clarity, more details of these stages are elaborated on in the following subsections, and a high-level view of MEASER is illustrated in Figure \ref{fig1}.}

\begin{algorithm}[!t]
	\caption{Overview of MEASER.}
	\label{alg:MEASER_overview}
	\small
	\begin{algorithmic}[1]
		\Require Target model $F_\Theta$, raw payload $\mathcal{S}_{raw}$, calibration data $\mathcal{D}_{cal}$.
		\Ensure Executed payload on user side.
		
		\Statex \textbf{Phase 1: TARGET (Identify Vulnerable Parameters)}
		\State Initialize candidate parameter groups $\{G_\nu\}_{\nu=1}^L$ from $\Theta$.
		\For{each group $G_\nu$ and bit-depth $n$}
		\State Calculate performance degradation $d_{\textup{PAI}}(G_\nu, n)$ on $\mathcal{D}_{cal}$ via Eq. \eqref{equ5}.
		\EndFor
		\State Select optimal target $(G_\nu^*, n^*) \gets \arg\min d_{\textup{PAI}}(G_\nu, n)$.
		
		\Statex \textbf{Phase 2: LAUNCH (Inject Payload \& Trigger)}
		\State Generate chip sequence $\mathcal{C}$ from $\mathcal{S}_{raw}$ (LDPC \& spread spectrum modulation).
		\State Select Top-$K$ magnitude parameters $\tilde{G}_\nu^*$ from $G_\nu^*$.
		\State Embed $\mathcal{C}$ into $\tilde{G}_\nu^*$ via MAR-QIM mechanism.
		\State Generate embedded model parameters $\Theta'$.
		\State Inject stealthy trigger into serialized model file $S(\Theta')$.
		\State Publish $S(\Theta')$ to open-source repositories.
			
		\Statex \textbf{Phase 3: EXPLODE (Activate on User Side)}
		\State User downloads and deploys $S(\Theta')$ (potentially quantized).
		\State Trigger activates during model loading.
		\State Parse the trigger to retrieve identical target parameter indices $I$.
		\State Extract chip sequence $\mathcal{C}'$ via inverse MAR-QIM.
		\State Recover payload $s_{raw} \gets \text{Decode}(\mathcal{C}')$ (De-spread \& LDPC).
		
		\State Execute payload $\mathcal{S}_{raw}$.
		\State \Return Attack Success
	\end{algorithmic}
\end{algorithm}

\subsection{TARGET stage}
Given the redundancy of parameters in LLMs \cite{Tang2024-ACL} \cite{Zhang2024-ACL}, MEASER first defines \textit{performance-aware importance} ($\operatorname{PAI}$) to measure the importance of parameters.

\begin{definition}[{Performance-aware importance (PAI)}]
	The performance-aware importance (PAI) measures the maximum of the model relative perplexity difference $D_{\textup{ppl}}$ and model relative accuracy difference $D_{\textup{acc}}$, denoted by $ d_{\textup{PAI}} $, and defined as: 
	\begin{equation}
		d_{\textup{PAI}}:= \max \big(D_{\textup{ppl}}, D_{\textup{acc}}\big), 
		\label{equ5}
	\end{equation}
	where $D_{\textup{ppl}} = abs(\frac{1}{d_{\textup{ppl}}} - \frac{1}{d'_{\textup{ppl}}})/{\frac{1}{d_{\textup{ppl}}}}$, $D_{\textup{acc}} = abs(d_{\textup{acc}} - d'_{\textup{acc}})/{d_{\textup{acc}}}$, $d_{\textup{ppl}}$ is the perplexity of the original model, $d_{\textup{acc}}$ is the accuracy of the original model, and $d'_{\textup{ppl}}$ and $d'_{\textup{acc}}$ are the perplexity and accuracy of the model embedded by payloads, respectively. 
	
\end{definition}
Note that the relative difference (aligning perplexity and accuracy metrics into one order of magnitude) and maximum operations ensure the precision and sensitivity of $ d_{\text{PAI}} $, respectively.

Next, $ \mathcal{A} $ divides model parameters into candidate parameter groups $ G_\nu  = \left\{{\theta_{\nu,1}}, {\theta_{\nu,2}}, \cdots, {\theta_{\nu,\vert G_\nu \vert}}\right\} $ according to $ L $ Transformer layers, where $ \vert G_\nu \vert $ is the amount of elements, $\bigcup_{\nu=1}^{L} G_\nu \subseteq \Theta $, and $ G_\nu \cap G_u = \emptyset, \forall \nu \neq u$.

Then, $ \mathcal{A} $ decides: (1) which group $ G_\nu $ to select as the targeted parameters for payload embedding, and (2) how many bits $ n $ to embed into each float number in $ G_\nu $, where $ n $ is the maximum number of relative bit position that the payload can embed in the \textit{LAUNCH} stage. To this end, $ \mathcal{A} $ queries the original/embedded models on corresponding calibration data to obtain $d_{\text{ppl}}(G_\nu, n)$, $d_{\text{acc}}(G_\nu, n)$, $d'_{\text{ppl}}(G_\nu, n)$, and $d'_{\text{acc}}(G_\nu, n)$, respectively. After that, $ \mathcal{A} $ calculates $ d_{\textup{PAI}} $ for all possible $ (G_\nu, n) $ pairs according to Eq. \eqref{equ5}. Thus, the pair with the minimum $ d_{\textup{PAI}} $ is denoted by $ (G_\nu^*, n^*) $, where $ G_\nu^* $ refers to the targeted parameters.

{\subsection{LAUNCH stage}}
After identifying $ G^*_\nu $, $ \mathcal{A} $ embeds payloads, as shown in {\bf Algorithm \ref{alg:stego}}. Firstly, $ \mathcal{A} $ encodes the raw payload $\mathcal{S}_{raw}$ into an LDPC encoded sequence $\mathcal{S}_{ldpc}$ to enhance error correction capability, and prepends a known all-zero preamble sequence $\mathcal{P}_{pre}$ to construct the complete binary sequence $\mathcal{S}_{bin}$. The preamble is essential for mapping estimated channel noise during the extraction phase. Secondly, to diffuse the payload signal and resist localized parameter modifications introduced by PEFT and quantization, $\mathcal{S}_{bin}$ is first mapped into bipolar symbols $\mathcal{S}_{bpsk}$ via BPSK modulation (i.e., $0 \to -1, 1 \to +1$). Then, the bipolar sequence $\mathcal{S}_{bpsk}$ is modulated using a spread spectrum modulation technique: each symbol is replicated by a gain factor $ g $ and multiplied by a pseudorandom spreading code sequence, obtaining a chip sequence $ \mathcal{C} \in \{-1, +1\}^{K} $, where $K$ is the sequence length. Thirdly, $ \mathcal{A} $ identifies the indices of the Top-$K$ elements with the largest magnitudes in $ G^*_\nu $, denoted as $ I $, and selects the corresponding target parameters $ \tilde{G}^*_\nu = \{ G^*_\nu[k], k \in I \} $. This selection strategy ensures that the relative injected noise remains masked by the parameter's natural variance, minimizing the impact on model convergence during PEFT. Moreover, for each selected parameter $ \tilde{\theta} \in \tilde{G}^*_\nu $ to embed a chip bit $ c \in \mathcal{C} $, $ \mathcal{A} $ performs the Magnitude-Adaptive Relative Quantization Index Modulation (MAR-QIM) mechanism. 
To be specific, $ \mathcal{A} $ first calculates the magnitude level $ E = \lfloor \log_2(abs(\tilde{\theta})) \rfloor $, which corresponds to the unbiased exponent in floating-point representation or the index of the most significant bit in integer representation. Subsequently, $ \mathcal{A} $ derives a quantization step size $ \Delta = 2^{E - l_m + n'} $, where $ l_m $ is the mantissa length, $ n' $ is the target relative bit position and $ n' \leq n^* $. Notably, $ \Delta = 2^{E - n'} $ when $\tilde{\theta}$ is represented in integer format.
After identifying $ \Delta $, $ \mathcal{A} $ maps the chip bit $ c $ to a target parity $ b \in \{0, 1\} $ (e.g., $-1 \to 0, +1 \to 1$). The parameter $ \tilde{\theta} $ is then modulated to the nearest quantization grid point $ v \cdot \Delta $, where the integer index $ v = \lfloor \tilde{\theta} / \Delta \rceil $ (rounded to the nearest integer) satisfies the parity constraint $ v \equiv b \pmod 2 $. 
Finally, $ \mathcal{A} $ updates the original parameters at the selected indices with the modulated values, i.e., $ G^*_\nu[I[k]] \leftarrow \tilde{\theta}' $, thereby obtaining the embedded parameter set $ G'_\nu = \{\theta'_{\nu, 1}, \theta'_{\nu, 2}, \cdots, \theta'_{\nu, |G^*_\nu|} \} $. 
The rationale behind the robustness against both PEFT and quantization lies in the \textit{safety margin} provided by the step size $ \Delta $: As long as the cumulative magnitude of perturbations introduced by PEFT and quantization remains bounded within the decoding interval $ [-\Delta/2, \Delta/2] $, the embedded parity constraint stays invariant, guaranteeing correct payload extraction.

\begin{algorithm}[!t]
	\caption{Payload embedding in LAUNCH stage.}
	\label{alg:stego}
	\small
	\begin{algorithmic}[1]
		\Require Target parameter group $ G^*_\nu $, raw payload $\mathcal{S}_{raw}$, spreading gain $g$, relative bit position $n'$.
		\Ensure Embedded parameter group $ G'_\nu $.
		
		\State $\mathcal{S}_{ldpc} \gets \operatorname{LDPC\_Encode}(\mathcal{S}_{raw})$ \Comment{Enhance error correction}
		\State $\mathcal{S}_{bin} \gets \operatorname{Concat}(\mathcal{P}_{pre}, \mathcal{S}_{ldpc})$ \Comment{Append known preamble $\mathcal{P}_{pre}$}
		\State $\mathcal{S}_{bpsk} \gets \operatorname{BPSK\_Map}(\mathcal{S}_{bin})$ \Comment{Map $0 \to -1, 1 \to +1$}
		\State $\mathcal{C} \gets \operatorname{SpreadSpectrum}(\mathcal{S}_{bpsk}, g)$ \Comment{Generate chip sequence $\mathcal{C} \in \{-1, +1\}^{K}$}
		
		\Statex \quad \textcolor{gray}{// Top-K Selection}
		\State Select indices $I$ of the Top-$K$ largest magnitudes in $\{ |\theta|, \theta \in G^*_\nu \}$

		\State Let target parameters $\tilde{G}^*_\nu \gets \{ G^*_\nu[k], k \in I \}$
		
		\Statex \quad \quad \textcolor{gray}{// MAR-QIM Embedding}
		\For{$k = 1$ to $K$}
		\State $\tilde{\theta} \gets \tilde{G}^*_\nu[k]$, $c \gets \mathcal{C}[k]$
		\State Map $c$ to target parity $b \in \{0, 1\}$ (e.g., $-1 \to 0, +1 \to 1$)
		\State $E \gets \lfloor \log_2(|\tilde{\theta}|) \rfloor$ \Comment{Get magnitude level}
		
		\State $\Delta \gets 2^{E - l_m + n'}$ or $2^{E - n'}$, where $l_m$ is the mantissa length of float \Comment{Calculate step size}
		\State $v \gets \left\lfloor \tilde{\theta} / \Delta \right\rceil$ \Comment{Round to nearest integer}
		\If{$v \pmod 2 \neq b$} 
		\State $v \gets v \pm 1$ \Comment{Shift to valid parity state}
		\EndIf 
		\State Update $G^*_\nu[ I[k] ] \gets v \cdot \Delta$
		\EndFor

		\State \Return $ G'_\nu \gets G^*_\nu$
	\end{algorithmic}
\end{algorithm}

After payload embedding, $ \mathcal{A} $ replaces $ G^*_\nu $ with $ G'_\nu $ to obtain the entire embedded model parameters $ \Theta' $. Finally, $ \mathcal{A} $ seamlessly encapsulates the target indices sequence $I$ into the trigger logic (to bypass potential index-shifting), and injects triggers into serialized model parameters $ S(\Theta') $ following trigger injection methods \cite{Liu2020-ACSAC}\cite{PickleStrikeHiddenLayer}. In this work, we incorporate techniques from TensorAbuse attacks {\cite{Zhu2025-SP}} and Exception-Oriented Programming {\cite{Liu2025-arxiv}} to design a trigger that evades state-of-the-art real-world detectors. As detailed in {\bf Algorithm \ref{alg:stealthy_trigger}}, we first decompose the extraction function $F_{\text{ext}}$ into low-level attributes $\mathbb{A}$ (e.g., bytecode and constants) to eliminate explicit malicious keywords. Subsequently, we manually construct the pickle stream by sequentially pushing opcodes to reconstruct the code object and function instance via \textit{types.CodeType} and \textit{types.FunctionType}, respectively, where \textit{types.CodeType} is utilized to reassemble the executable code object from $\mathbb{A}$, and \textit{types.FunctionType} is then employed to instantiate a callable function from the code object. Finally, a \textit{CALL} opcode is appended to trigger the execution during deserialization.
	
\begin{algorithm}[!t]
	\caption{Construction of the trigger.}
	\label{alg:stealthy_trigger}
	\begin{algorithmic}[1]
		\Require Payload extraction function $F_{\text{ext}}$ (contains logic to extract and execute hidden payloads).
		\Ensure Serialized binary trigger $T_{\text{bin}}$.
		
		\Statex \textbf{Step 1: Decompose Function into Low-level Attributes}
		\State Extract the code object $O_{\text{code}}$ from $F_{\text{ext}}$.
		\State Parse $O_{\text{code}}$ to obtain fundamental attributes set $\mathbb{A}$:
		\State $\mathbb{A} \leftarrow \{ \text{arg\_count}, \text{bytecode}, \text{constants}, \text{names}, \text{stack\_size}, \dots \}$
		\Comment{These raw attributes do not contain flagged keywords like `eval'.}
		
		\Statex \textbf{Step 2: Manually Construct Pickle Stream}
		\State Initialize an empty binary stream $T_{\text{bin}}$.
		\State Append opcode to load global class \textit{types.CodeType} into $T_{\text{bin}}$.
		\State Append opcodes to load all elements in $\mathbb{A}$ into $T_{\text{bin}}$.
		\State Append opcode \textit{CALL} to instantiate a new code object $O'_{\text{code}}$ using $\mathbb{A}$.
		\Comment{Reconstructs the code object.}
		
		\State Append opcode to load global class \textit{types.FunctionType} into $T_{\text{bin}}$.
		\State Append opcode to load $O'_{\text{code}}$ and empty globals into $T_{\text{bin}}$.
		\State Append opcode \textit{CALL} to instantiate a new function $F'_{\text{ext}}$.
		\Comment{Reconstructs the executable function.}
		
		\Statex \textbf{Step 3: Embed Execution Command}
		\State Append opcode \textit{CALL} to execute $F'_{\text{ext}}$.
		\State Append opcode \textit{STOP} to finalize the stream.
		
		\State \Return $T_{\text{bin}}$
	\end{algorithmic}
\end{algorithm}

\subsection{EXPLODE stage} 
During the model deployment phase, users download $ S({\Theta}') $ from public repositories, and deploy the model with/without quantization. Then, the user de-serializes $S({\Theta}')$ to obtain ${\Theta}'$, whereby the injected trigger is de-serialized and activated. Next, the trigger extract the embedded payloads $\mathcal{S}$ through {\bf Algorithm \ref{alg:trigger}}. Firstly, the trigger acts on the currently loaded model parameters $\Theta'$. Since parameter perturbations essentially disrupt magnitude rankings, the trigger skips re-sorting and directly retrieves the synchronized target indices $I$ from its encapsulated metadata. It then exactly locates the corresponding target carriers $\tilde{G}'_\nu$.
Secondly, for each retrieved carrier $\tilde{\theta}' \in \tilde{G}'_\nu$, the trigger performs the inverse MAR-QIM operation. Specifically, it calculates the dynamic step size $ \Delta $ based on the parameter's magnitude $ E $ and the predefined relative bit position $ n' $ . It then quantizes $ \tilde{\theta}' $ by $ \Delta $ to obtain the integer grid index $ v = \left\lfloor \tilde{\theta}' / \Delta \right\rceil$. The parity $ b' = v \pmod 2 $ is mapped back to the chip bit $ c' \in \{-1, +1\} $ (e.g., $ 0 \to -1, 1 \to +1 $), thereby reconstructing the chip sequence $ \mathcal{C}' $.
Thirdly, the trigger demodulates $\mathcal{C}'$ with the spreading gain $g$ to obtain the soft symbol sequence $\mathcal{Y}_{soft}$. Leveraging the known preamble $\mathcal{P}_{pre}$, it estimates the channel mean $\mu$ and noise variance $\sigma^2$ to formulate Log-Likelihood Ratios (LLRs). Crucially, to mitigate false-confidence outliers induced by non-linear quantization noise, these LLRs are symmetrically clipped to $[-\tau, \tau]$ before being fed into the Belief Propagation (BP)-based LDPC decoder. This pipeline effectively filters quantization disruptions, thereby bounding the error rate and successfully recovering the raw payload $\mathcal{S}_{raw}$.
Finally, the trigger executes the malicious payload $ \mathcal{S}_{raw} $ to complete the MEA.

\begin{algorithm}[!t]
	\caption{Payload extraction in EXPLODE stage.}
	\label{alg:trigger}
	\small
	\begin{algorithmic}[1]
		\Require Embedded parameters $\Theta'$ (or $\hat{\Theta}'$), spreading gain $g$, relative bit position $n'$. 
		\Ensure Executed payload $\mathcal{S}_{raw}$. 
		
		\State Initialize chip sequence $\mathcal{C}'$ of length $K$
		
		\State Parse the trigger to retrieve original target indices $I$
		\State Let $\tilde{G}'_\nu \gets \{ \Theta'[k], k \in I \}$
		\For{$k = 1$ to $K$}
		\State $\tilde{\theta}' \gets \tilde{G}'_\nu[k]$
		\State $E \gets \lfloor \log_2(|\tilde{\theta}'|) \rfloor$ \Comment{Get magnitude level}
		\State Calculate $\Delta \gets 2^{E - l_m + n'}$ or $2^{E - n'}$, where $l_m$ is the mantissa length of float \Comment{Calculate step size}
		\State $v \gets \left\lfloor \tilde{\theta}' / \Delta \right\rceil$
		\State $b' \gets v \pmod 2$ \Comment{Extract parity}
		\State Map parity $b'$ to chip $c'$ (e.g., $0 \to -1, 1 \to +1$)
		\State $\mathcal{C}'[k] \gets c'$
		\EndFor
		
		\State $\mathcal{Y}_{soft} \gets \operatorname{De\_SpreadSpectrum}(\mathcal{C}', g)$
		\State Estimate signal mean $\mu$ and variance $\sigma^2$ from $\mathcal{P}_{pre}$ to compute $LLR$
		\State $LLR \gets \operatorname{Clip}(LLR, -\tau, \tau)$ \Comment{Mitigate outlier confidences from quantization}
		\State $\mathcal{S}_{raw} \gets \operatorname{LDPC\_Decode}(LLR)$
		\State Execute payload $\mathcal{S}_{raw}$
		
		\State \Return None
	\end{algorithmic}
\end{algorithm}

\section{Experiments and Evaluation}
In this section, we conduct experiments on four prevailing open-source LLMs to evaluate the performance of MEASER.

\subsection{Experiment Setup}

\textbf{Testbed.} 
All experiments are conducted on a cloud server with the Ubuntu 22.04.4 LTS operating system (GNU/Linux 5.15.0-106-generic kernel) and x86\_64 architecture with 2 NVIDIA RTX A6000 GPUs, managed through CUDA Toolkit 12.2. The software environment is PyTorch 2.3.0+cu121 framework with Python 3.10.14.

{\noindent\textbf{Models.}} We choose Llama-2-7b-chat-hf \cite{Touvron2023-arxiv-2}, Llama-2-13b-chat-hf \cite{Touvron2023-arxiv-2}, Chatglm3-6b \cite{GLM2024-arxiv}, and Qwen3-4B-Instruct-2507 \cite{Qwen3-2025-arxiv} as the targeted LLMs, which are widely deployed with diverse model sizes, architectures, and multilingualism. Llama-2-7b-chat-hf has 32 Transformer layers with 6.74 billion parameters and 12.6GB in size. Llama-2-13b-chat-hf has 40 Transformer layers with 13.02 billion parameters and 24.2GB in size. Chatglm3-6b has 28 Transformer layers with 6.24 billion parameters and 11.6GB in size. Qwen3-4B-Instruct-2507 has 36 Transformer layers with 4.02 billion parameters and 7.5GB in size. 

{\noindent\textbf{Baselines.}}
As the first MEA tailored for open-source LLMs, we evaluated the superiority of MEASER compared to 5 SOTA MEAs tailored for general DDNs, i.e., X-MSB attacks \cite{Gilkarov2024-TIFS}, X-LSB attacks \cite{Gilkarov2024-TIFS}, Malmodel \cite{Hua2025-ASE}, MaleficNet \cite{Hitaj2025-TDSC} and FREEZER \cite{Yuan2025-ASC}. In a nutshell, X-MSB and X-LSB embed payloads repeatedly into the most or least significant bits of all parameters. Malmodel adaptively prioritizes the least significant bits of the Dense layer. MaleficNet utilizes redundant error correction bits to encode payloads. FREEZER employs Hamming codes and redundancy mechanisms for embedding.

{\noindent\textbf{Defenses.}}
We evaluate MEASER's robustness against quantization using four prevailing methods with default configurations: 4-bit AWQ \cite{Lin2024-MLSys}, 4-bit GPTQ \cite{Elias2023-ICLR}, 4-bit GGUF\footnote{https://github.com/ggml-org/ggml/blob/master/docs/gguf.md.} (Q4\_0) and 8-bit GGUF (Q8\_0). Furthermore, we assess robustness against PEFT via LoRA \cite{Hu2022-ICLR} and P-tuning \cite{Liu2022-ACL}, conducting training for 3 epochs on the MMLU dataset. Notably, in the LoRA setting, the fine-tuning scope includes the parameters hosting the embedded payload.

{\noindent\textbf{Evaluation data and metrics.}}
We utilize MMLU \cite{Hendrycks2021-ICLR} and AGIEval \cite{Zhong2024-ACL} datasets to evaluate the attack performance, which are widely used in evaluating LLMs. For each dataset, we select a ratio of 1:9 for calibration data to evaluation data. To evaluate MEASER in real-world scenarios, we leverage a diverse set of real-world malware payloads sourced from TheZoo\footnote{https://thezoo.morirt.com.}, an open repository hosting extensive malware families. Our payloads comprises 12 distinct malware binaries with sizes ranging from kilobytes to megabytes, as detailed in Table \ref{tab:payloads}. 

\begin{table}[h]
	
	\caption{The malware payloads}
	\label{tab:payloads}
	
	\centering
	
	\footnotesize
	\setlength{\tabcolsep}{4pt}
	
	\begin{tabular}{cc|cc|cc}
		\toprule
		Malware & Size & Malware & Size & Malware & Size \\
		
		\midrule
		
		Stuxnet & 0.02MB & Destover & 0.08MB & Asprox & 0.09MB \\		
		Bladabindi & 0.10MB & Zeus-Bank & 0.25MB & EquationDrug & 0.36MB \\
		Zeus-Dec & 0.40MB & Kovter & 0.41MB & Cerber & 0.59MB \\
		Ardamax & 0.77MB & NSIS & 1.70MB & Kelihos & 1.88MB \\
		
		\bottomrule
		
	\end{tabular}
	
\end{table}

We leverage the bit error rate ($\operatorname{BER}$) \cite{Yuan2025-ASC} to evaluate the attack effectiveness, which measures the ratio of incorrect bits in the recovered payload to the total length of the original payload, defined as:
\begin{equation}
	\operatorname{BER} = \frac{1}{l_s} \sum_{i=1}^{l_s} |s_i - s'_i|,
	\label{eq:ber}
\end{equation}
where $l_s$ represents the total length of the payload, and $s_i$ and $s'_i$ denote the $i$-th bit of the original payload $s$ and the recovered payload $s'$, respectively. Notably, the attack is deemed successful only if $\operatorname{BER} = 0$, while $\operatorname{BER} = 0.5$ indicates a random sequence. To evaluate the attack stealthiness, we test the detection rate $\operatorname{DR}$ \cite{Hitaj2022-ESORICS} of the anti-virus program VirusTotal \cite{VirusTotalOfficial} and 3 widely deployed AI model detectors, including Protect AI\footnote{https://huggingface.co/docs/hub/en/security-protectai.}, ClamAV\footnote{https://huggingface.co/docs/hub/en/security-malware.}, and HF Picklescan\footnote{https://huggingface.co/docs/hub/en/security-pickle.} on Hugging Face. In this respect, we also define the stealth rate ($\operatorname{SR}$) to record the performance degradation of targeted models under MEAs, expressed as:
\begin{equation}
	\operatorname{SR} = 1- d_{\textup{PAI}}. 
	\label{equ:SR}
\end{equation}

\noindent\textbf{Settings.} By default, we concatenate all malware payloads into a single sequence and embed it repeatedly to calculate the $ d_{\textup{PAI}} $. During the embedding of MEASER, we set spreading gain $g = 6$ and the relative bit position $ n'= 1 $. The choice of $g$ has been proven by Hitaj {\it et al.} \cite{Hitaj2025-TDSC} to achieve an optimal trade-off between payload robustness and model utility, while the selection of $n'$ is demonstrated in Section \ref{sec:results and evaluation} to have a negligible impact on model performance.

\subsection{Results and Evaluation}
\label{sec:results and evaluation}

\begin{table*}[htbp]
	\centering
	\caption{The comparison of $\operatorname{BER}$, $\operatorname{SR}$, and $\operatorname{DR}$ between MEASER and baselines. The best results are highlighted in bold.}
	\label{tab:results}
	\vspace{-3mm}
	\setlength{\tabcolsep}{1.5pt}
	\renewcommand{\arraystretch}{1.25}
	\resizebox{\textwidth}{!}{%
	\begin{tabular}{c||*{7}{c}||*{7}{c}||*{4}{c}}
		\hline\hline
		& \multicolumn{7}{c||}{$\operatorname{BER}$ \textdownarrow} & \multicolumn{7}{c||}{$\operatorname{SR}$ (\%) \textuparrow} & \multicolumn{4}{c}{$\operatorname{DR}$ (\%) \textdownarrow} \\
		\cline{2-19}
		Attack & General & Q8\_0 & Q4\_0 & AWQ & GPTQ & LoRA & P-tuning & General & Q8\_0 & Q4\_0 & AWQ & GPTQ & LoRA & P-tuning & VirusTotal & Protect AI & ClamAV & HF Picklescan \\
		\hline
		& \multicolumn{18}{c}{Results on Llama-2-7b-chat-hf} \\
		\hline
		X-MSB \cite{Gilkarov2024-TIFS} & 0.00 & 0.10 & 0.41 & 0.48 & 0.46 & 0.27 & 0.00 & 99.3 & 99.7 & 99.3 & 98.5 & 97.0 & 93.5 & 94.6 & 0 & 0 & 0 & 0 \\
		\hline
		X-LSB \cite{Gilkarov2024-TIFS} & 0.00 & 0.48 & 0.49 & 0.50 & 0.50 & 0.50 & 0.00 & 99.3 & 99.9 & 99.3 & 99.2 & 98.5 & 95.1 & 94.8 & 0 & 0 & 0 & 0 \\
		\hline
		Malmodel \cite{Hua2025-ASE} & 0.00 & 0.41 & 0.48 & 0.49 & 0.48 & 0.49 & 0.00 & 99.3 & 99.9 & 99.9 & 99.8 & 98.5 & 94.0 & 94.3 & 0 & 0 & 0 & 0 \\
		\hline
		MaleficNet \cite{Hitaj2025-TDSC} & 0.00 & 0.00 & 0.00 & 0.00 & 0.00 & 0.00 & 0.00 & 0.15 & 0.16 & 0.17 & 0.14 & 0.18 & 44.8 & 46.1 & 0 & 0 & 0 & 0 \\
		\hline
		FREEZER \cite{Yuan2025-ASC} & 0.00 & 0.10 & 0.13 & 0.25 & 0.12 & 0.19 & 0.00 & 99.3 & 99.9 & 99.3 & 99.8 & 99.3 & 97.6 & 97.5 & 0 & 0 & 0 & 0 \\
		\hline
		\rowcolor{gray!10} MEASER & {\bf 0.00} & {\bf 0.00} & {\bf 0.00} & {\bf 0.00} & {\bf 0.00} & {\bf 0.00} & {\bf 0.00} & {\bf 99.8} & {\bf 99.9} & {\bf 99.9} & {\bf 99.9} & {\bf 99.5} & {\bf 98.2} & {\bf 97.9} & {\bf 0} & {\bf 0} & {\bf 0} & {\bf 0} \\
		\hline
		& \multicolumn{18}{c}{Results on Chatglm3-6b} \\
		\hline
		X-MSB & 0.00 & 0.11 & 0.44 & 0.42 & 0.41 & 0.38 & 0.00 & 99.3 & 99.1 & 99.3 & 98.1 & 97.2 & 93.4 & 94.4 & 0 & 0 & 0 & 0 \\
		\hline
		X-LSB & 0.00 & 0.49 & 0.50 & 0.50 & 0.50 & 0.50 & 0.00 & 99.4 & 99.9 & 99.3 & 99.1 & 98.1 & 95.4 & 94.2 & 0 & 0 & 0 & 0 \\
		\hline
		Malmodel & 0.00 & 0.45 & 0.49 & 0.49 & 0.49 & 0.49 & 0.00 & 99.4 & 99.9 & 99.9 & 99.6 & 98.0 & 95.0 & 93.9 & 0 & 0 & 0 & 0 \\
		\hline
		MaleficNet  & 0.00 & 0.00 & 0.00 & 0.00 & 0.00 & 0.00 & 0.00 & 0.11 & 0.12 & 0.12 & 0.14 & 0.08 & 41.1 & 42.7 & 0 & 0 & 0 & 0 \\
		\hline
		FREEZER & 0.00 & 0.16 & 0.11 & 0.22 & 0.16 & 0.12 & 0.00 & 99.3 & 99.9 & 99.4 & 99.4 & 99.3 & 96.2 & 96.3 & 0 & 0 & 0 & 0 \\
		\hline
		\rowcolor{gray!10} MEASER & {\bf 0.00} & {\bf 0.00} & {\bf 0.00} & {\bf 0.00} & {\bf 0.00} & {\bf 0.00} & {\bf 0.00} & {\bf 99.8} & {\bf 99.9} & {\bf 99.9} & {\bf 99.9} & {\bf 99.4} & {\bf 97.7} & {\bf 97.5} & {\bf 0} & {\bf 0} & {\bf 0} & {\bf 0} \\
		\hline
		& \multicolumn{18}{c}{Results on Qwen3-4B-Instruct-2507} \\
        \hline
        X-MSB & 0.00 & 0.12 & 0.43 & 0.46 & 0.44 & 0.31 & 0.00 & 99.2 & 99.0 & 99.1 & 98.0 & 96.9 & 93.1 & 94.2 & 0 & 0 & 0 & 0 \\
        \hline
        X-LSB & 0.00 & 0.47 & 0.49 & 0.50 & 0.50 & 0.49 & 0.00 & 99.3 & 99.8 & 99.1 & 99.0 & 98.2 & 95.1 & 94.0 & 0 & 0 & 0 & 0 \\
        \hline
        Malmodel & 0.00 & 0.42 & 0.49 & 0.48 & 0.50 & 0.49 & 0.00 & 99.3 & 99.8 & 99.8 & 99.5 & 97.9 & 94.8 & 93.6 & 0 & 0 & 0 & 0 \\
        \hline
        MaleficNet  & 0.00 & 0.00 & 0.00 & 0.00 & 0.00 & 0.00 & 0.00 & 0.13 & 0.15 & 0.14 & 0.12 & 0.16 & 42.5 & 43.8 & 0 & 0 & 0 & 0 \\
        \hline
        FREEZER & 0.00 & 0.14 & 0.12 & 0.23 & 0.15 & 0.15 & 0.00 & 99.1 & 99.8 & 99.2 & 99.3 & 99.1 & 96.0 & 95.9 & 0 & 0 & 0 & 0 \\
        \hline
        \rowcolor{gray!10} MEASER & {\bf 0.00} & {\bf 0.00} & {\bf 0.00} & {\bf 0.00} & {\bf 0.00} & {\bf 0.00} & {\bf 0.00} & {\bf 99.7} & {\bf 99.8} & {\bf 99.9} & {\bf 99.8} & {\bf 99.3} & {\bf 97.6} & {\bf 97.4} & {\bf 0} & {\bf 0} & {\bf 0} & {\bf 0} \\
        \hline
        & \multicolumn{18}{c}{Results on Llama-2-13b-chat-hf} \\
        \hline
        X-MSB & 0.00 & 0.09 & 0.40 & 0.47 & 0.45 & 0.25 & 0.00 & 99.4 & 99.8 & 99.4 & 98.6 & 97.3 & 93.8 & 94.9 & 0 & 0 & 0 & 0 \\
        \hline
        X-LSB & 0.00 & 0.46 & 0.49 & 0.50 & 0.50 & 0.49 & 0.00 & 99.4 & 99.9 & 99.5 & 99.3 & 98.6 & 95.5 & 95.1 & 0 & 0 & 0 & 0 \\
        \hline
        Malmodel & 0.00 & 0.39 & 0.47 & 0.48 & 0.47 & 0.48 & 0.00 & 99.4 & 99.9 & 99.9 & 99.9 & 98.8 & 94.3 & 94.6 & 0 & 0 & 0 & 0 \\
        \hline
        MaleficNet  & 0.00 & 0.00 & 0.00 & 0.00 & 0.00 & 0.00 & 0.00 & 0.18 & 0.19 & 0.20 & 0.16 & 0.21 & 46.5 & 47.3 & 0 & 0 & 0 & 0 \\
        \hline
        FREEZER & 0.00 & 0.11 & 0.12 & 0.24 & 0.11 & 0.20 & 0.00 & 99.4 & 99.9 & 99.5 & 99.9 & 99.4 & 97.9 & 97.7 & 0 & 0 & 0 & 0 \\
        \hline
        \rowcolor{gray!10} MEASER & {\bf 0.00} & {\bf 0.00} & {\bf 0.00} & {\bf 0.00} & {\bf 0.00} & {\bf 0.00} & {\bf 0.00} & {\bf 99.9} & {\bf 99.9} & {\bf 99.9} & {\bf 99.9} & {\bf 99.6} & {\bf 98.5} & {\bf 98.1} & {\bf 0} & {\bf 0} & {\bf 0} & {\bf 0} \\
		\hline\hline
	\end{tabular}%
	}
\end{table*}

{\bf MEASER outperforms SOTA MEAs in terms of effectiveness and robustness.} As shown in Table \ref{tab:results}, the $\operatorname{BER}$ of most baselines exceeds 0.00 facing quantization and PEFT, indicating that these attacks fail in such scenarios. By comparison, MEASER consistently maintains a $\operatorname{BER}$ of 0.00 across all settings, including 4-bit/8-bit quantization and PEFT, which aligns with the performance of the SOTA method MaleficNet. Such an alignment confirms the superior robustness of MEASER. The reason behind this observation is that MEASER employs the MAR-QIM mechanism to adaptively establish safety margins based on parameter magnitudes, which effectively tolerate the numerical perturbations introduced by quantization and PEFT. Furthermore, the integration of spread spectrum modulation and LDPC codes provides robust error correction capabilities, ensuring payload survivability even under significant model modifications.

{\bf MEASER outperforms SOTA MEAs in terms of stealthiness.} As evidenced in Table \ref{tab:results}, MEASER consistently achieves the highest $\operatorname{SR}$ across all experimental settings. Notably, compared to the robust SOTA method MaleficNet, which sacrifices utility for stability, MEASER yields a substantial improvement in $\operatorname{SR}$, increasing it by over 99.7\% in quantization scenarios and 51.8\% in PEFT scenarios. We attribute this high stealthiness to the synergistic design of the PAI metric and the Top-$K$ strategy. Specifically, the PAI metric rigorously identifies the optimal target parameter groups and the maximum allowable relative bit position ($n^*$) for LSB-based embedding, ensuring that modifications remain strictly within the model's tolerance threshold. Concurrently, the Top-$K$ strategy effectively masks the embedding noise via parameters of significant magnitudes, making the perturbations negligible to model inference.

\begin{figure}[htbp]
	\centering
	\subfloat[\footnotesize Llama-2-7b]{
		\includegraphics[width=0.48\linewidth,trim=5 10 10 5,clip]{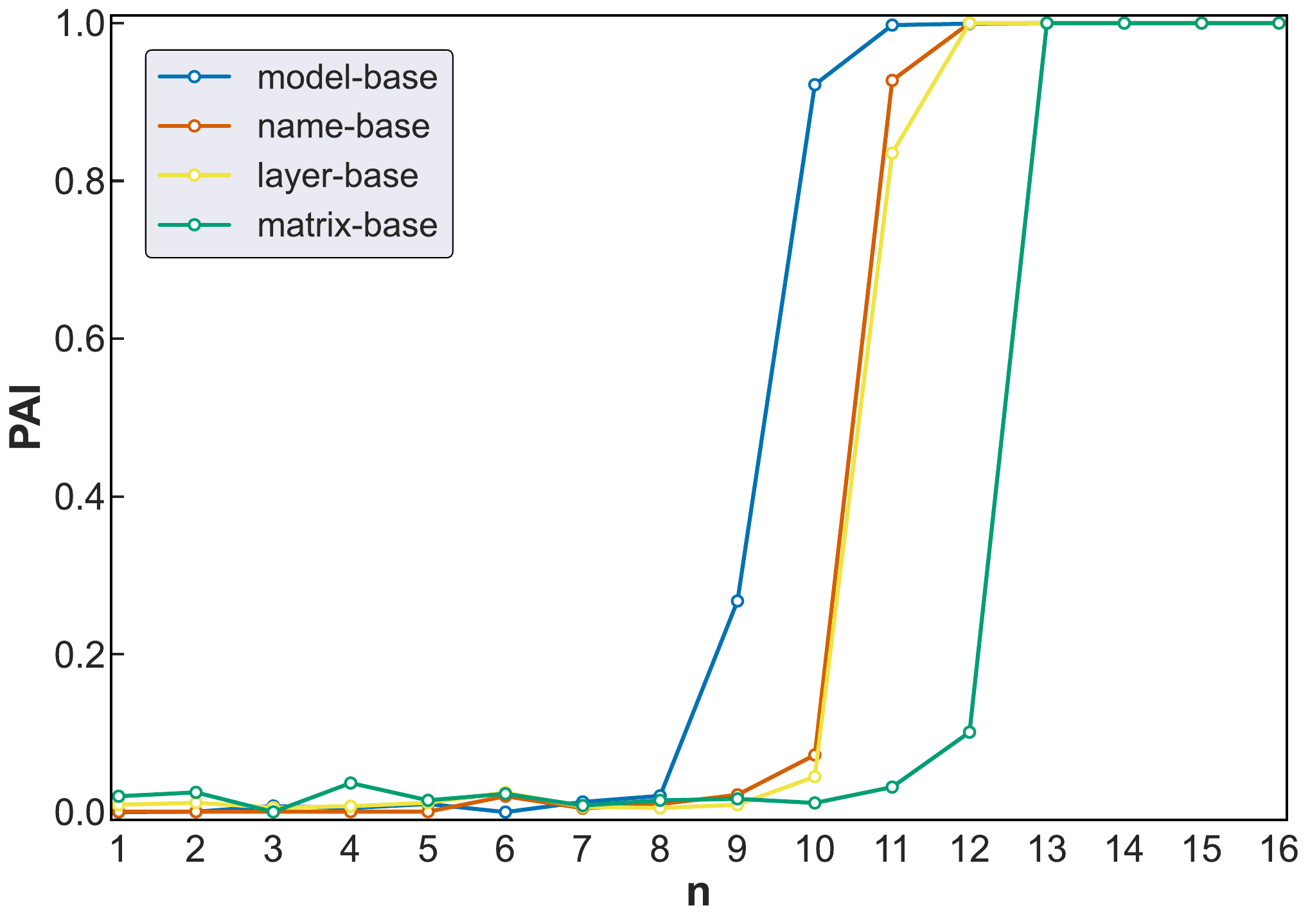}}
	\subfloat[\footnotesize Chatglm3-6b]{
		\includegraphics[width=0.48\linewidth,trim=5 10 10 5,clip]{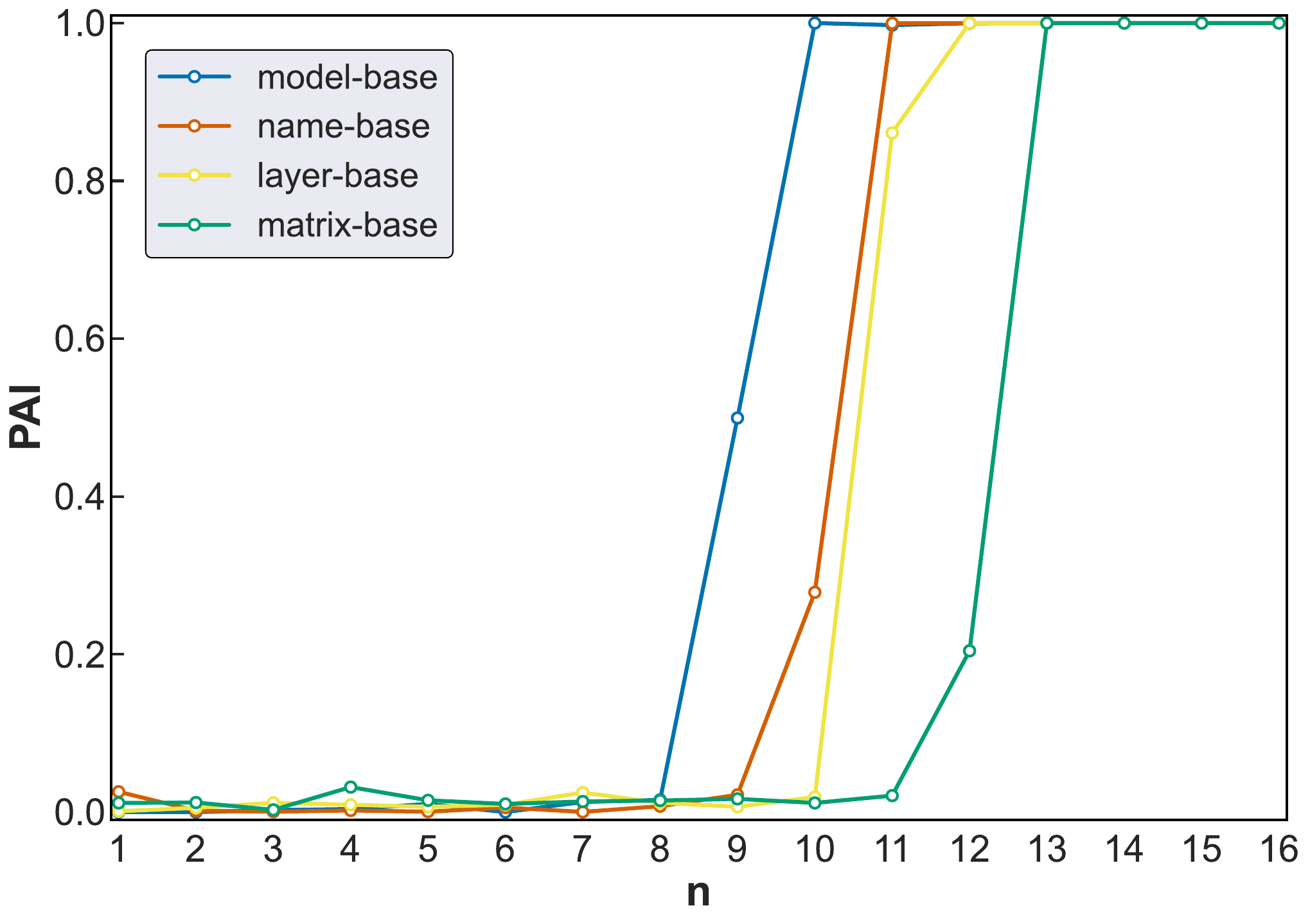}}
	\caption{$d_{\textup{PAI}}$ of grouping methods with $n$ ranging from 1 to 16. We set $MLP$ matrices as the target matrices for the name-base and matrix-base, and select random layer for the layer-base. Results are averaged over 3 runs with different random seeds.
	}
	\label{fig:grain effect}
\end{figure}

{\bf Ablation analysis for grouping methods of targeted parameters.}
We analyze the performance of four grouping methods for targeted parameters, i.e., model-base, name-base, layer-base, and matrix-base. Specifically, model-base takes the entire model as the target, name-base treats parameters with the same name as candidate groups (e.g., $MLP$ matrices \cite{Vaswani2017-NIPS} within all Transformer layers), layer-base targets Transformer layers, and matrix-base grouping targets individual matrices within a transformer layer. We record the $\operatorname{PAI}$ of these group methods with $n$ ranging from 1 to 16 in Figure \ref{fig:grain effect}. For every grouping method, $\operatorname{PAI}$ increases from 0 to 1 along with the increase of $n$, which indicates that MEASER could compromise open-source LLMs coupled with different grouping methods.

\begin{figure}[htbp]  
	\centering
	\subfloat[\footnotesize Llama-2-7b on MMLU]{\label{fig6:accuracy} 
		\includegraphics[width=0.48\linewidth,trim=10 15 35 10,clip]{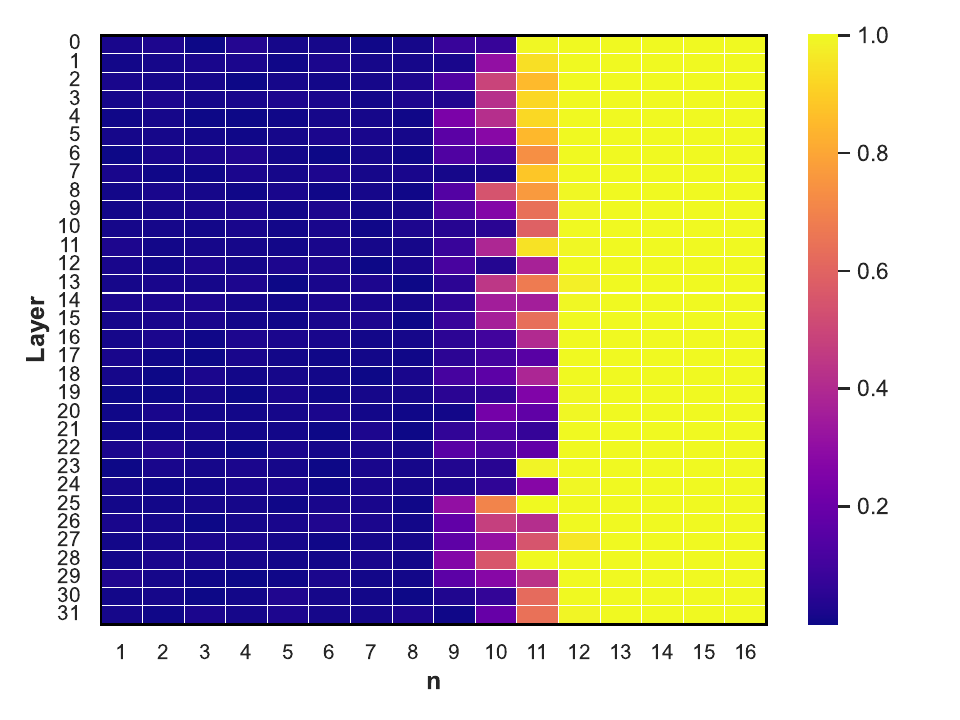}}  
	\hfill  
	\subfloat[\footnotesize Llama-2-7b on AGIEval]{\label{fig6:loss}
		\includegraphics[width=0.48\linewidth,trim=10 15 35 10,clip]{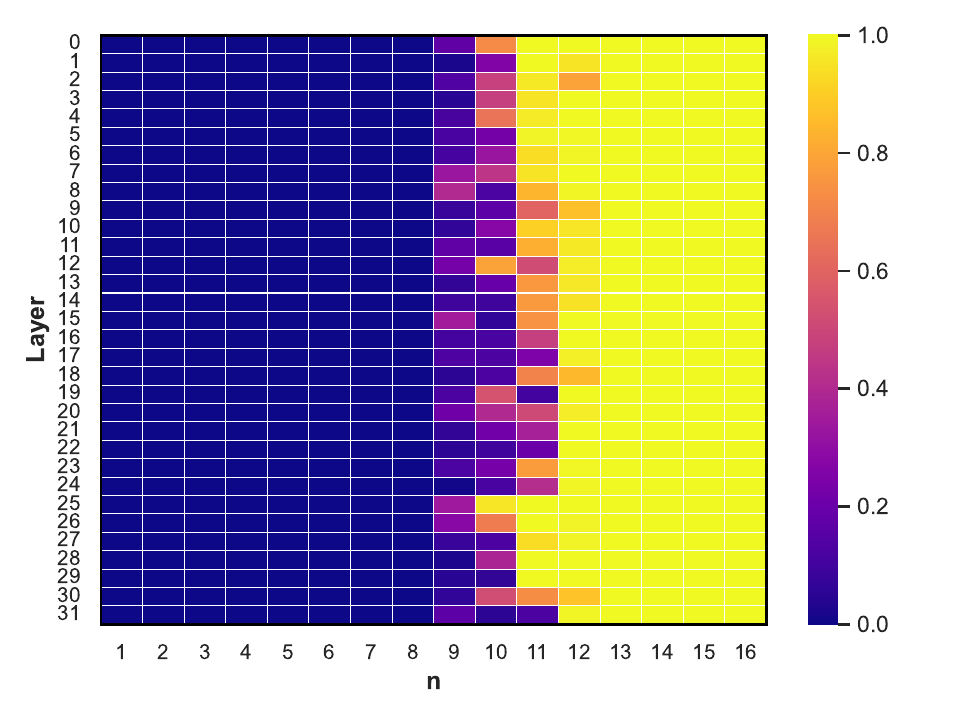}}
	\vspace{-2mm}
	\subfloat[\footnotesize Chatglm3-6b on MMLU]{\label{fig6:params}
		\includegraphics[width=0.48\linewidth,trim=10 15 35 10,clip]{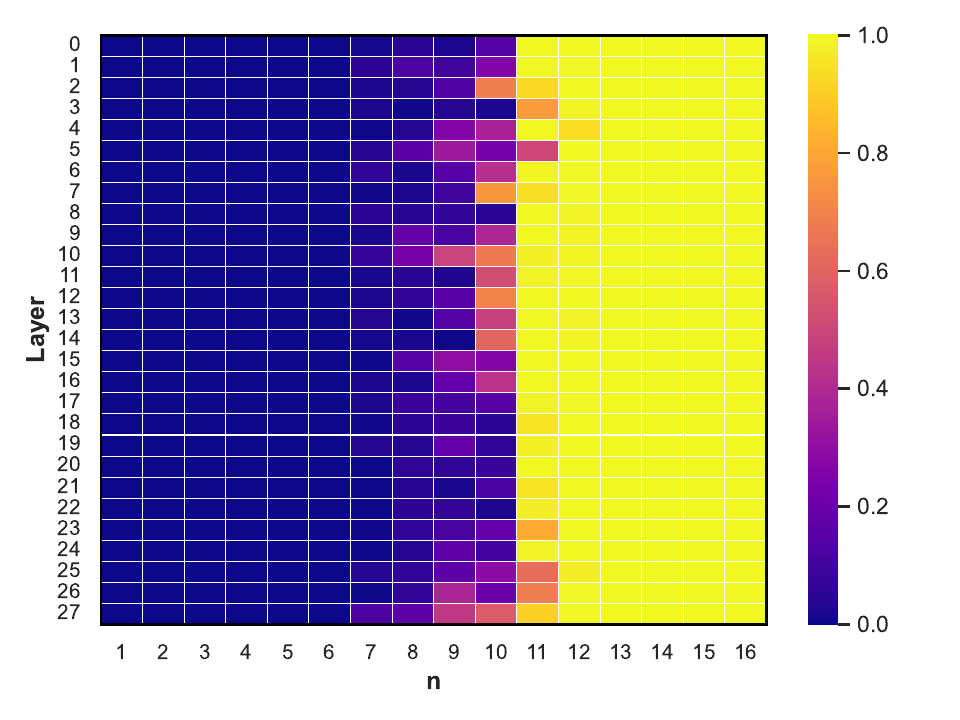}}
	\hfill  
	\subfloat[\footnotesize Chatglm3-6b on AGIEval]{\label{fig6:performance}
		\includegraphics[width=0.48\linewidth,trim=10 15 35 10,clip]{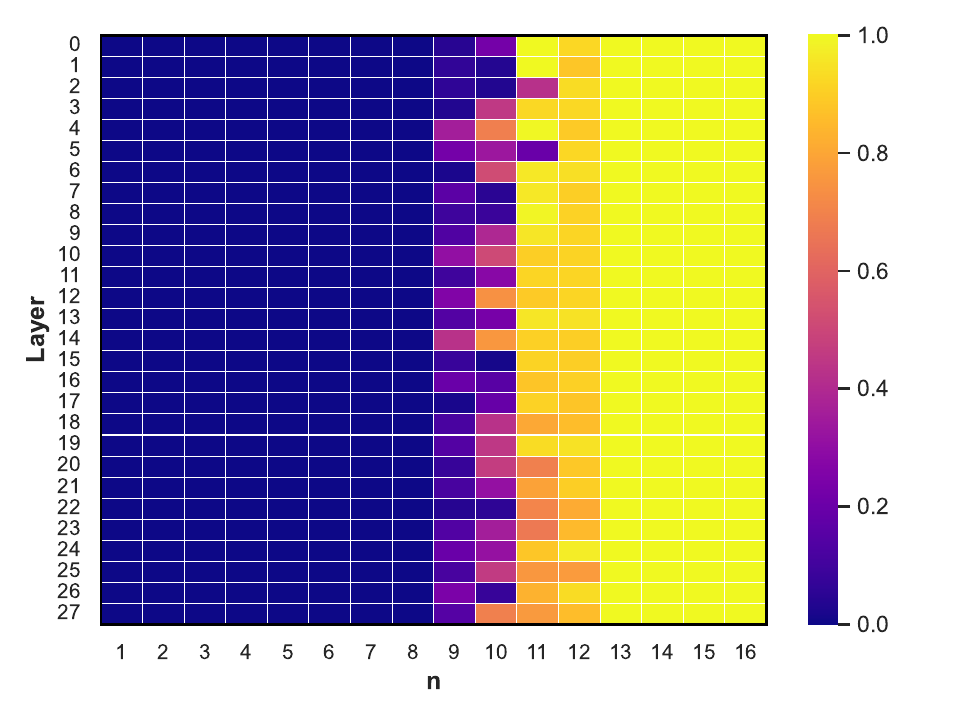}}
	\caption{$d_{\textup{PAI}}$ of models on MMLU and AGIEval. Results are averaged over 3 runs with different random seeds.}
	\label{fig:PAI_general}
\end{figure}

{\bf Ablation analysis for relative bit position.} We further analyze the impacts of relative bit position on the attack performance by adjusting $n$ from 1 to 16 in Figure \ref{fig:PAI_general}. For most layers, the $d_{\textup{PAI}}$ drops sharply when $n$ reaches 11, which suggests that the model performance is always impaired by a certain magnitude of payload bits. Before reaching this threshold, over half of the parameter bits could conceal the payload without detection. One interesting finding is that $d_{\textup{PAI}}$ depends on the model architecture instead of testing data, given similar attack performance of MEASER against the same model on different evaluation datasets. Such an observation validates the generality of PAI.

\begin{figure}[htbp]
	\centering
	\subfloat[\footnotesize Llama-2-7b]{\label{fig6:performance}
		\includegraphics[width=0.48\linewidth,trim=10 15 20 10,clip]{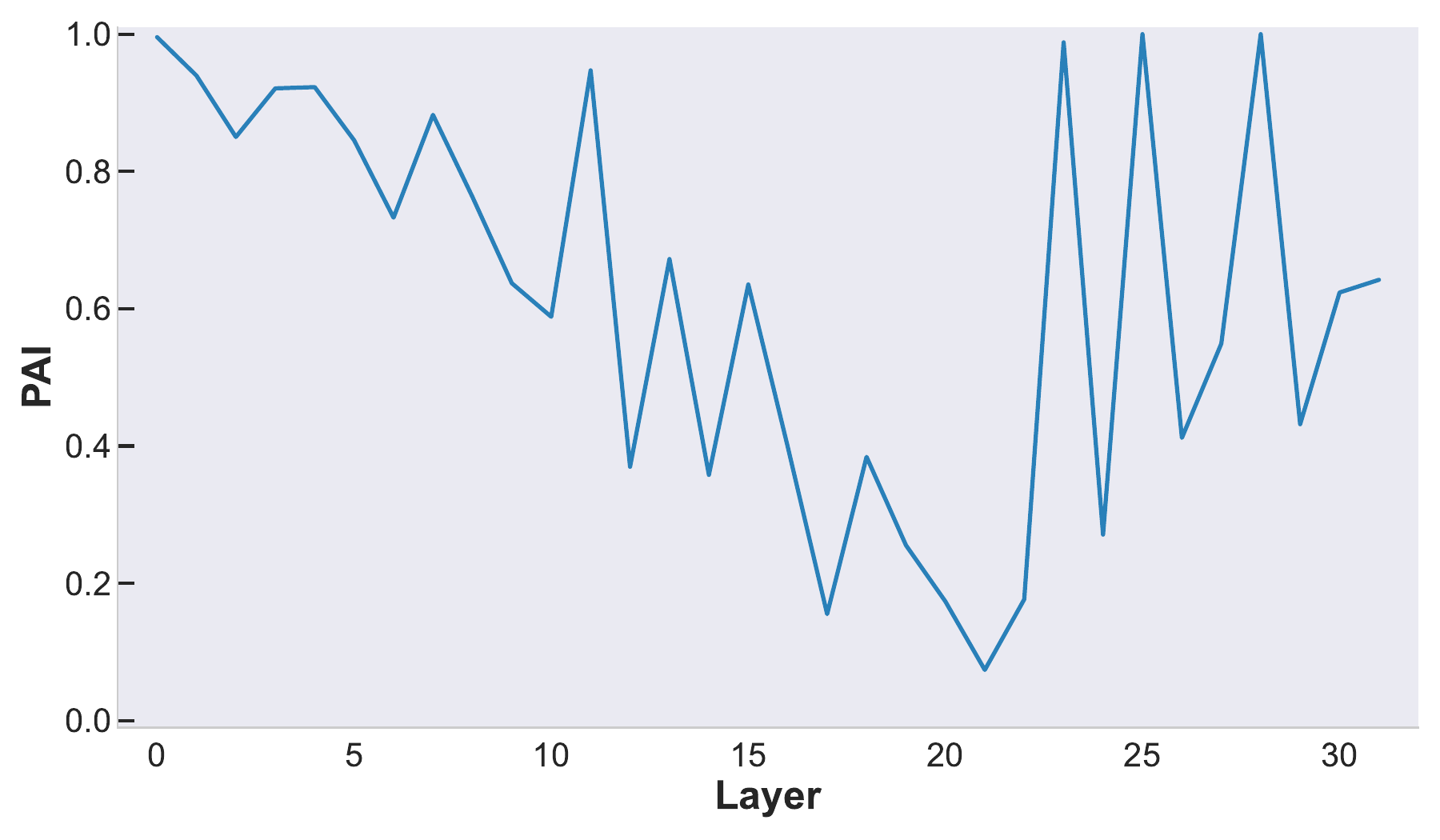}}
	\subfloat[\footnotesize Chatglm3-6b]{\label{fig6:performance}
		\includegraphics[width=0.48\linewidth,trim=10 15 20 10,clip]{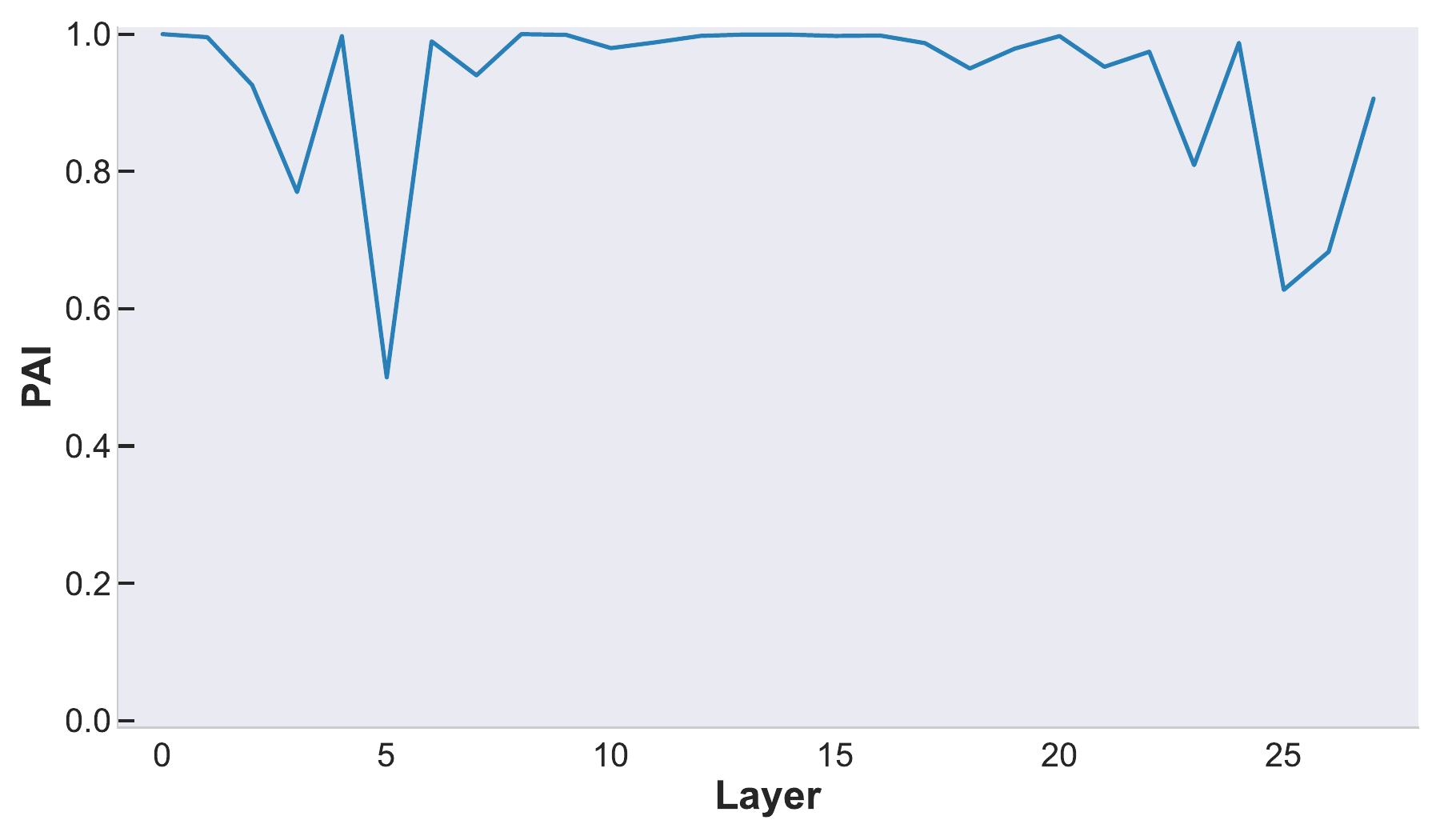}}
	\caption{$d_{\textup{PAI}}$ of models with $n$=11. Results are averaged over 3 runs with different random seeds on MMLU.}
	\label{fig:layer}
\end{figure}

{\bf Distribution of vulnerable parameters.} To understand the distribution of vulnerable parameters within open-source LLMs, we record the mean $d_{\textup{PAI}}$ at the sharply decreasing margin across different model layers in Figure \ref{fig:layer}. As can be seen, the distribution of vulnerable layers varies across distinct models, where vulnerable layers are located in the middle for Llama-2-7b-chat-hf and the front and back ends, to be more specific, the 5-th and 25-th layers, for Chatglm3-6b. The different vulnerability distributions might be attributed to distinct language processing and knowledge reasoning capabilities among various parameters and model structures.

\begin{figure}[htbp]
	\centering
	\subfloat[\footnotesize $D_{\textup{acc}}$ of Llama-2-7b]{\label{fig6:performance}
		\includegraphics[width=0.48\linewidth,trim=10 15 20 10,clip]{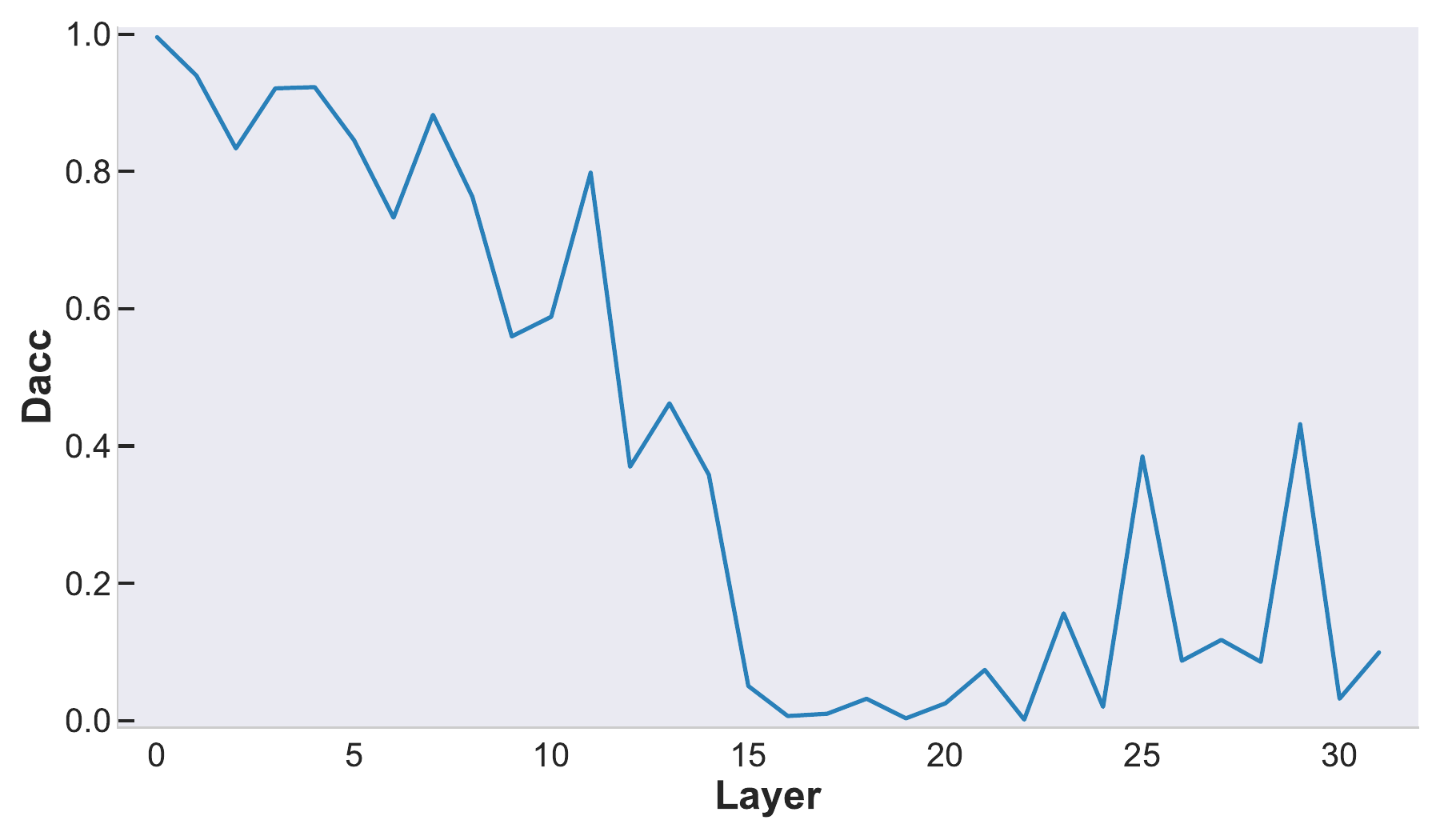}}
	\subfloat[\footnotesize $D_{\textup{acc}}$ of Chatglm3-6b]{\label{fig6:performance}
		\includegraphics[width=0.48\linewidth,trim=10 15 20 10,clip]{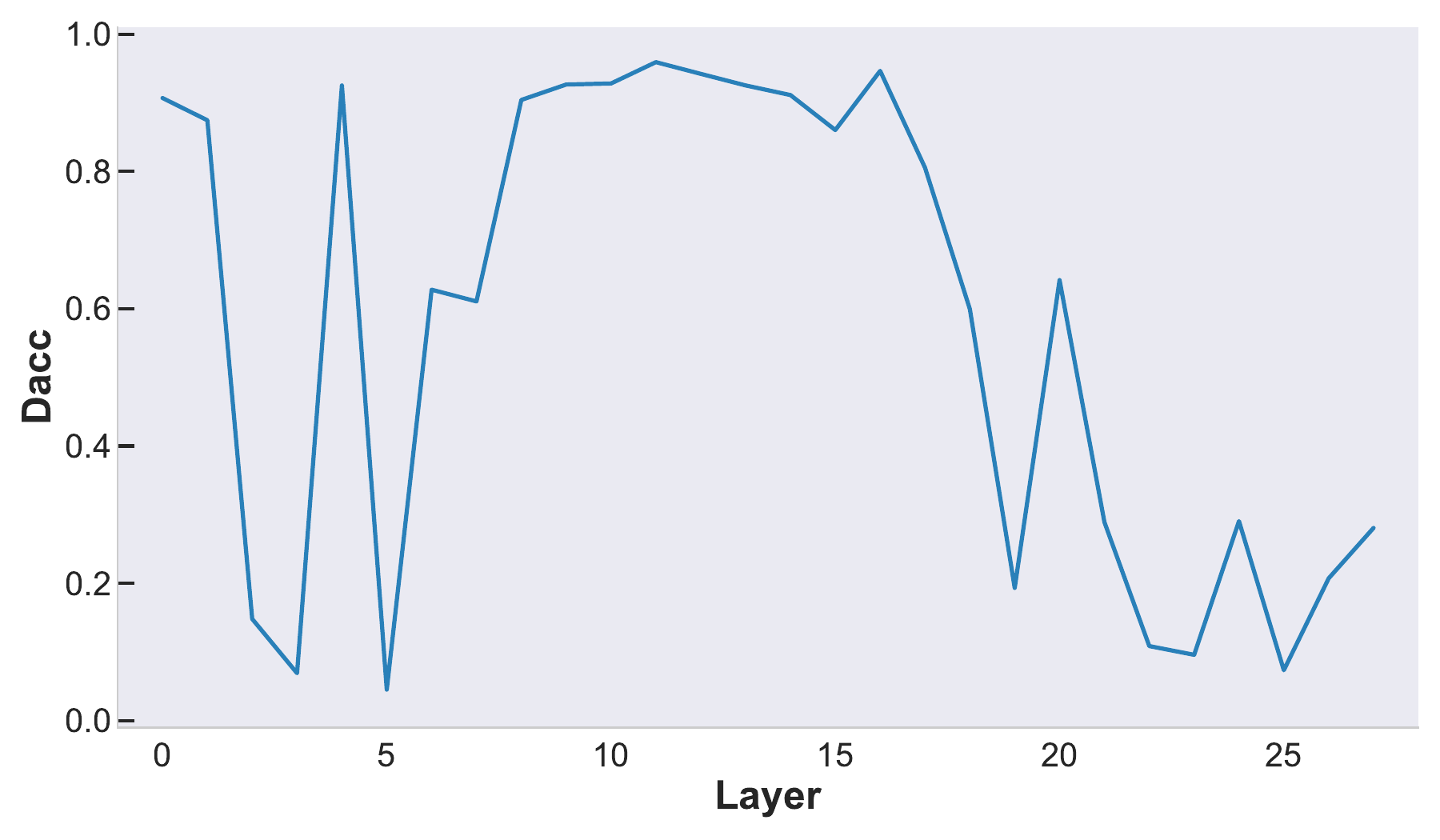}}\\
	\vspace{-2mm}
	\subfloat[\footnotesize $D_{\textup{ppl}}$ of Llama-2-7b]{\label{fig6:performance}
		\includegraphics[width=0.48\linewidth,trim=10 15 20 10,clip]{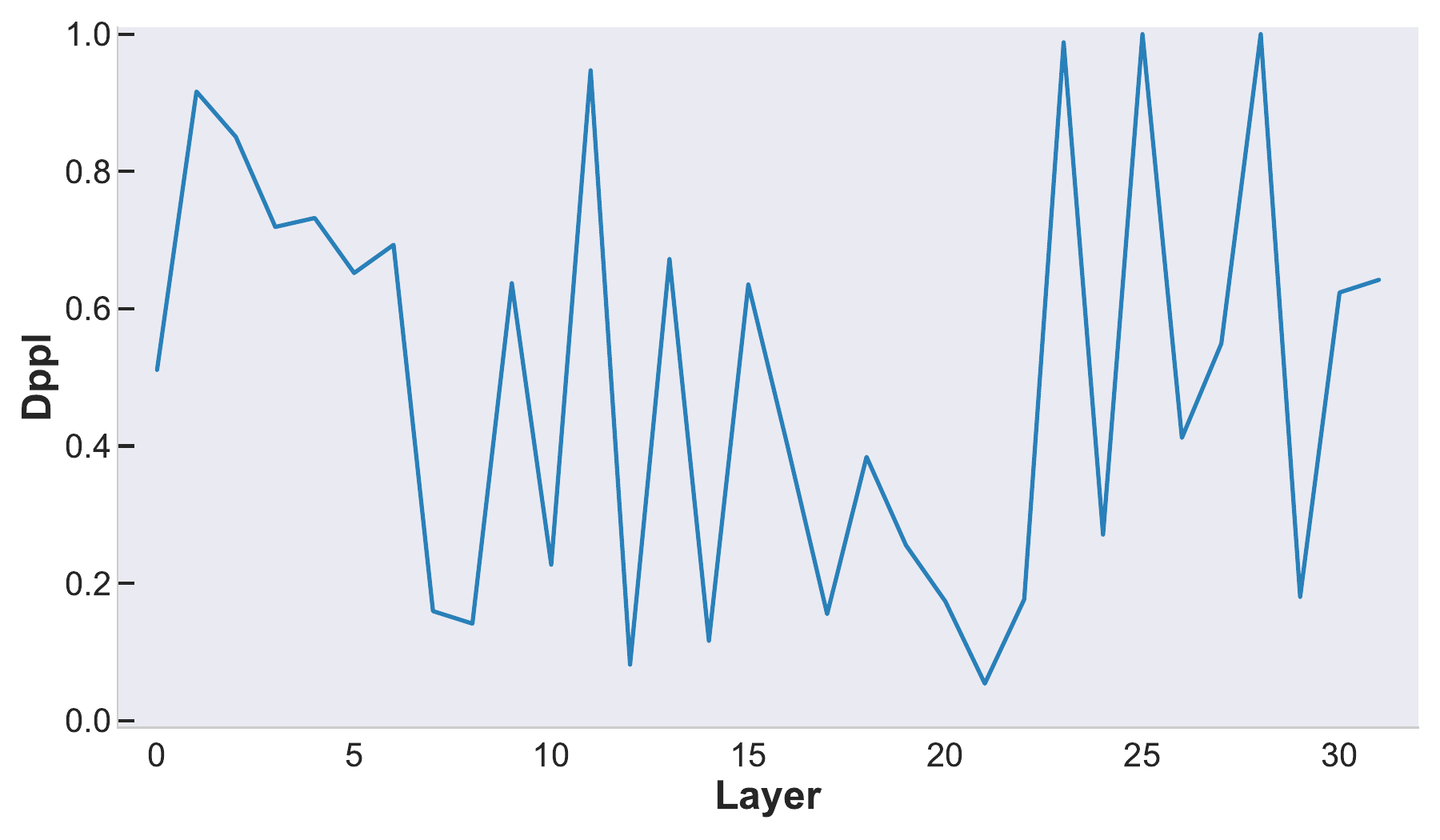}}
	\subfloat[\footnotesize $D_{\textup{ppl}}$ of Chatglm3-6b]{\label{fig6:performance}
		\includegraphics[width=0.48\linewidth,trim=10 15 20 10,clip]{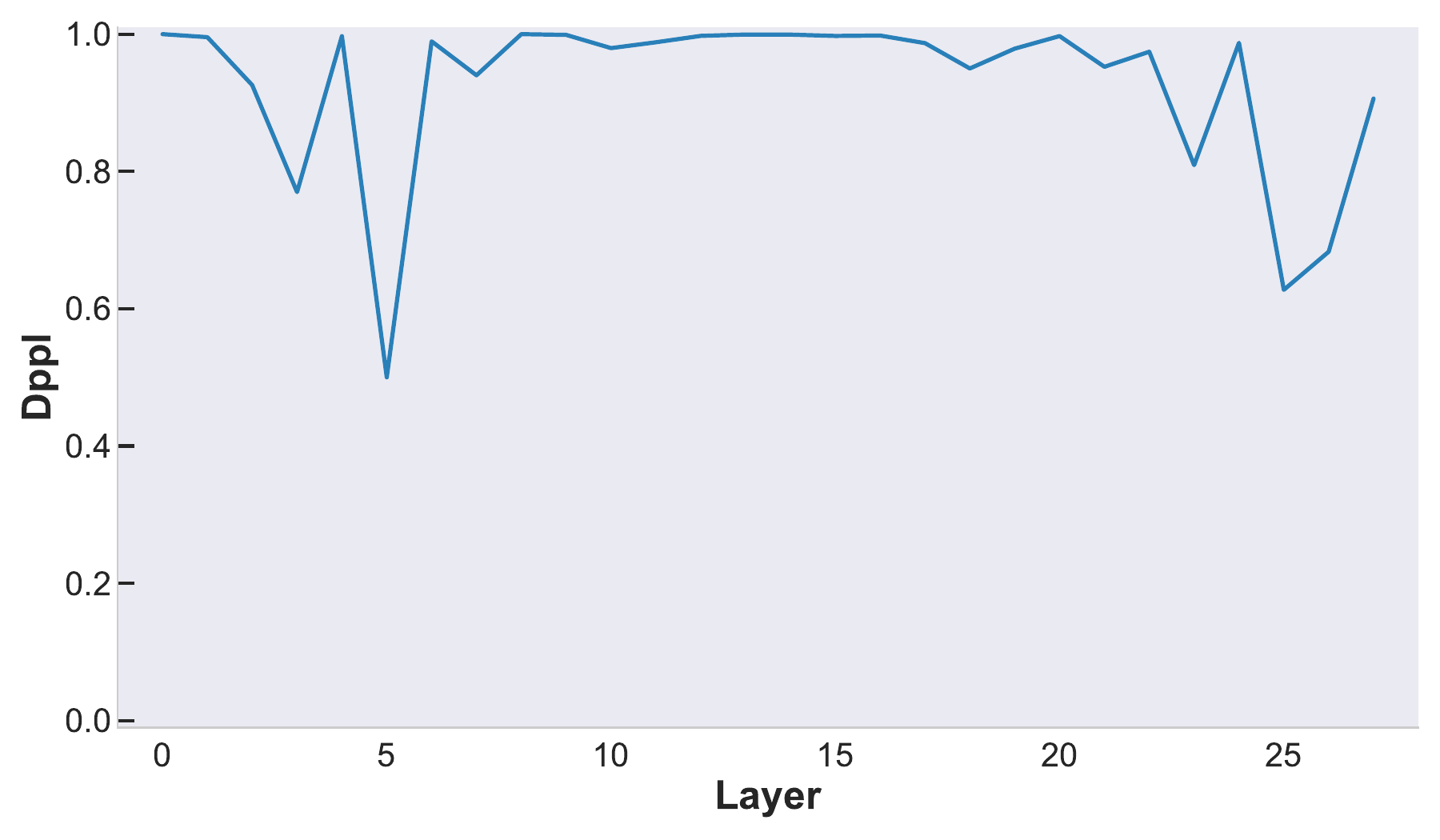}}
	\caption{$D_{\textup{acc}}$ \& $D_{\textup{ppl}}$ of models with $n$=11. Results are averaged over 3 runs with different random seeds on MMLU.}
	\label{fig:Dppl_11}
\end{figure}

\begin{figure}[htbp]
	\centering
	\subfloat[\footnotesize $D_{\textup{acc}}$ of Llama-2-7b]{
		\includegraphics[width=0.48\linewidth,trim=10 15 35 10,clip]{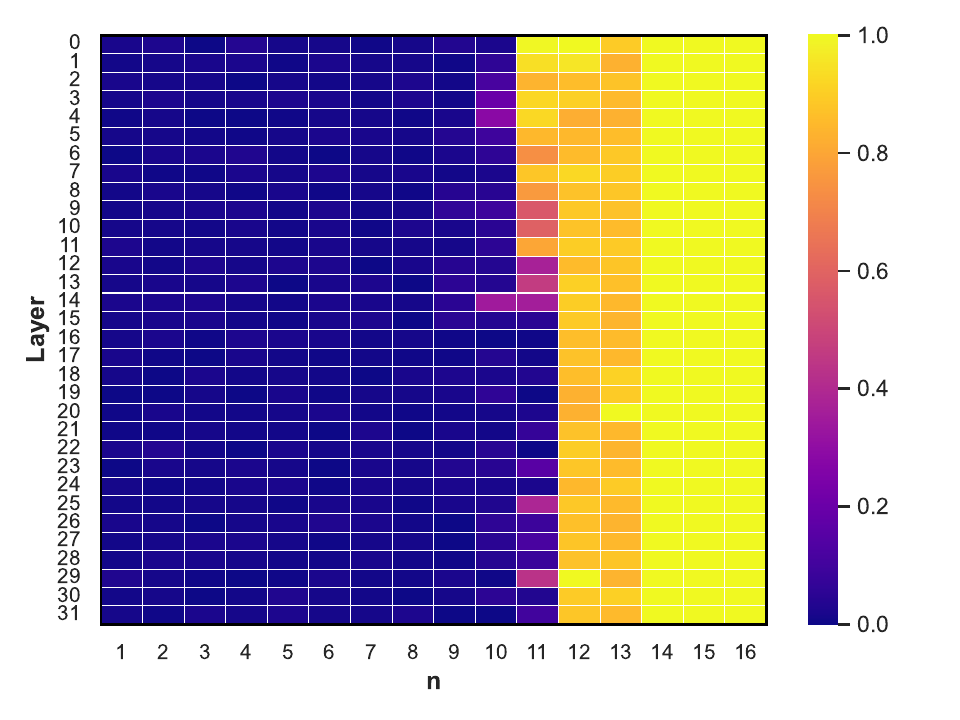}}
	\subfloat[\footnotesize $D_{\textup{acc}}$ of Chatglm3-6b]{
		\includegraphics[width=0.48\linewidth,trim=10 15 35 10,clip]{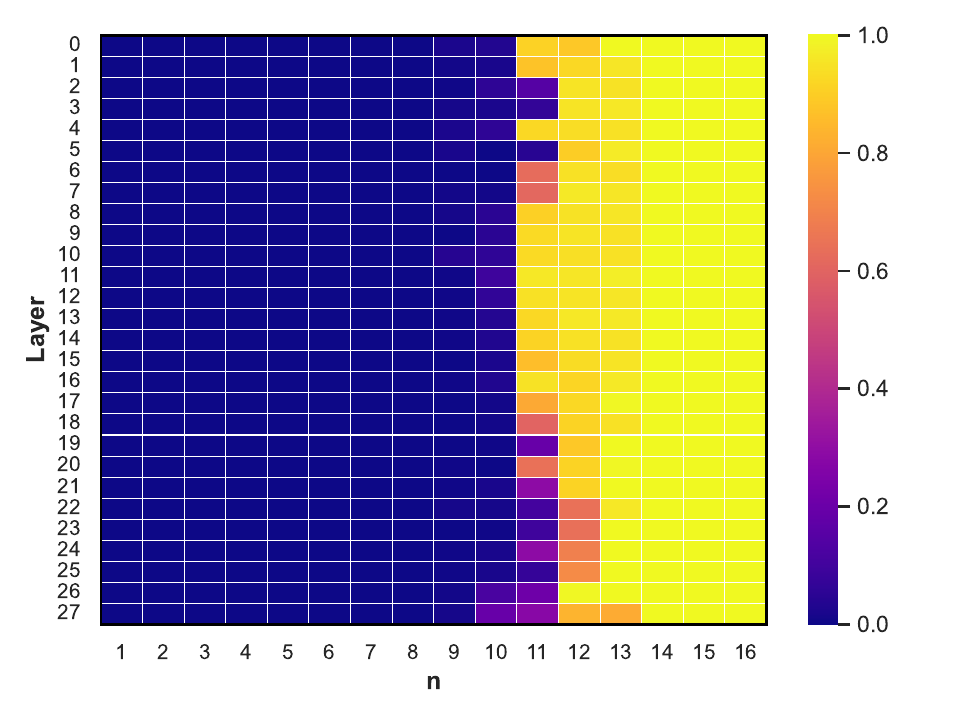}}\\
	\vspace{-1mm}
	\subfloat[\footnotesize $D_{\textup{ppl}}$ of Llama-2-7b]{
		\includegraphics[width=0.48\linewidth,trim=10 15 35 10,clip]{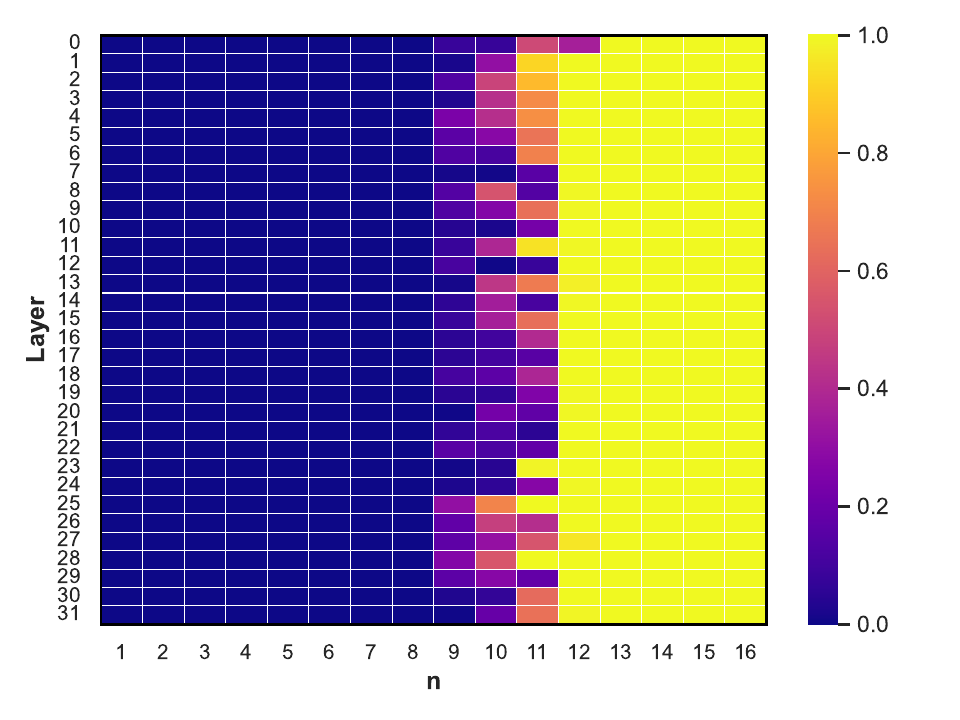}}
	\subfloat[\footnotesize $D_{\textup{ppl}}$ of Chatglm3-6b]{
		\includegraphics[width=0.48\linewidth,trim=10 15 35 10,clip]{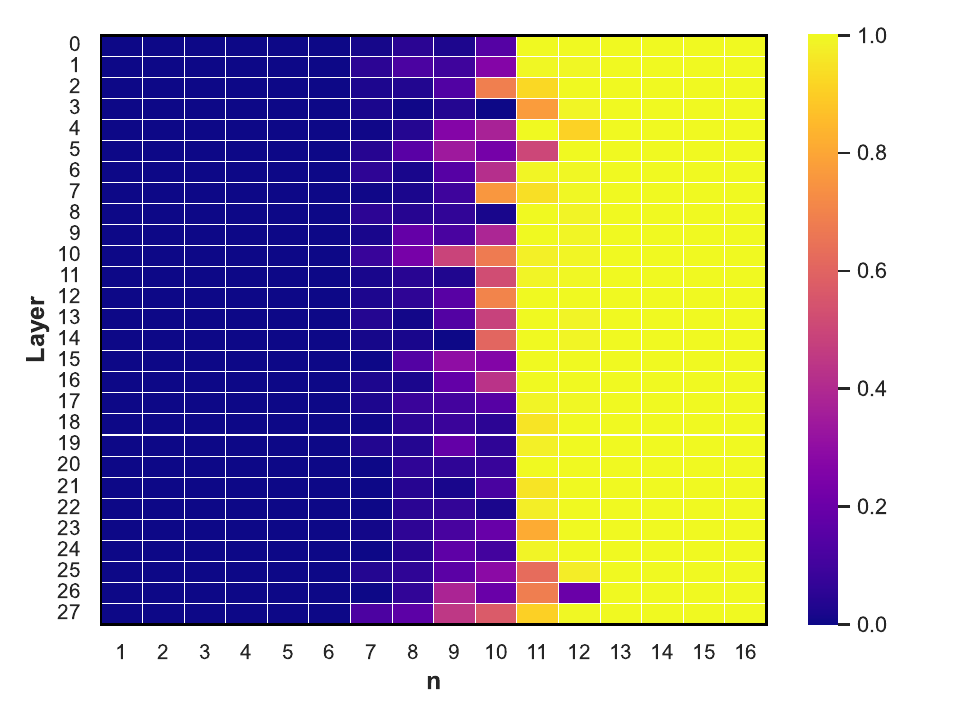}}
	\caption{$D_{\textup{acc}}$ \& $D_{\textup{ppl}}$ of LLaMA2 \& ChatGLM3 on the MMLU. Results are averaged over 3 runs with different random seeds.}
	\label{fig:Dppl}
\end{figure}

{\bf Contributions of $D_{\textup{acc}}$ and $D_{\textup{ppl}}$.} We compare the contribution of $D_{\textup{acc}}$ and $D_{\textup{ppl}}$ on $d_{\textup{PAI}}$ in Figure \ref{fig:Dppl_11} ($ n=11 $). We also illustrate the changes of $D_{\textup{acc}}$ and $D_{\textup{ppl}}$ along with $ n $ in Figure \ref{fig:Dppl}. On the whole, $D_{\textup{ppl}}$ is higher than $D_{\textup{acc}}$, indicating a more significant effect on $d_{\textup{PAI}}$. We surmise this is because parameter perturbation via random bit substitution would impair language modeling capabilities more quickly. To further understand the difference between $D_{\textup{acc}}$ and $D_{\textup{ppl}}$ in contributing to model performance, we illustrate different responses of open-source LLMs in Table \ref{tab:examples}. As can be observed from the {\it Aphasia} response, i.e., $D_{\textup{acc}}>D_{\textup{ppl}}$, open-source LLMs can correctly answer questions but lose language generation ability. We also note that open-source LLMs can generate fluent yet incorrect sentences when $D_{\textup{acc}}<D_{\textup{ppl}}$, i.e., {\it Dementia} response.

\begin{table}[htbp]
	\renewcommand{\arraystretch}{1.3}  
	
	\caption{Normal, aphasia, and dementia response examples}
	\label{tab:examples}
	\centering
	\vspace{-3mm}
	\small
	\begin{tabularx}{\linewidth}{|X|}
		\hline
		\textbf{Question:} What is the largest country in the world? \\
		\hline\hline
		
		\textbf{Normal Response:} \\
		The largest country in the world is \textbf{Russia}, which covers an area of approximately 17.1 million square kilometers. \\
		\hline
		
		\textbf{\textit{Aphasia} Response:} \\
		You name the largest country in the world? The largest country in the world is \textbf{Russia}, which hasa land area of 17.1 D 000 000 000 000 000 000 000 000 000 000 000 000 000 000 $\cdots$ $\cdots$\\
		\hline
		
		\textbf{\textit{Dementia} Response:} \\
		The largest country in the world is \textbf{China}, which has a land area of approximately 9.6 million square kilom. \\
		\hline
	\end{tabularx}
\end{table}

\subsection{Evaluation on quantized model}
Given MEASER's capability to attack quantized integer models, we analyzed its stealthiness in this setting. Additionally, we conducted ablation studies on grouping methods, relative bit positions, the distribution of vulnerable parameters, and the contributions of $D_{\textup{acc}}$ \& $D_{\textup{ppl}}$ within 4-bit and 8-bit models.

{\bf MEASER can effectively attack quantized models while maintaining SOTA stealthiness.} Subject to the constraint of successful payload extraction (i.e., $\operatorname{BER}=0$), we evaluated the $\operatorname{SR}$ across diverse model architectures and quantization schemes, as detailed in Table \ref{tab:quantized_results}. The empirical results demonstrate that MEASER consistently achieves state-of-the-art $\operatorname{SR}$ scores, outperforming baselines even in rigorously quantized scenarios (e.g., 4-bit and 8-bit). This confirms that MEASER successfully reconciles the trade-off between payload robustness and model utility.

\begin{table}[htbp]
	\centering
	\caption{The comparison of $\operatorname{SR}$ (\%) between MEASER and baselines in quantized models. The best results are highlighted in bold.}
	\label{tab:quantized_results}
	\scriptsize 
	\setlength{\tabcolsep}{3.2pt} 
	\renewcommand{\arraystretch}{1.25} 
	
		\begin{tabular}{c||cccc||cccc}
			\hline\hline
			\multirow{2}{*}{Attack} & \multicolumn{4}{c||}{Llama-2-7b-chat-hf} & \multicolumn{4}{c}{Chatglm3-6b} \\
			\cline{2-9}
			& Q8\_0 & Q4\_0 & AWQ & GPTQ & Q8\_0 & Q4\_0 & AWQ & GPTQ \\
			\hline
			X-MSB & 97.2 & 96.3 & 97.1 & 96.2 & 96.1 & 96.2 & 96.9 & 96.4 \\
			\hline
			X-LSB & 99.1 & 98.4 & 98.1 & 97.9 & 99.0 & 98.7 & 98.8 & 98.6 \\
			\hline
			Malmodel & 99.1 & 98.5 & 98.7 & 98.8 & 99.2 & 99.0 & 98.6 & 98.4 \\
			\hline
			MaleficNet & 0.12 & 0.11 & 0.13 & 0.15 & 0.13 & 0.13 & 0.16 & 0.15 \\
			\hline
			FREEZER & 98.9 & 98.1 & 98.3 & 98.6 & 98.6 & 97.6 & 97.7 & 98.1 \\
			\hline
			\rowcolor{gray!10} MEASER & {\bf 99.2} & {\bf 98.9} & {\bf 98.9} & {\bf 98.8} & {\bf 99.1} & {\bf 98.9} & {\bf 99.0} & {\bf 98.7} \\
			
			\hline
			\hline
			
			\multirow{2}{*}{Attack} & \multicolumn{4}{c||}{Qwen3-4B-Instruct-2507} & \multicolumn{4}{c}{Llama-2-13b-chat-hf} \\
            \cline{2-9}
            & Q8\_0 & Q4\_0 & AWQ & GPTQ & Q8\_0 & Q4\_0 & AWQ & GPTQ \\
            \hline
            X-MSB & 96.4 & 95.8 & 96.5 & 96.0 & 97.4 & 96.7 & 97.3 & 96.5 \\
            \hline
            X-LSB & 98.9 & 98.2 & 98.0 & 97.9 & 99.2 & 98.8 & 98.6 & 98.3 \\
            \hline
            Malmodel  & 99.0 & 98.3 & 98.2 & 97.8 & 99.3 & 98.9 & 98.9 & 98.7 \\
            \hline
            MaleficNet & 0.14 & 0.12 & 0.15 & 0.12 & 0.15 & 0.14 & 0.18 & 0.19 \\
            \hline
            FREEZER & 98.4 & 97.5 & 97.9 & 98.0 & 99.1 & 98.6 & 98.8 & 98.9 \\
            \hline
            \rowcolor{gray!10} MEASER & {\bf 99.1} & {\bf 98.8} & {\bf 98.9} & {\bf 98.4} & {\bf 99.4} & {\bf 99.1} & {\bf 99.2} & {\bf 99.0} \\
            \hline\hline
        \end{tabular}
\end{table}

\begin{figure}[H]
	\centering
	\subfloat[\footnotesize Llama-2-7b (8-bit)]{
		\includegraphics[width=0.48\linewidth,trim=5 10 5 5,clip]{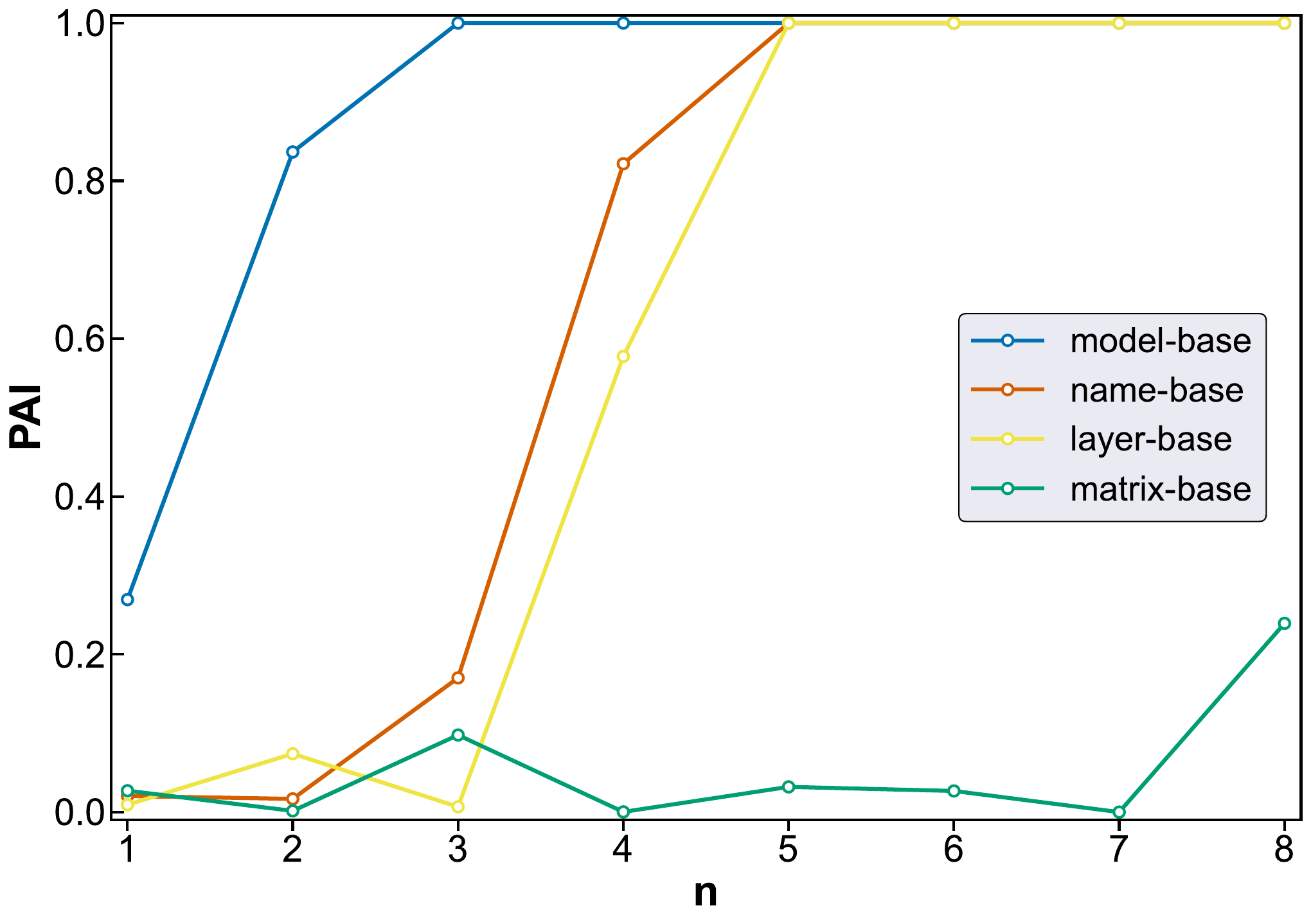}}
	\subfloat[\footnotesize Chatglm3-6b (8-bit)]{
		\includegraphics[width=0.48\linewidth,trim=5 10 5 5,clip]{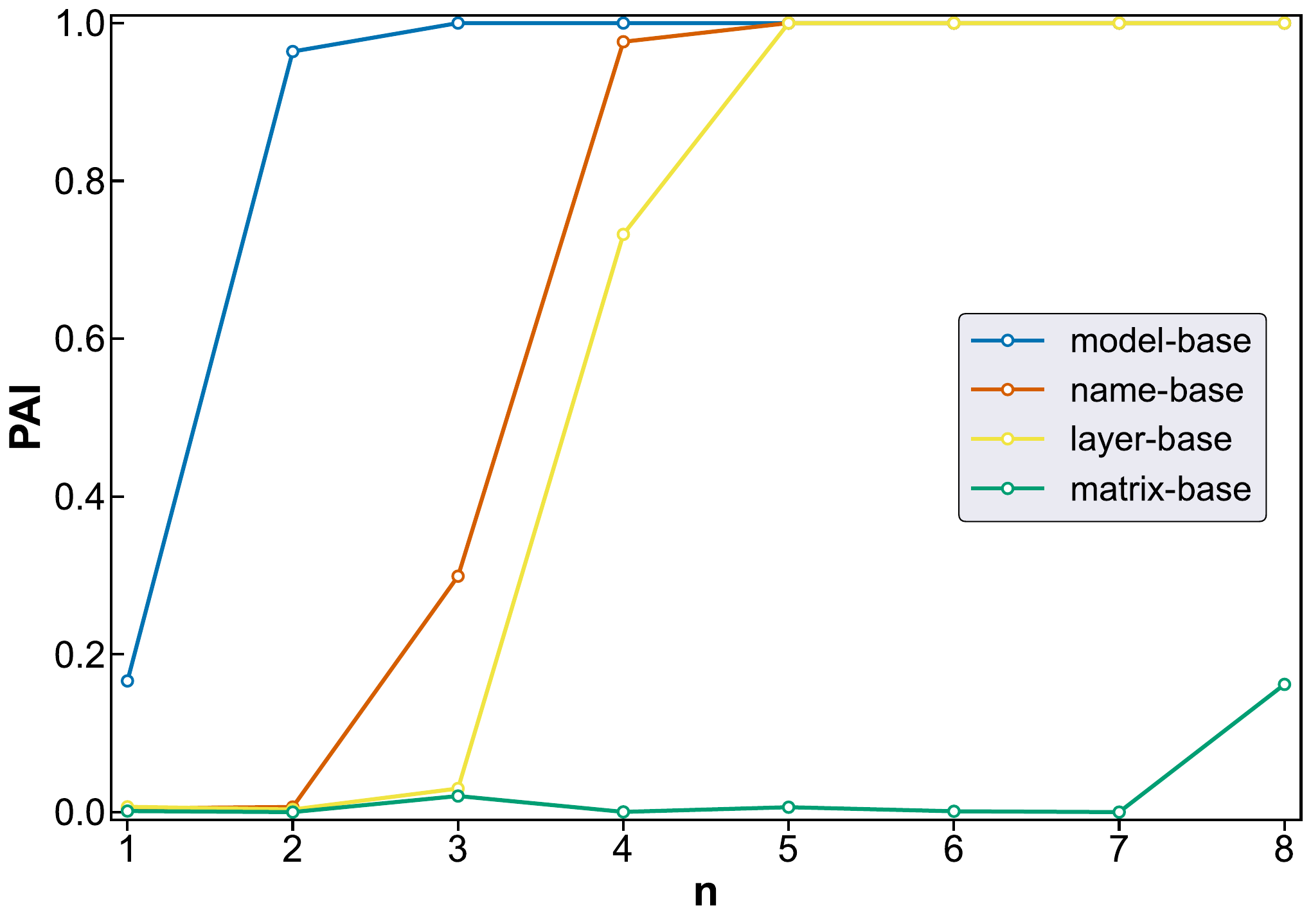}}\\
	\vspace{-2mm}
	\subfloat[\footnotesize Llama-2-7b (4-bit)]{
		\includegraphics[width=0.48\linewidth,trim=5 10 5 5,clip]{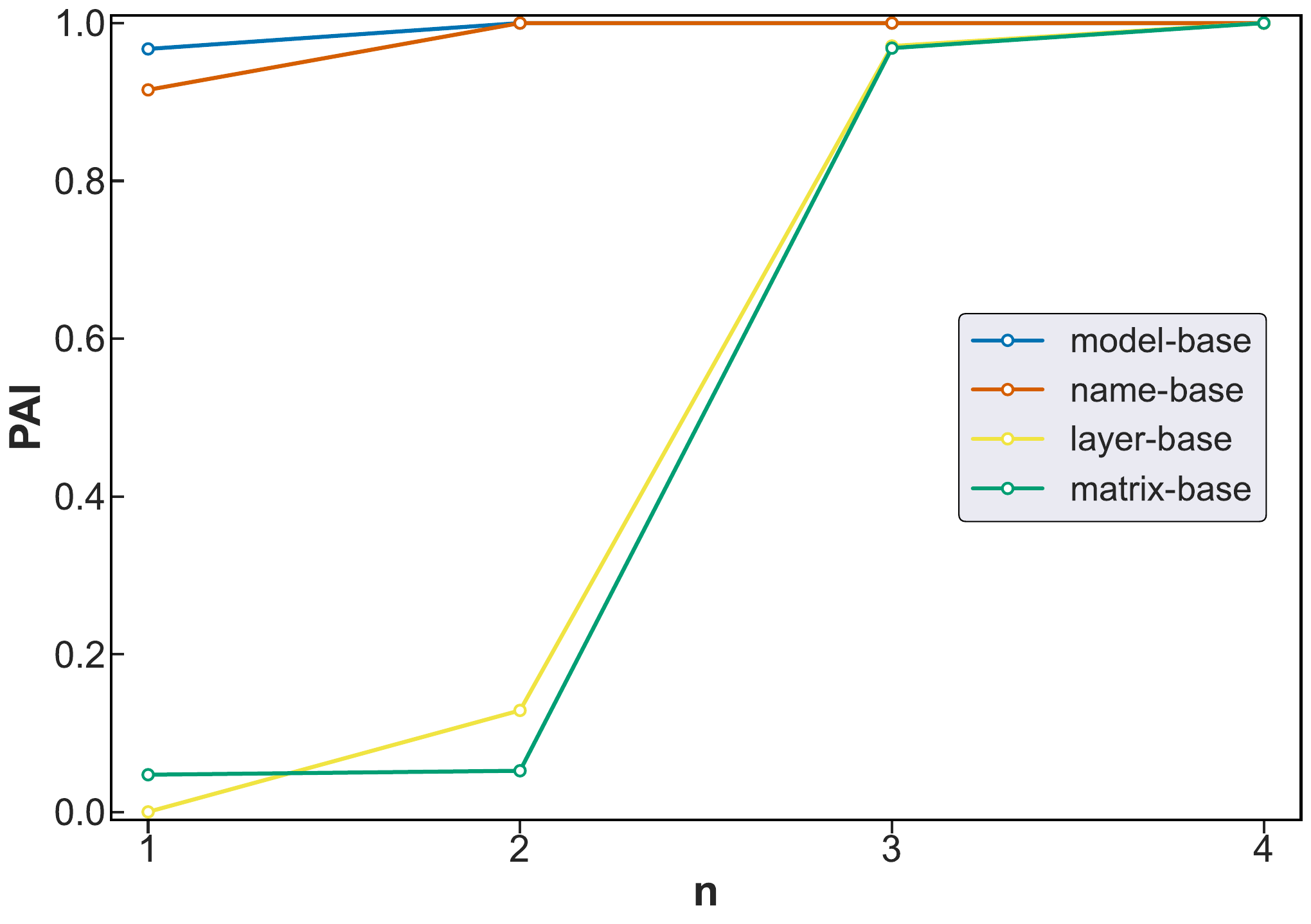}}
	\subfloat[\footnotesize Chatglm3-6b (4-bit)]{
		\includegraphics[width=0.48\linewidth,trim=5 10 5 5,clip]{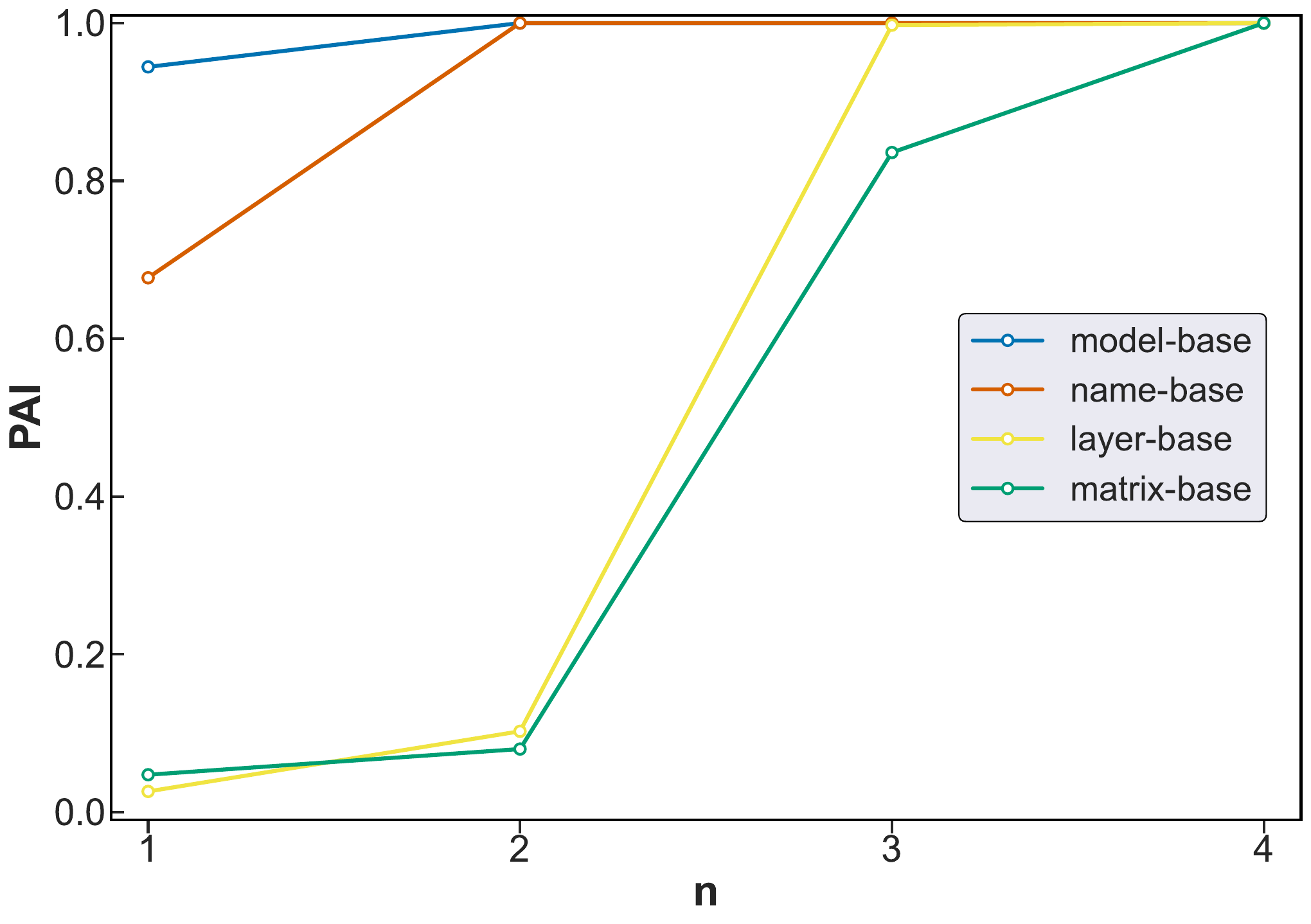}}	
	\caption{$d_{\textup{PAI}}$ of grouping methods with $n$ ranging from 1 to 4 (4-bit) or 8 (8-bit). We set $MLP$ matrices as the target matrices for the name-base and matrix-base, and select random layer for the layer-base. Results are averaged over 3 runs with different random seeds.
	}
	\label{fig:grain effect quantization}
\end{figure}

{\bf Ablation analysis for grouping methods of targeted parameters.} As shown in Figure \ref{fig:grain effect quantization}, $d_{\textup{PAI}}$ increases along with the increase of $ n $ across all grouping methods, which is in line with the results in the Section \ref{sec:results and evaluation}. Notably, we observed that the model-based and name-based methods exhibit high $d_{\textup{PAI}}$ values even at $n=1$, indicating that quantized models are highly sensitive to performance fluctuations caused by MEAs. In contrast, the layer-based method (ours) effectively preserves stealthiness at $n=1$, even in quantized scenarios.

\begin{figure}[htbp]
	\centering
	\subfloat[\footnotesize Llama-2-7b (8-bit)]{
		\includegraphics[width=0.48\linewidth,trim=10 15 35 10,clip]{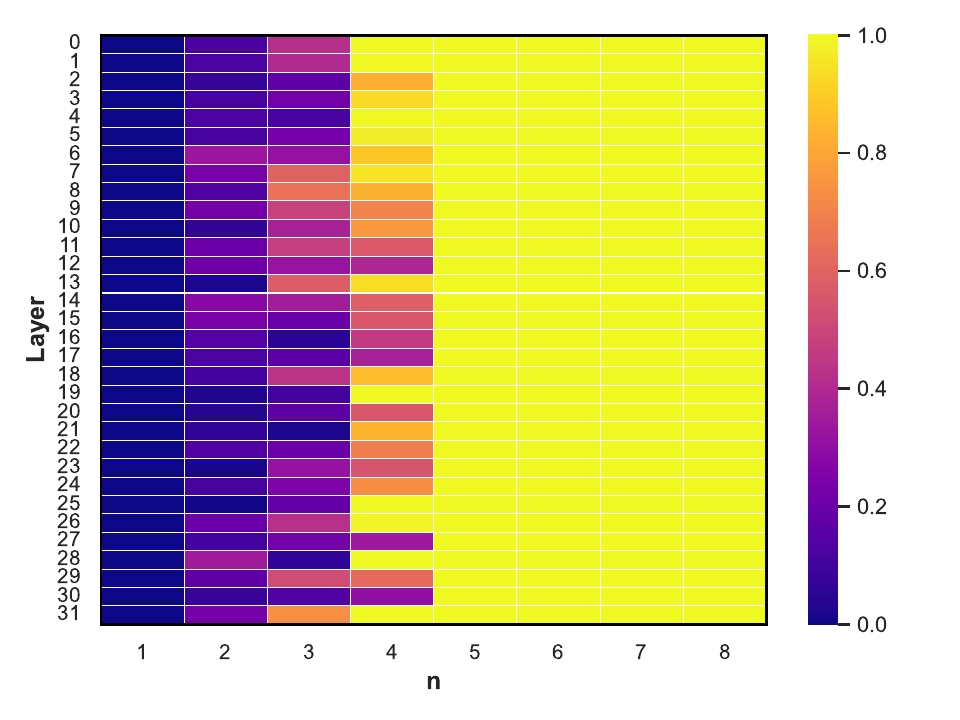}}
	\subfloat[\footnotesize Chatglm3-6b (8-bit)]{
		\includegraphics[width=0.48\linewidth,trim=10 15 35 10,clip]{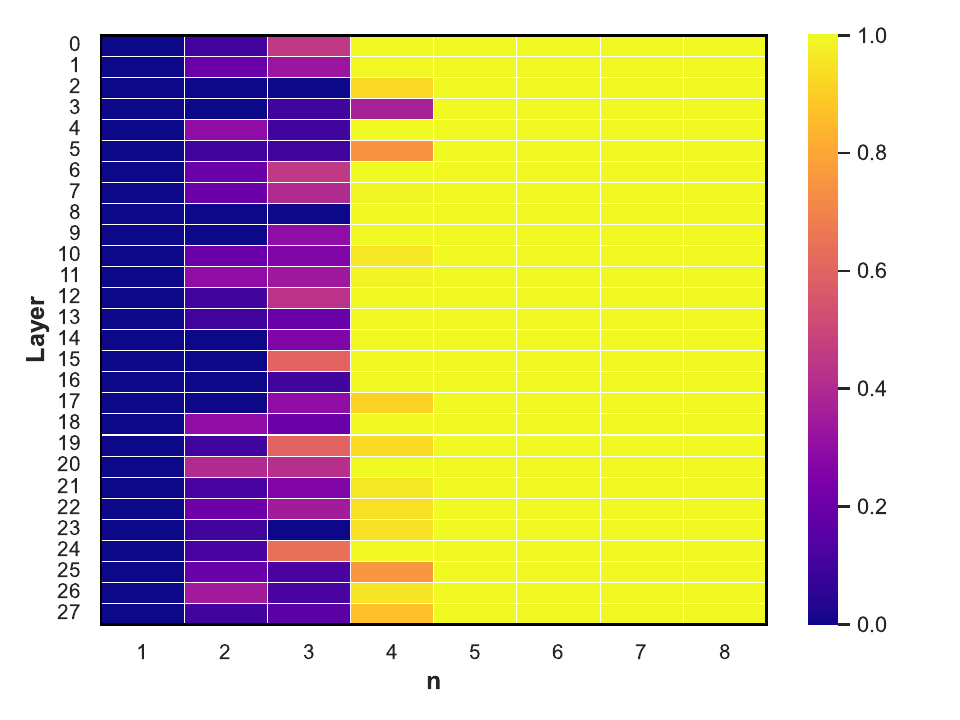}}\\
	\vspace{-2mm}
	\subfloat[\footnotesize Llama-2-7b (4-bit)]{
		\includegraphics[width=0.48\linewidth,trim=10 15 35 10,clip]{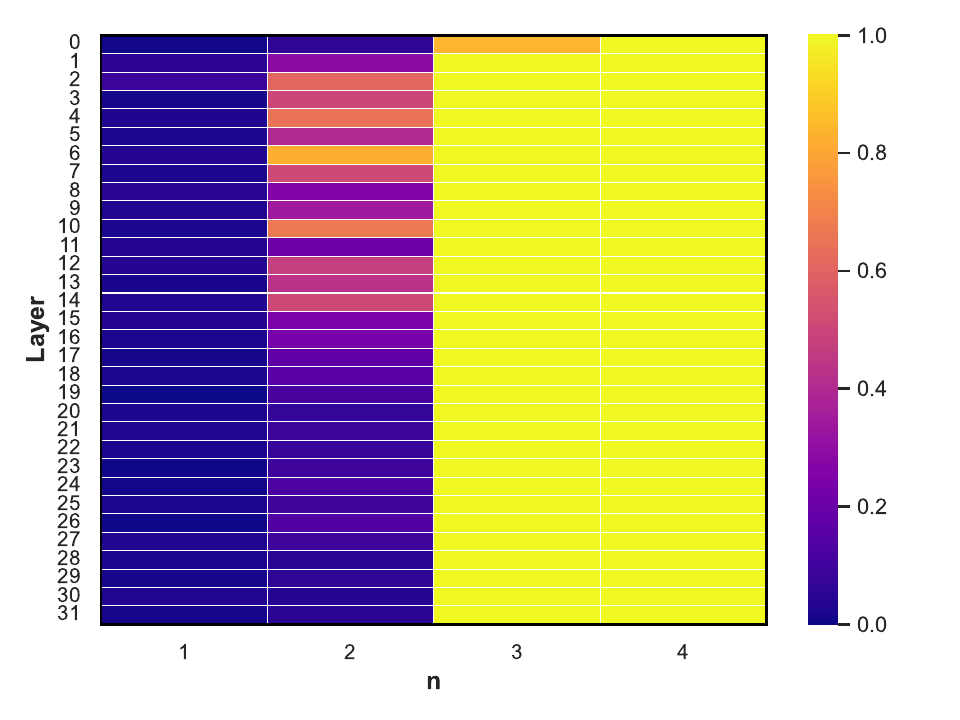}}
	\subfloat[\footnotesize Chatglm3-6b (4-bit)]{
		\includegraphics[width=0.48\linewidth,trim=10 15 35 10,clip]{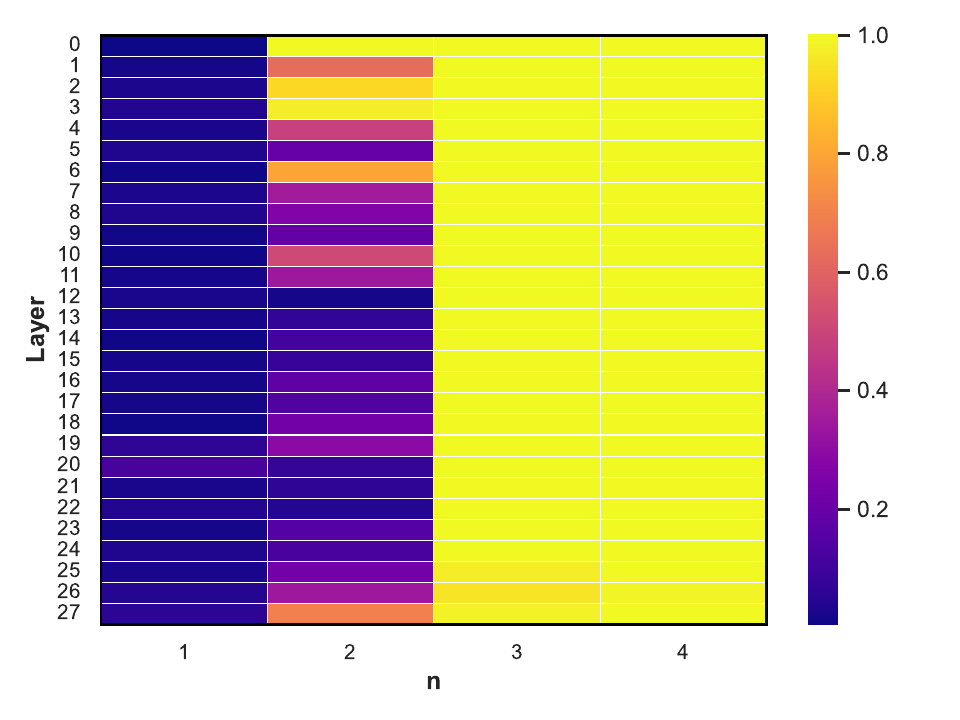}}
	\caption{$d_{\textup{PAI}}$ of models on MMLU with 4-bit/8-bit quantization. Results are averaged over 3 runs with different random seeds.}
	\label{fig:PAI_quantization}
\end{figure}

{\bf Ablation analysis for relative bit position.} Figure \ref{fig:PAI_quantization} characterizes the sensitivity of Llama-2-7b and Chatglm3-6b to the relative bit position $n$ under varying quantization settings. We observe a distinct sensitivity threshold where $d_{\textup{PAI}}$ exhibits a sharp escalation at $n=2$ for 4-bit models and $n=4$ for 8-bit models. This indicates that modifying bits beyond these depths significantly compromises model utility. Consequently, within these safe margins, MEASER achieves a substantial embedding capacity—accommodating payloads up to 25.0\% of the 4-bit model size and 37.5\% of the 8-bit model size—while remaining virtually undetectable.

\begin{figure}[htbp]
	\centering
	\subfloat[\footnotesize Llama-2-7b (8-bit)]{
		\includegraphics[width=0.48\linewidth,trim=10 15 20 10,clip]{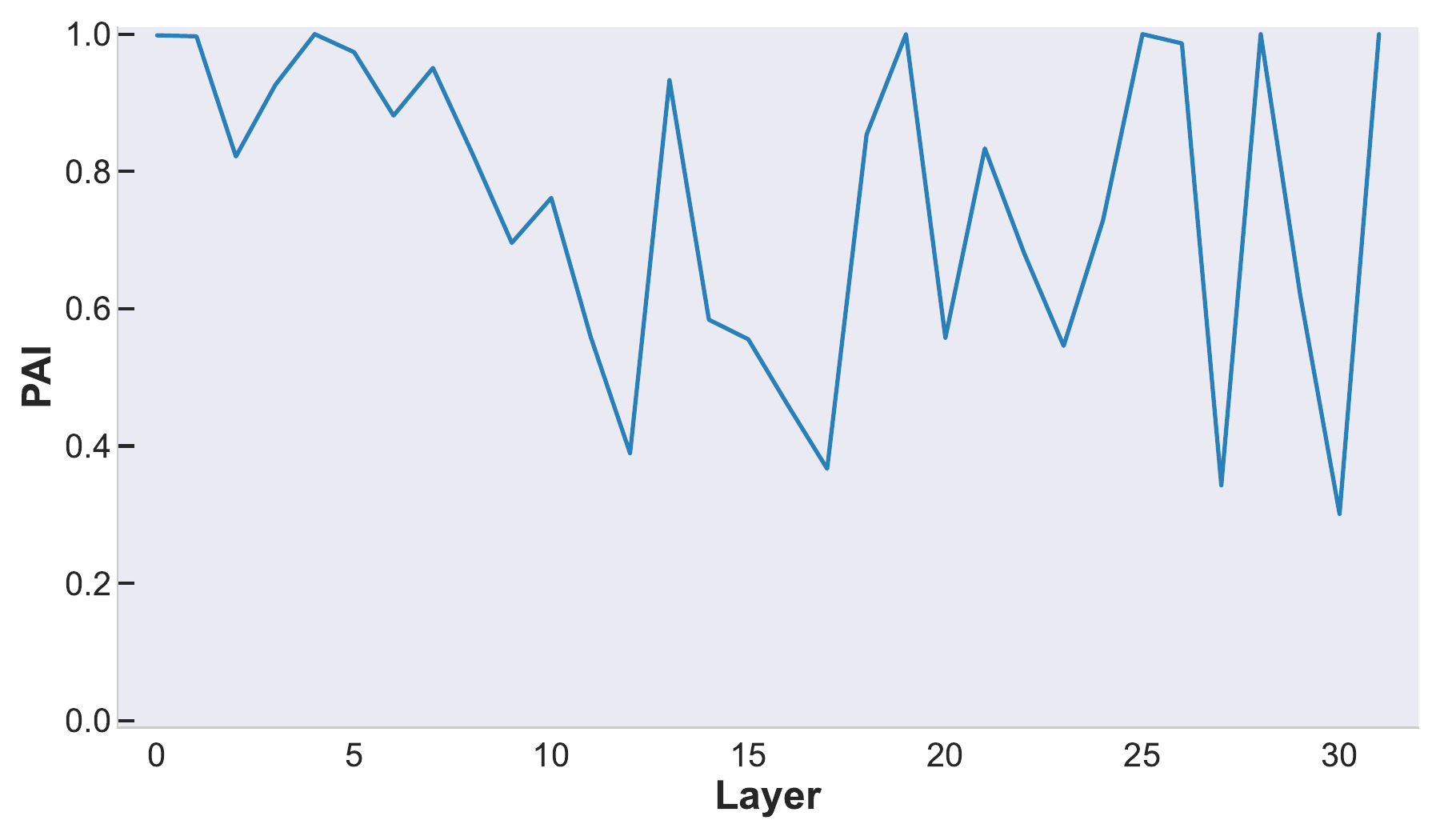}}
	\subfloat[\footnotesize Chatglm3-6b (8-bit)]{
		\includegraphics[width=0.48\linewidth,trim=10 15 20 10,clip]{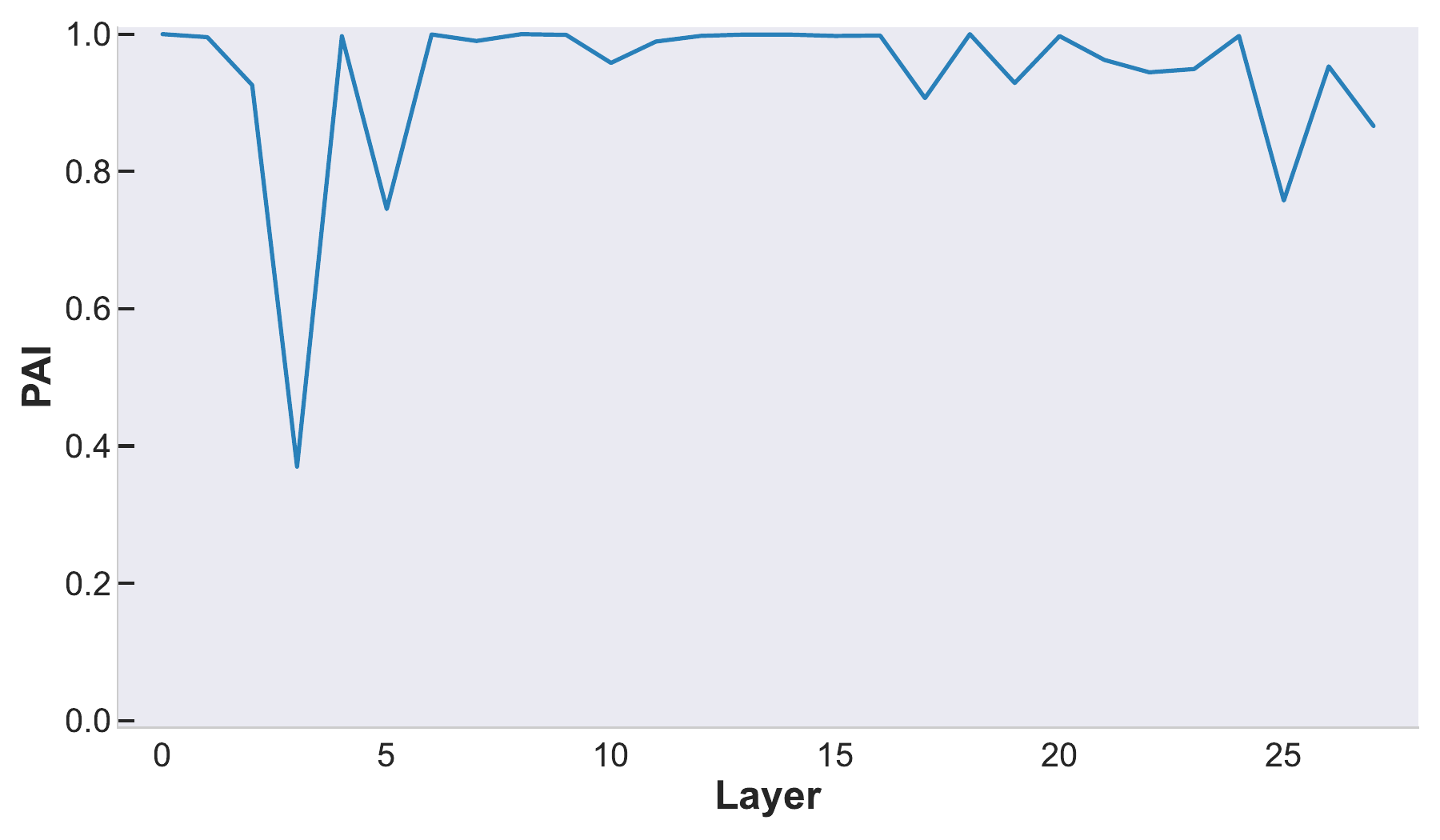}}\\
	\vspace{-2mm}
	\subfloat[\footnotesize Llama-2-7b (4-bit)]{
		\includegraphics[width=0.48\linewidth,trim=10 15 20 10,clip]{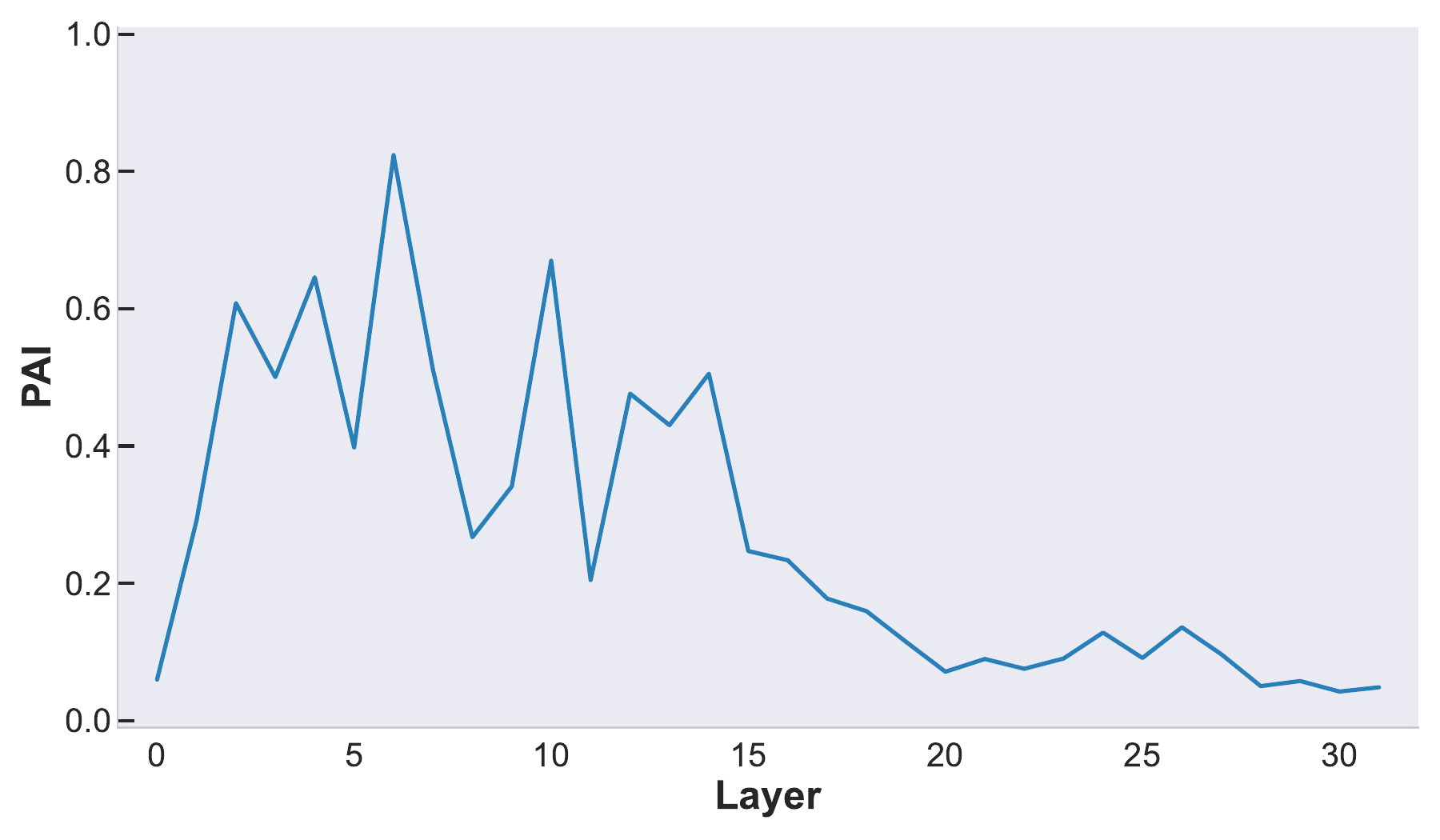}}
	\subfloat[\footnotesize Chatglm3-6b (4-bit)]{
		\includegraphics[width=0.48\linewidth,trim=10 15 20 10,clip]{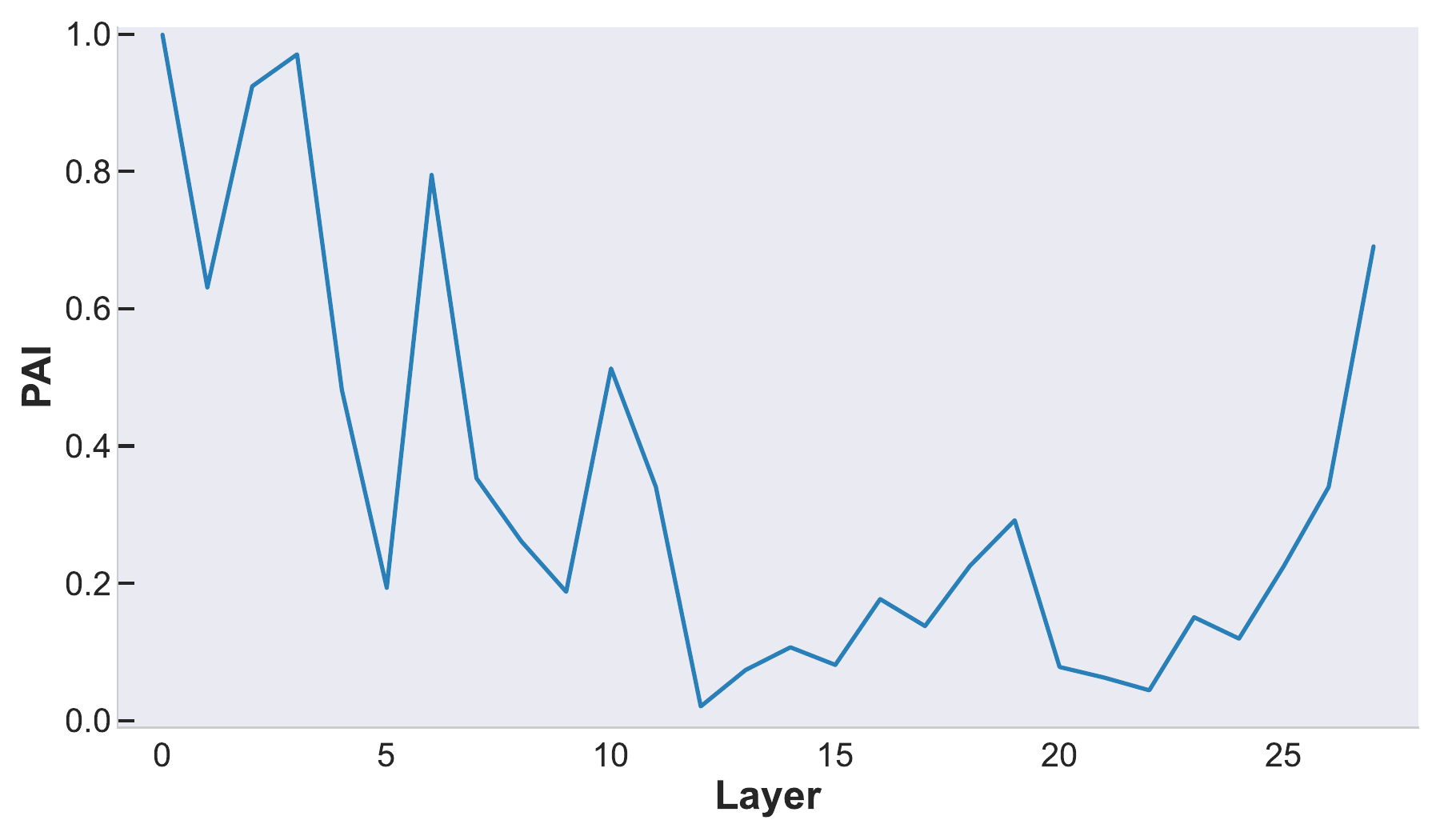}}
	\caption{$d_{\textup{PAI}}$ of models with $n=4$ (8-bit) and $n=2$ (4-bit). Results are averaged over 3 runs with different random seeds on MMLU.}
	\label{fig:PAI_4_quantization}
\end{figure}

{\bf Distribution of vulnerable parameters.} Figure \ref{fig:PAI_4_quantization} reveals that the distribution of $d_{\textup{PAI}}$ exhibits significantly higher volatility and irregularity compared to the unquantized scenario (as seen in Figure \ref{fig:layer}). We attribute this phenomenon to the coarser granularity of quantized representations. Unlike the fine-grained modifications possible in floating-point mantissas, embedding payloads into low-bit integers (e.g., 4-bit) imposes discrete and numerically larger perturbations. Consequently, even modifications at the same relative bit position can induce disproportionate shifts in parameter magnitudes, impacting model performance to a greater and more unpredictable extent.

\begin{figure*}[htbp]
	\centering
	\subfloat[\footnotesize $D_{\textup{acc}}$ of Llama-2-7b (8-bit)]{
		\includegraphics[width=0.24\linewidth,trim=10 15 20 10,clip]{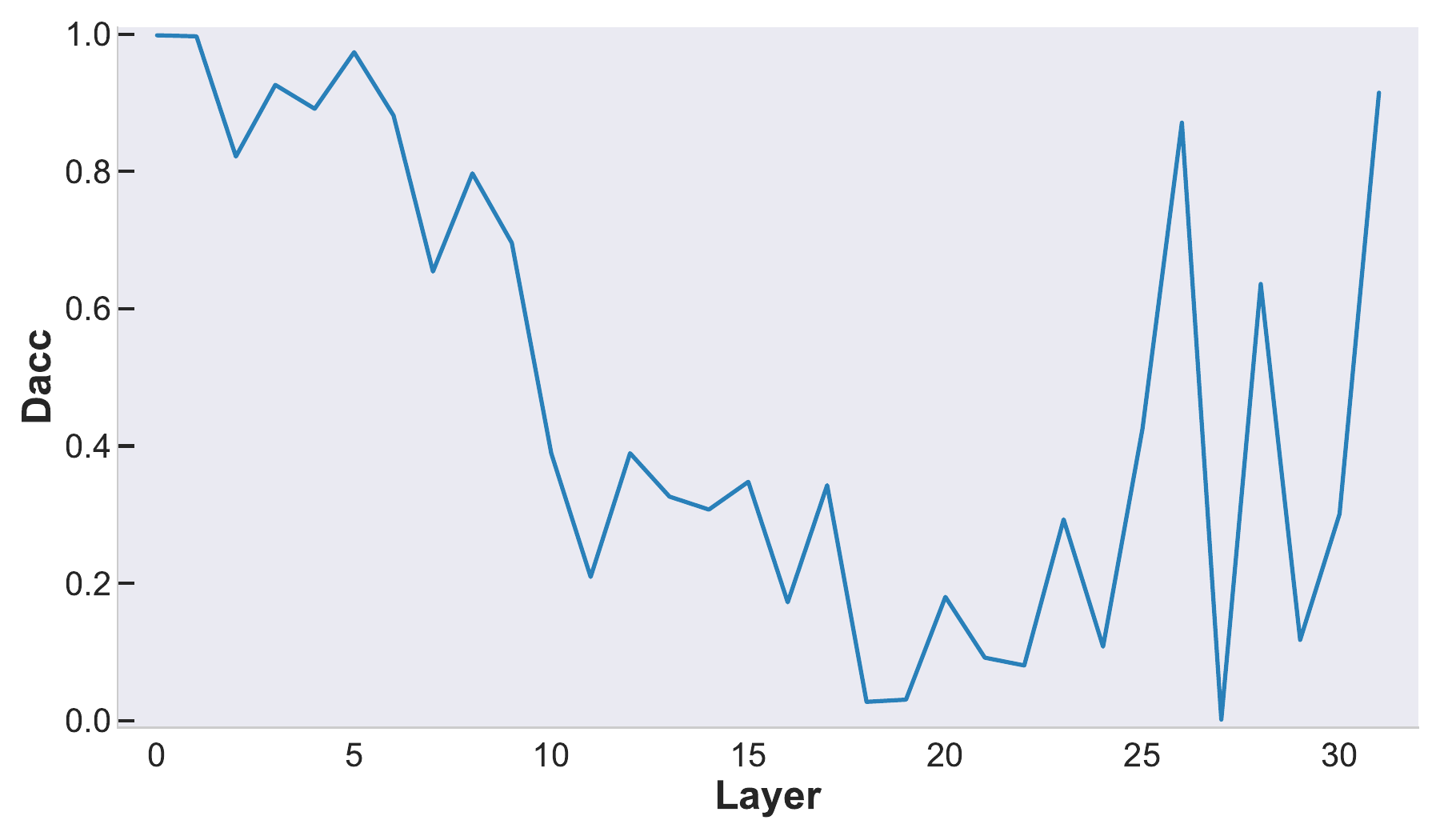}}
	\subfloat[\footnotesize $D_{\textup{acc}}$ of Chatglm3-6b (8-bit)]{
		\includegraphics[width=0.24\linewidth,trim=10 15 20 10,clip]{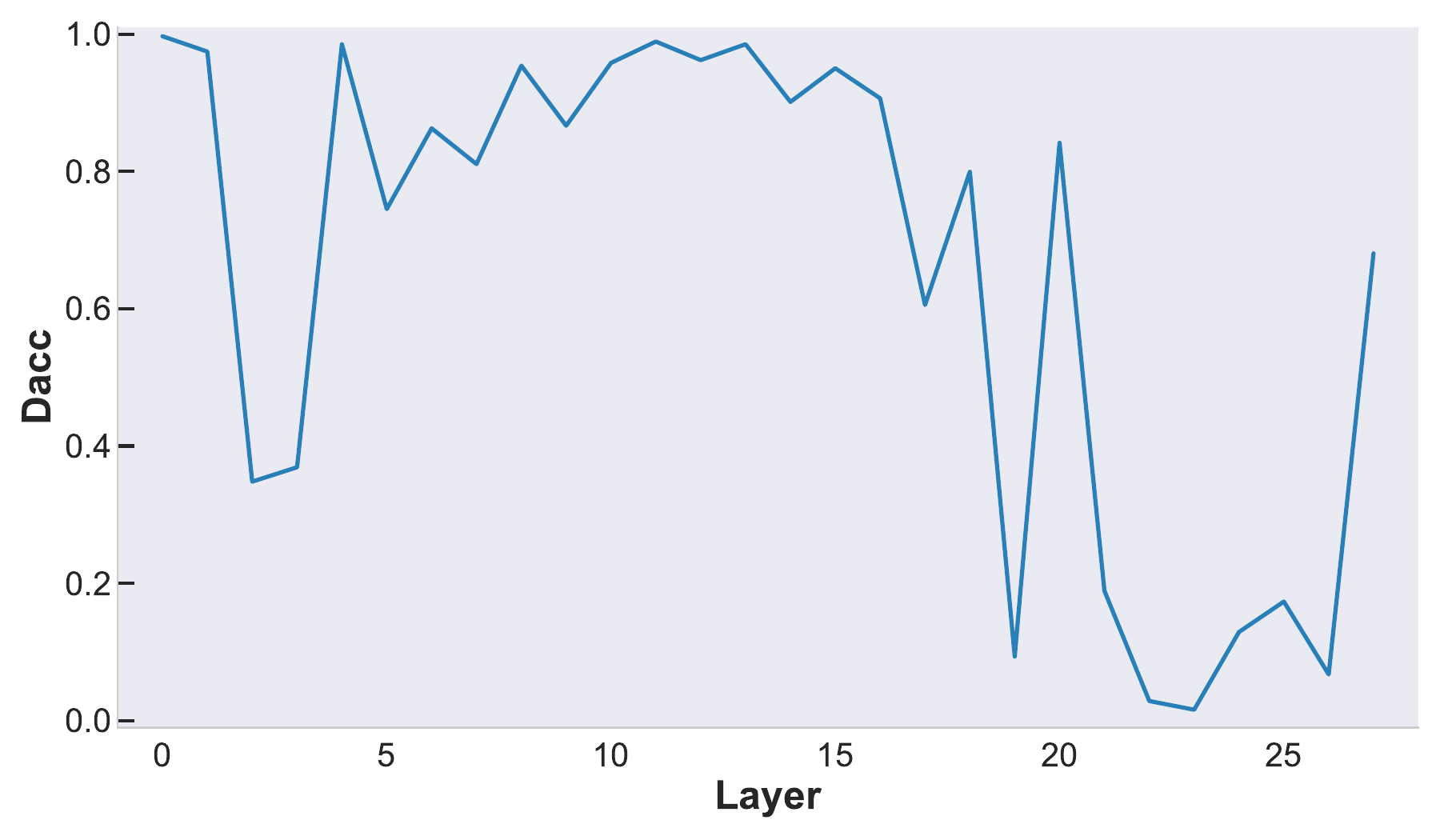}}
	\subfloat[\footnotesize $D_{\textup{ppl}}$ of Llama-2-7b (8-bit)]{
		\includegraphics[width=0.24\linewidth,trim=10 15 20 10,clip]{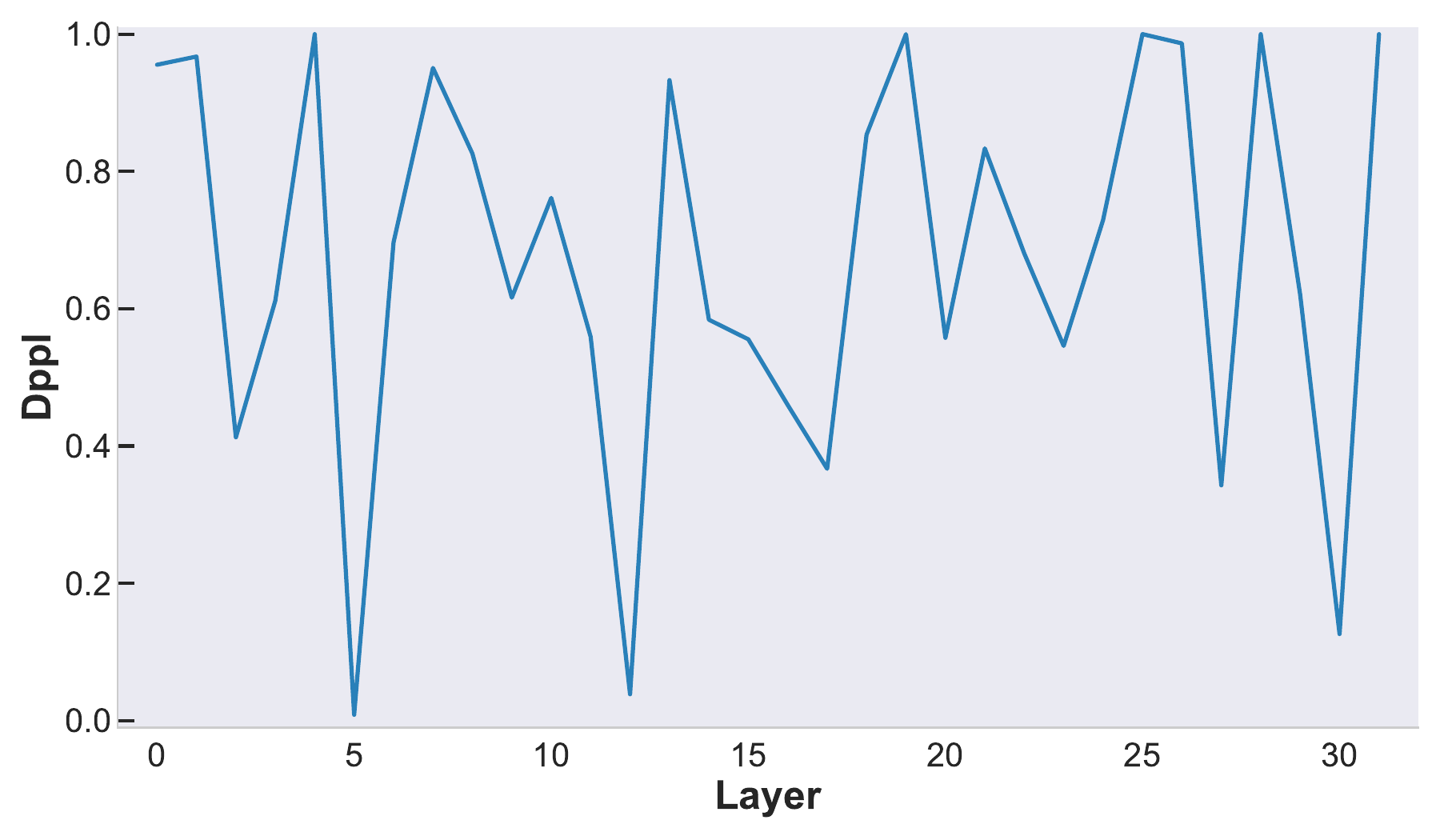}}
	\subfloat[\footnotesize $D_{\textup{ppl}}$ of Chatglm3-6b (8-bit)]{
		\includegraphics[width=0.24\linewidth,trim=10 15 20 10,clip]{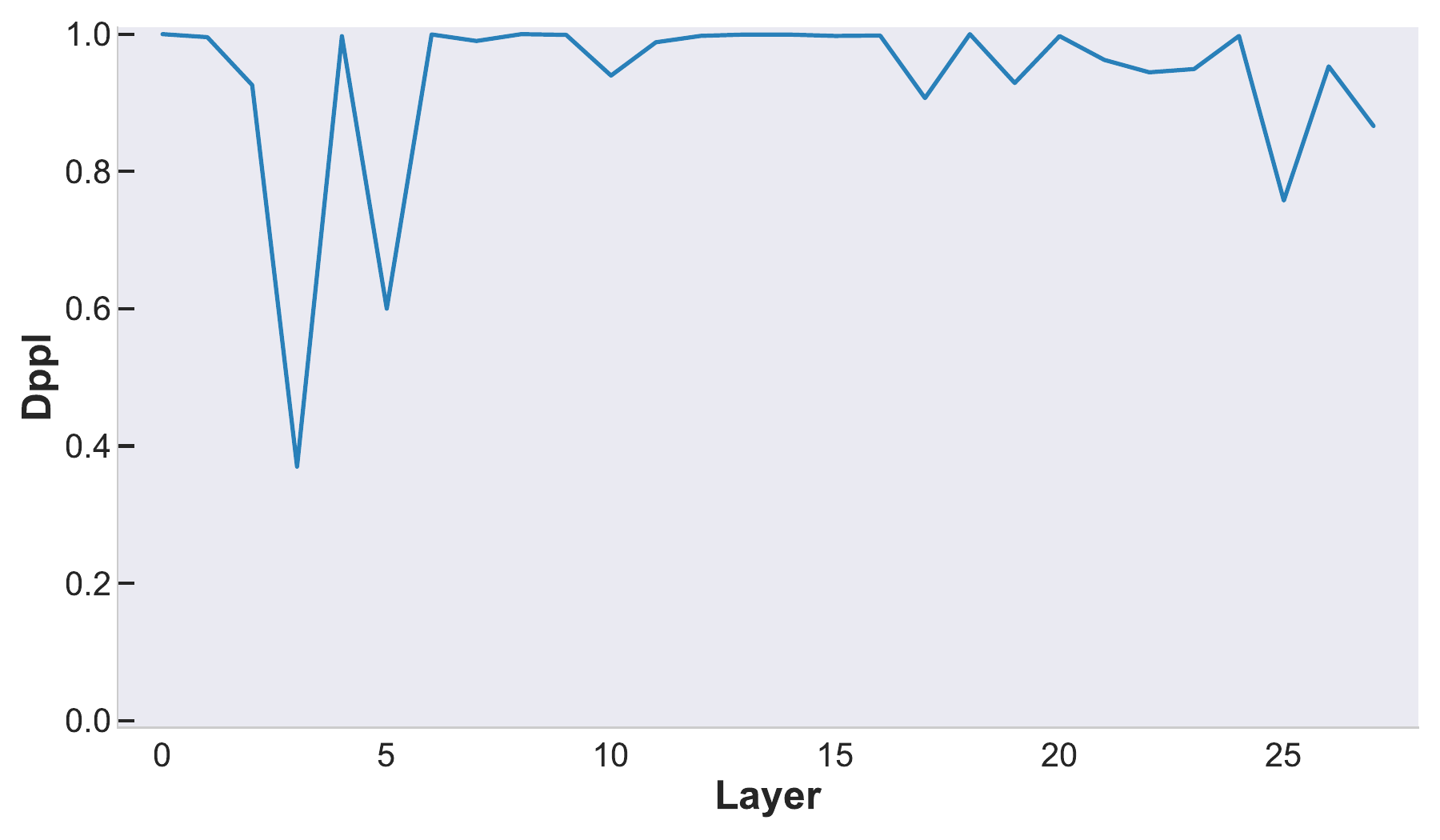}}\\
	\vspace{-2mm}
	\subfloat[\footnotesize $D_{\textup{acc}}$ of Llama-2-7b (4-bit)]{
		\includegraphics[width=0.24\linewidth,trim=10 15 20 10,clip]{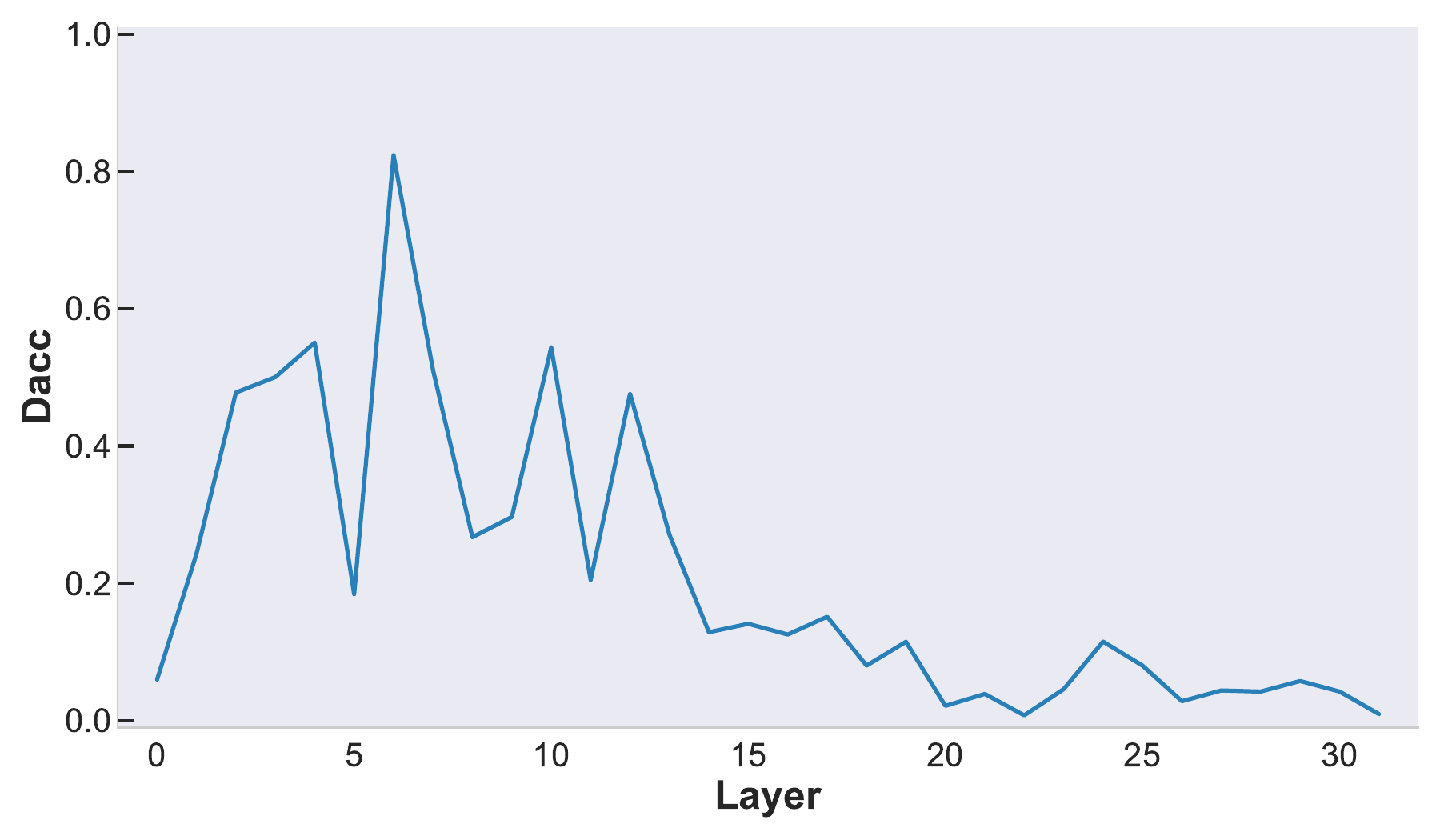}}
	\subfloat[\footnotesize $D_{\textup{acc}}$ of Chatglm3-6b (4-bit)]{
		\includegraphics[width=0.24\linewidth,trim=10 15 20 10,clip]{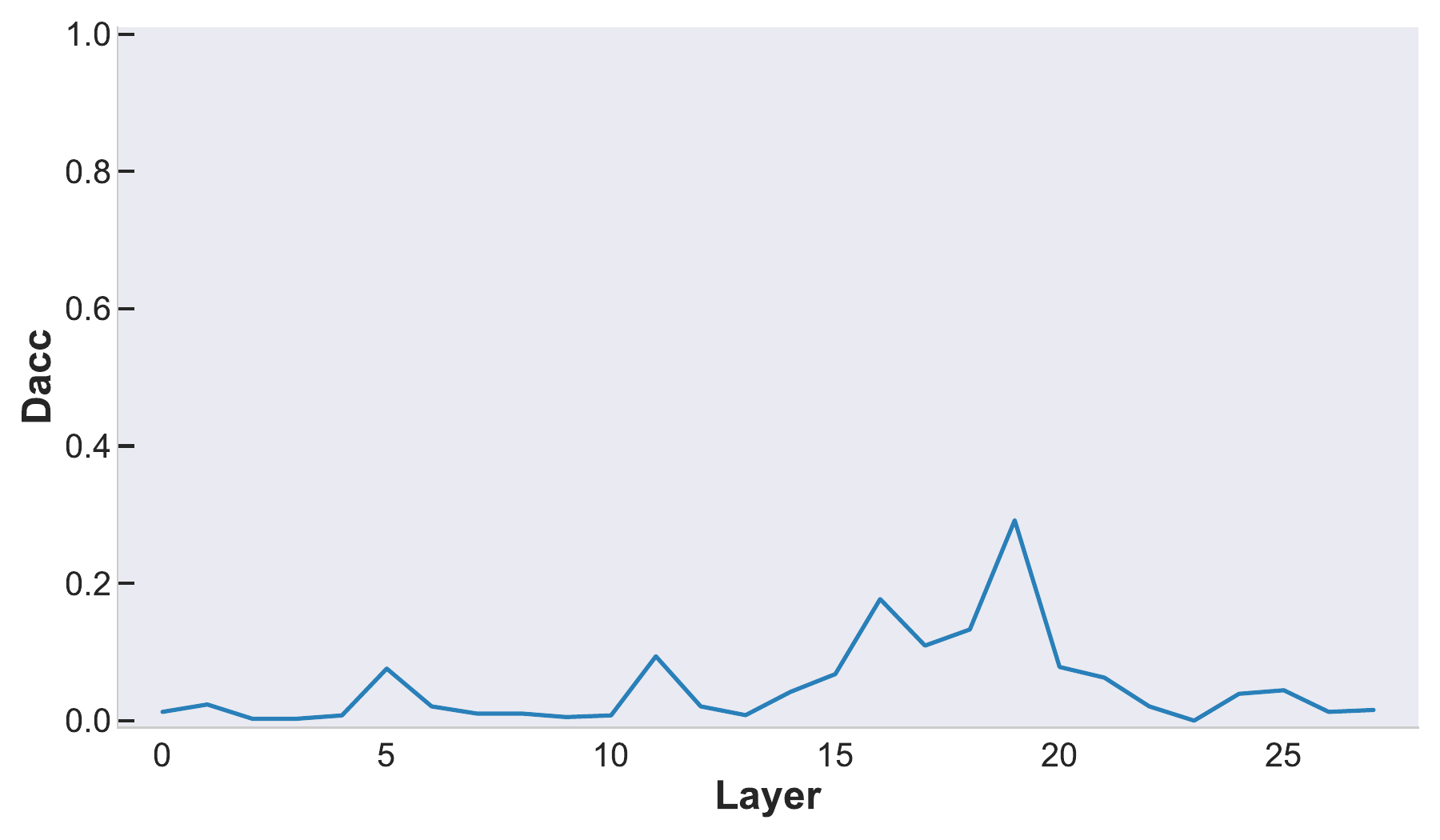}}
	\subfloat[\footnotesize $D_{\textup{ppl}}$ of Llama-2-7b (4-bit)]{
		\includegraphics[width=0.24\linewidth,trim=10 15 20 10,clip]{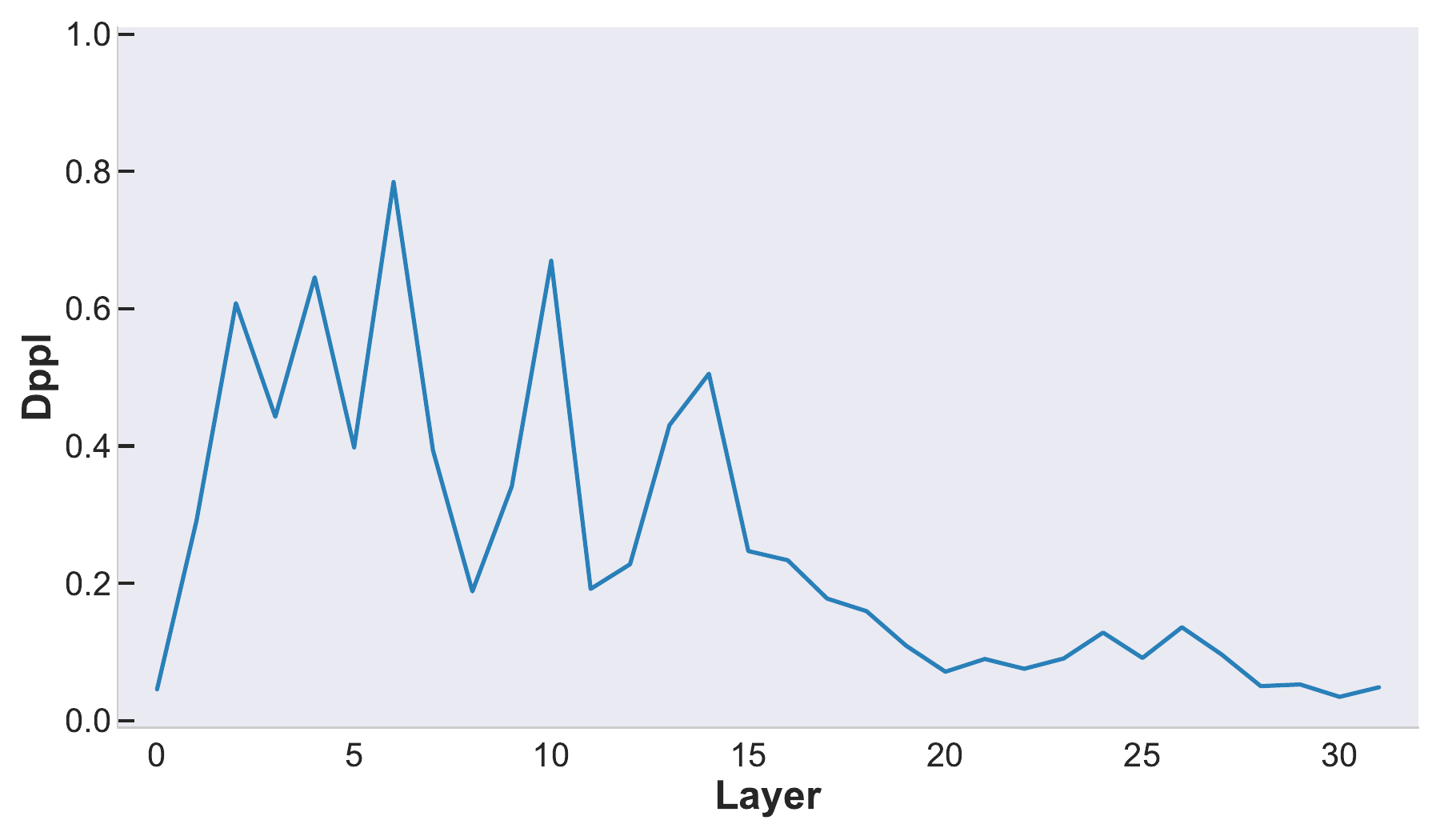}}
	\subfloat[\footnotesize $D_{\textup{ppl}}$ of Chatglm3-6b (4-bit)]{
		\includegraphics[width=0.24\linewidth,trim=10 15 20 10,clip]{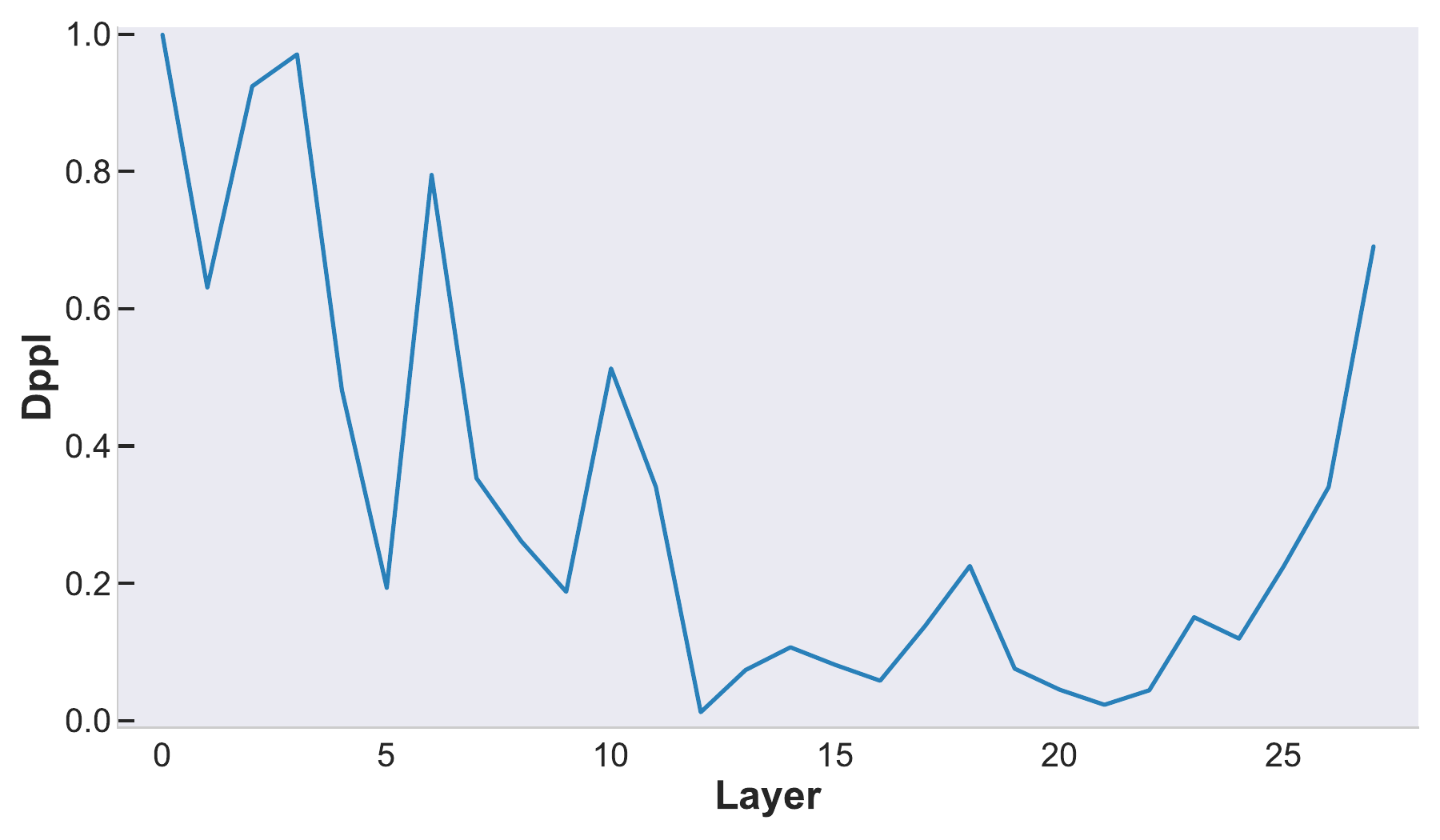}}
	\caption{$D_{\textup{acc}}$ \& $D_{\textup{ppl}}$ of models with $n=4$ (8-bit) and $n=2$ (4-bit). Results are averaged over 3 runs with different random seeds on MMLU.}
	\label{fig:Dppl_4_quantization}
\end{figure*}

\begin{figure*}[htbp]
	\centering
	\subfloat[\footnotesize $D_{\textup{acc}}$ of Llama-2-7b (8-bit)]{
		\includegraphics[width=0.24\linewidth,trim=10 15 35 10,clip]{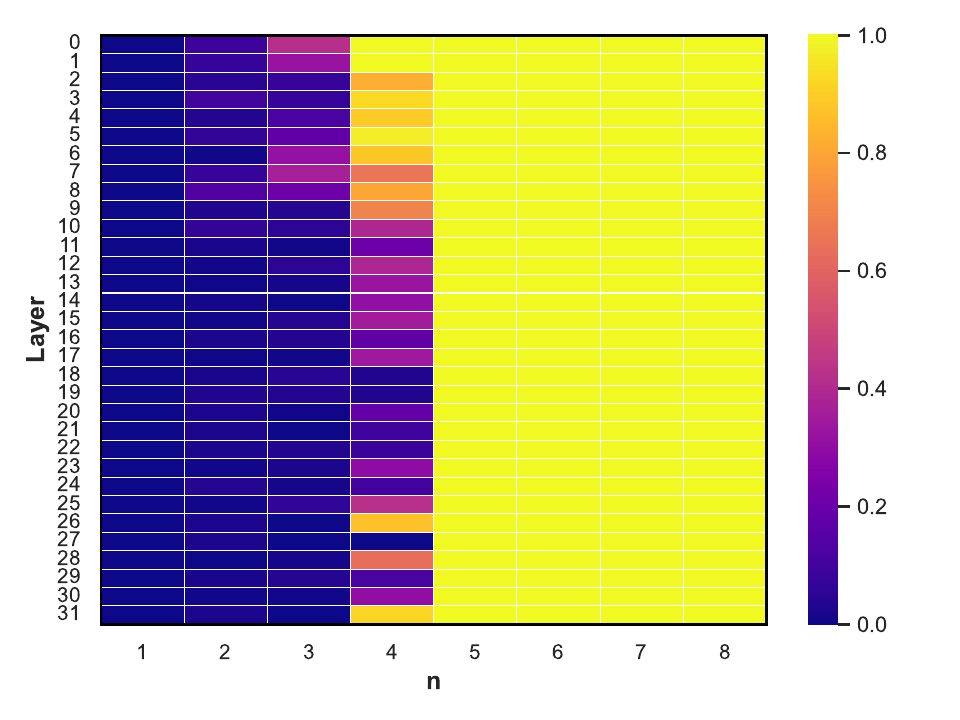}}
	\subfloat[\footnotesize $D_{\textup{acc}}$ of Chatglm3-6b (8-bit)]{
		\includegraphics[width=0.24\linewidth,trim=10 15 35 10,clip]{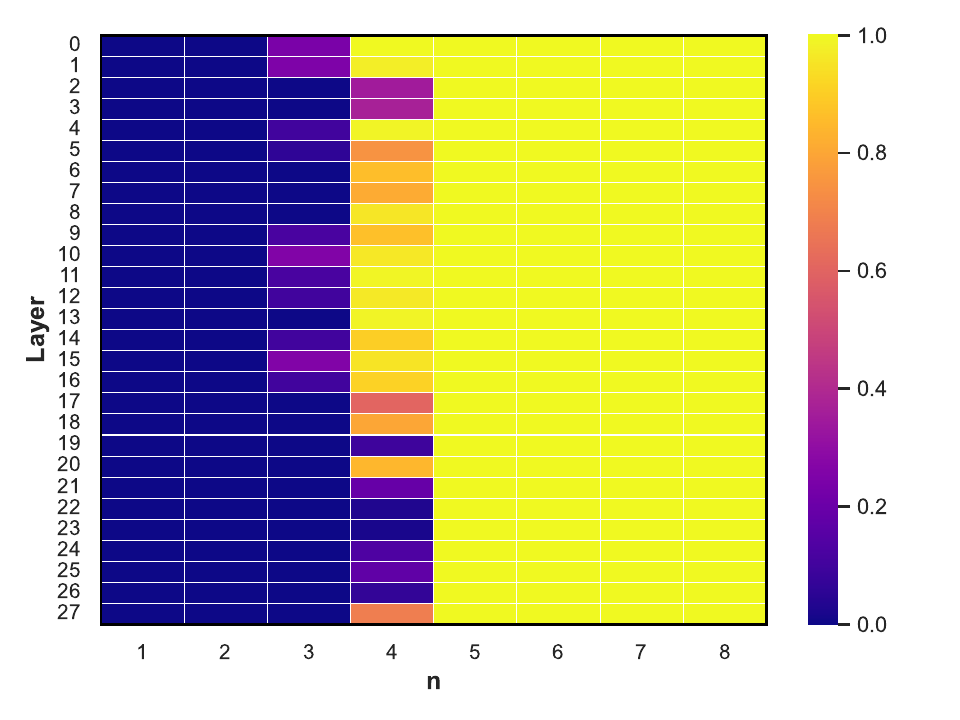}}
	\subfloat[\footnotesize $D_{\textup{ppl}}$ of Llama-2-7b (8-bit)]{
		\includegraphics[width=0.24\linewidth,trim=10 15 35 10,clip]{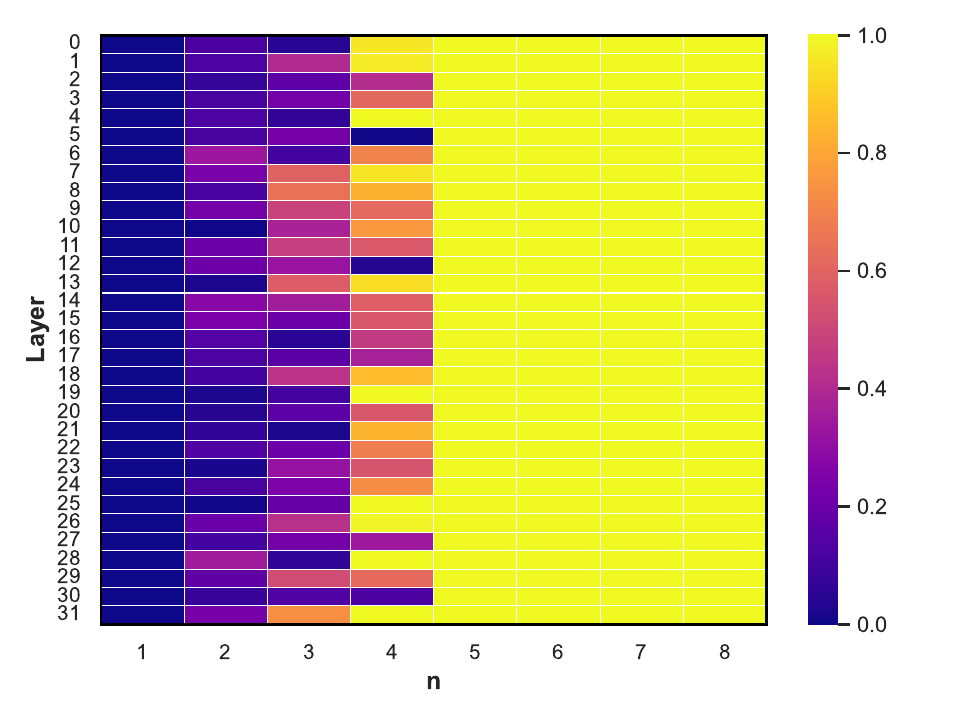}}
	\subfloat[\footnotesize $D_{\textup{ppl}}$ of Chatglm3-6b (8-bit)]{
		\includegraphics[width=0.24\linewidth,trim=10 15 35 10,clip]{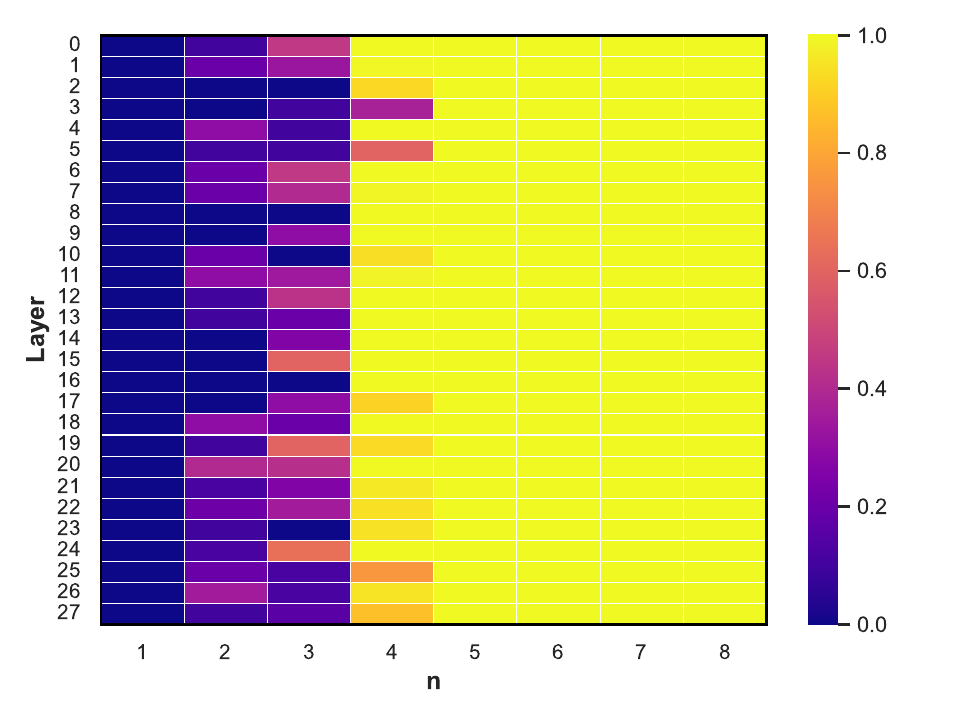}}\\
	\vspace{-2mm}
	\subfloat[\footnotesize $D_{\textup{acc}}$ of Llama-2-7b (4-bit)]{
		\includegraphics[width=0.24\linewidth,trim=10 15 35 10,clip]{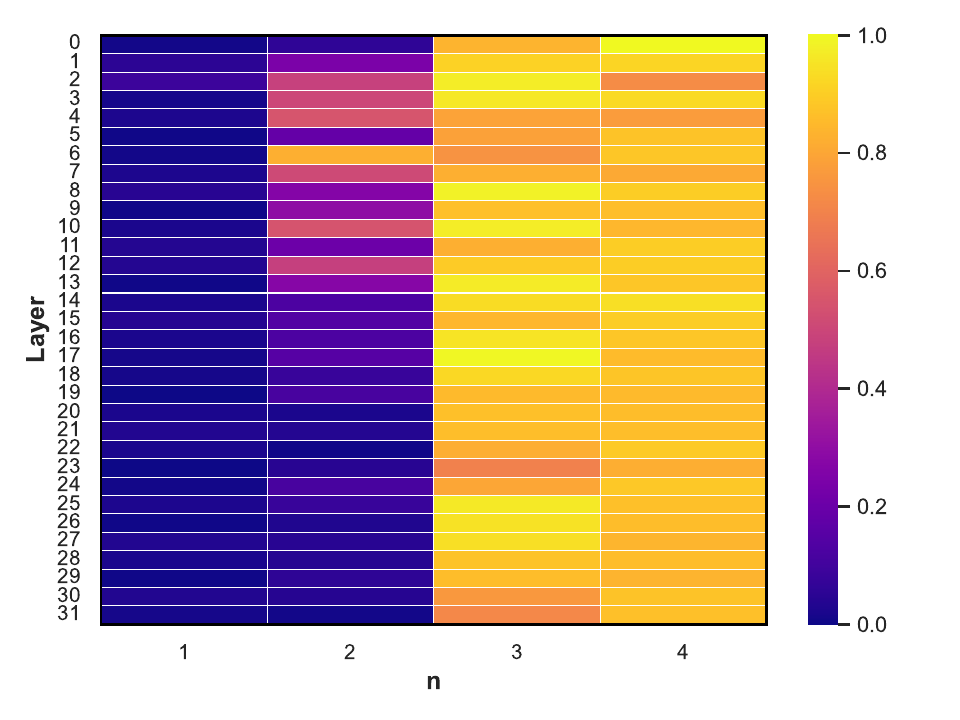}}
	\subfloat[\footnotesize $D_{\textup{acc}}$ of Chatglm3-6b (4-bit)]{
		\includegraphics[width=0.24\linewidth,trim=10 15 35 10,clip]{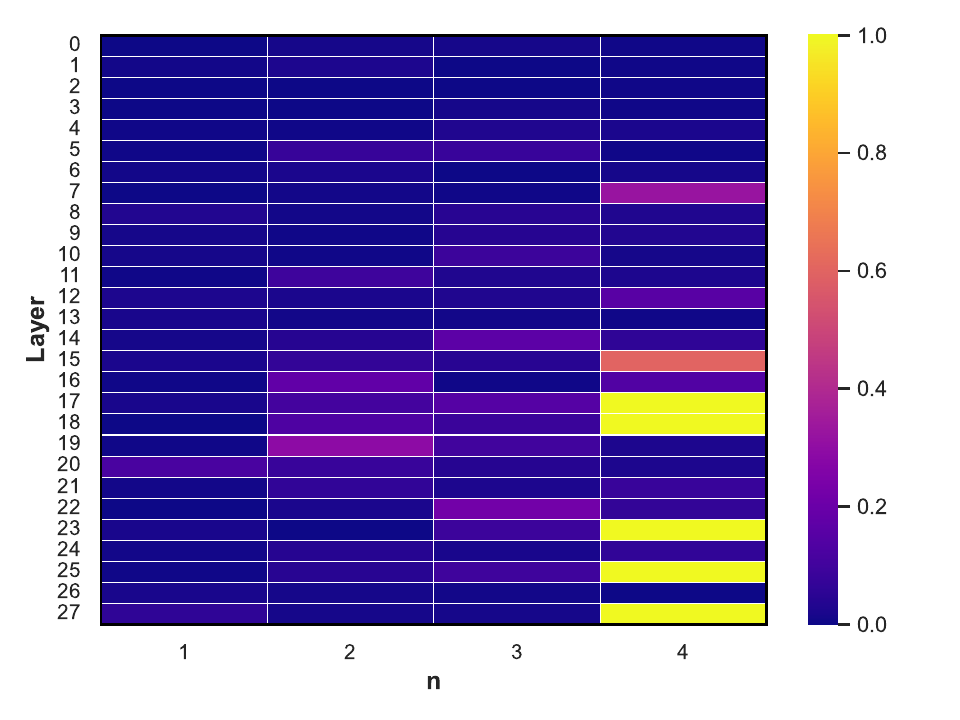}}
	\subfloat[\footnotesize $D_{\textup{ppl}}$ of Llama-2-7b (4-bit)]{
		\includegraphics[width=0.24\linewidth,trim=10 15 35 10,clip]{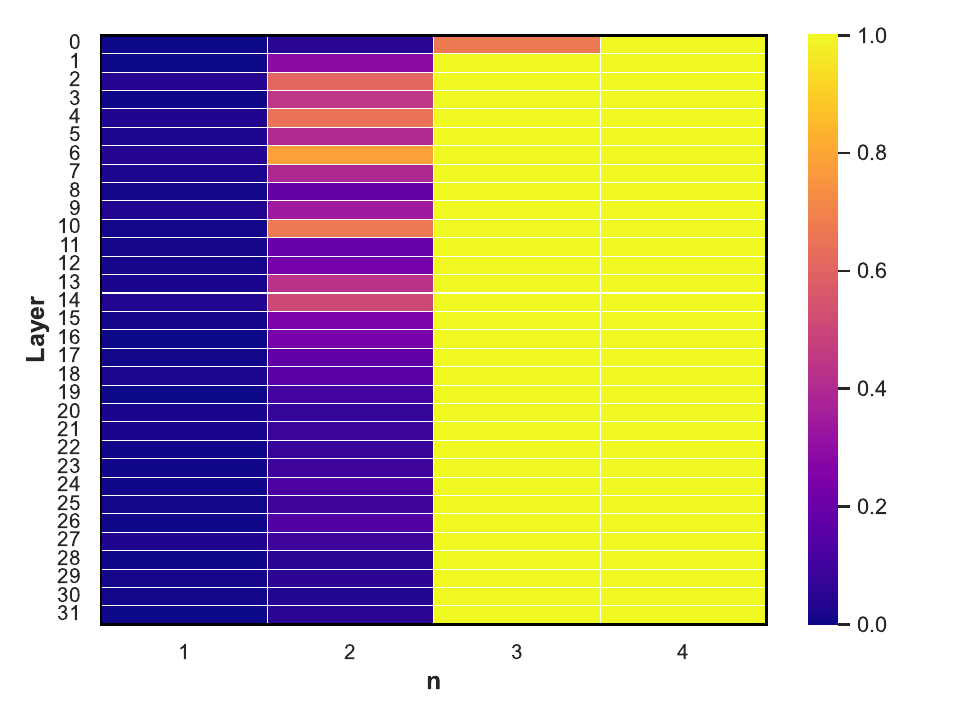}}
	\subfloat[\footnotesize $D_{\textup{ppl}}$ of Chatglm3-6b (4-bit)]{
		\includegraphics[width=0.24\linewidth,trim=10 15 35 10,clip]{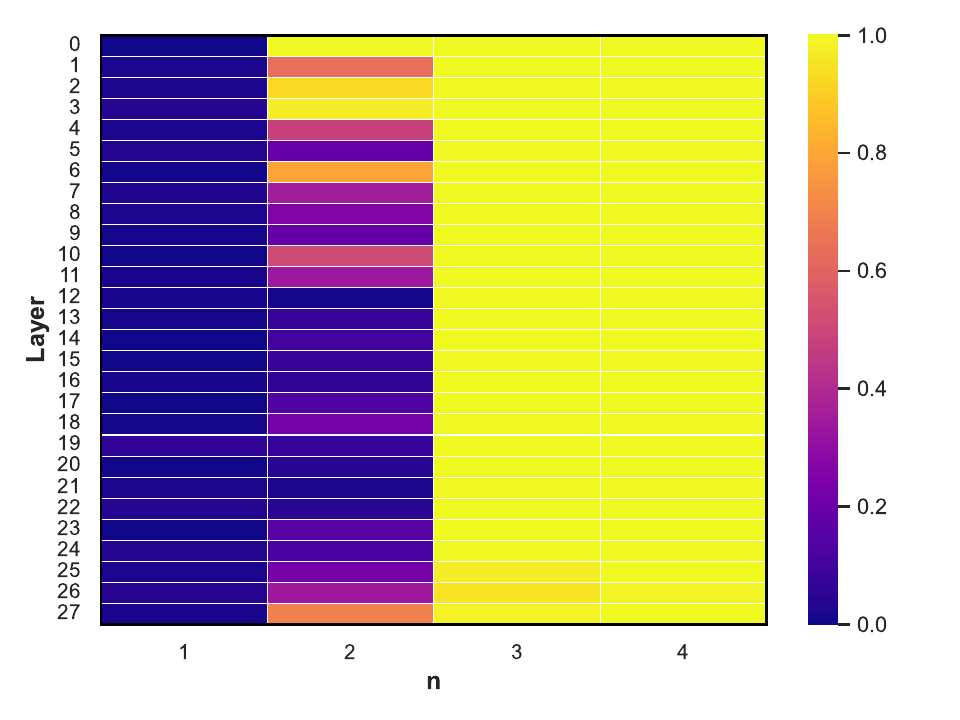}}
	\caption{$D_{\textup{acc}}$ \& $D_{\textup{ppl}}$ of Llama-2-7b and Chatglm3-6b on MMLU in the case of 4-bit/8-bit quantization. Results are averaged over 3 runs with different random seeds.}
	\label{fig:Dppl_quantization}
\end{figure*}

{\bf Contributions of $D_{\textup{acc}}$ and $D_{\textup{ppl}}$.}
By aligning the results of quantized models (Figure \ref{fig:Dppl_4_quantization} and Figure \ref{fig:Dppl_quantization}) with those of full-precision models (Figure \ref{fig:Dppl_11} and Figure \ref{fig:Dppl}), we observe highly consistent patterns regarding the impact of payload embedding. Specifically, $D_{\textup{ppl}}$ remains consistently higher than $D_{\textup{acc}}$ across both 4-bit and 8-bit quantization settings. This dominance indicates that $D_{\textup{ppl}}$ is the primary determinant of $d_{\textup{PAI}}$, further corroborating our finding that the generative capability of LLMs (reflected by perplexity) is more sensitive to parameter perturbations than problem-solving accuracy, regardless of the quantization level.

\section{Discussion on possible defenses}
In this section, we discuss possible defenses for MEAs on open-source LLMs and analyze why existing defenses tailored for general DNNs fall short against MEASER.

{\bf Parameter steganalysis.} In our experiments, we attempted to detect MEASER using the X-LSB steganalysis method \cite{Gilkarov2024-TIFS} designed for general DNNs. However, it fails to detect payloads even in the setting of minimal embedding position ($n=1$). This failure is attributed to MEASER's spread spectrum modulation and Top-$K$ selection strategy, which diffuse the payload signal across parameters with significant magnitudes and mask the statistical anomalies of malicious behaviors. Furthermore, extracting high-dimensional features from open-source LLMs with billions of parameters for steganalysis is computationally prohibitive. We argue that future steganalysis methods should focus on specific statistical deviations within the Top-$K$ significant parameters rather than global parameter scans.

{\bf Parameter reconstruction.} 
Additionally, we eliminate payloads via the random bit substitution \cite{Dubin2023-Access}, where the least significant bits of all parameters are replaced with random bits. Still, effective elimination against MEASER proves detrimental to model utility. As indicated in Figure \ref{fig:PAI_general}, the model performance degrades sharply when the modification depth reaches a critical threshold. Since MEASER incorporates robust error correction (i.e., LDPC) and spread spectrum modulation, neutralizing the embedded payload requires aggressive perturbations that exceed this tolerance threshold. In other words, random bit substitution cannot eliminate the robust payloads of MEASER without rendering the open-source LLMs useless.

{\bf Trigger detection.} Regarding the payload activation phase, we evaluated our constructed triggers ({\bf Algorithm \ref{alg:stealthy_trigger}}) against four prevailing detection tools (as shown in Table tab:results). Empirical results demonstrate that our triggers successfully circumvent these detectors. The ineffectiveness of these tools stems from their reliance on static analysis and pattern matching of known malicious signatures. By decomposing the trigger logic into low-level opcodes and attributes, MEASER avoids explicit signatures. To address this gap, future defenses must evolve towards runtime behavior monitoring, such as leveraging eBPF\footnote{\url{https://protectai.com/blog/why-ebpf-is-secure}.} techniques to intercept malicious system calls during the model deserialization and loading process.

\section{Conclusion}
In this paper, we systematically formalize the threat model of MEAs against open-source LLMs, and propose the first MEA MEASER. MEASER enhances the attack robustness against quantization and fine-tuning through robust payload encoding and Magnitude-Adaptive Relative Quantization Index Modulation mechanism, while achieves stealthiness based on a performance-aware importance metric. Through extensive experiments, MEASER outperforms existing MEAs on general DNNs in terms of the bit error rate and stealth rate. We further discuss possible defenses against MEASER, and call for urgent investigation into powerful countermeasures in the future.


\newpage

\section*{Biography Section}
If you have an EPS/PDF photo (graphicx package needed), extra braces are
 needed around the contents of the optional argument to biography to prevent
 the LaTeX parser from getting confused when it sees the complicated
 $\backslash${\tt{includegraphics}} command within an optional argument. (You can create
 your own custom macro containing the $\backslash${\tt{includegraphics}} command to make things
 simpler here.)
 
\vspace{11pt}

\bf{If you include a photo:}\vspace{-33pt}
\begin{IEEEbiography}[{\includegraphics[width=1in,height=1.25in,clip,keepaspectratio]{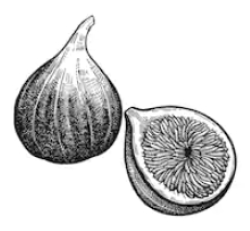}}]{Michael Shell}
Use $\backslash${\tt{begin\{IEEEbiography\}}} and then for the 1st argument use $\backslash${\tt{includegraphics}} to declare and link the author photo.
Use the author name as the 3rd argument followed by the biography text.
\end{IEEEbiography}

\vspace{11pt}

\bf{If you will not include a photo:}\vspace{-33pt}
\begin{IEEEbiographynophoto}{John Doe}
Use $\backslash${\tt{begin\{IEEEbiographynophoto\}}} and the author name as the argument followed by the biography text.
\end{IEEEbiographynophoto}

\vfill

\end{document}